\documentclass[preprint,review,12pt,compress]{elsarticle}

\usepackage{amssymb}
\usepackage{xcolor}
\usepackage{longtable,booktabs}
\usepackage{geometry}

\usepackage{graphicx}
\usepackage{subcaption}
\usepackage{wrapfig}
\usepackage{float}
\usepackage{placeins}
\usepackage{amsmath}
\usepackage{lipsum}
\usepackage{breakurl}
\usepackage{multicol}
\usepackage{multirow}
\usepackage{url}
\usepackage[breaklinks,colorlinks,
citecolor=blue,
urlcolor=blue,
pdfsubject={LaTeX}]{hyperref}

\usepackage{soul}

\makeatletter
\renewcommand{\MaketitleBox}{%
  \resetTitleCounters
  \def\baselinestretch{1}%
  \begin{center}
    \def\baselinestretch{1}%
    \Large \@title \par
    \vskip 18pt
    \normalsize\elsauthors \par
    \vskip 10pt
    \footnotesize \itshape \elsaddress \par
  \end{center}
  \vskip 12pt
}
\makeatother

\begin{document}

\journal{}

\begin{frontmatter}



\title{Phase-field modeling of radiation-induced composition redistribution: An application to additively manufactured austenitic Fe-Cr-Ni \tnoteref{footnote}}

\author[inst1]{Sourabh B. Kadambi\corref{cor1}}
\author[inst1]{Daniel Schwen}
\author[inst1]{Jia-Hong Ke}
\author[inst1,inst2]{Lingfeng He}
\author[inst1]{Andrea Jokisaari}

\cortext[cor1]{Email: SourabhBhagwan.Kadambi@inl.gov}
\tnotetext[footnote]{Published version: \href{https://doi.org/10.1016/j.commatsci.2025.113895}{https://doi.org/10.1016/j.commatsci.2025.113895}}

\affiliation[inst1]{organization={Idaho National Laboratory},
            city={Idaho Falls},
            state={Idaho},
            country={USA}}
            
\affiliation[inst2]{organization={North Carolina State University},
            city={Raleigh},
            state={North Carolina},
            country={USA}}
            

\begin{abstract}

Multicomponent alloys undergoing irradiation damage develop radiation-induced composition redistribution at point defect sinks such as grain boundaries (GBs) and dislocations.
Such redistribution results in undesired changes to their mechanical behavior and corrosion resistance.
Additively manufactured alloys proposed for future nuclear applications are expected to demonstrate a distinct response to irradiation owing to their unique microstructure with as-solidified dislocation density and chemical microsegregation.  
To capture the composition redistribution in such systems, we develop a mesoscale model with coupled evolution of atomic and point defect components in the presence of dislocation density, dislocation heterogeneity, and thermodynamic interactions at the GB.
The model is parameterized for an FCC Fe-Cr-Ni alloy as a representative system for austenitic stainless steels, and simulations are performed in 1D and 2D as a function of irradiation temperature, dose, dislocation density, and grain size.
Radiation-induced segregation (RIS) characterized by Cr depletion and Ni enrichment is predicted at both the GB and the dislocation cell wall, with RIS being lower in magnitude but wider at the cell wall.
Strongly biased absorption of self-interstitials by dislocations is found to suppress Ni enrichment but slightly enhance Cr depletion under certain conditions.
Thermodynamic segregation at the GB is predicted to be narrower and opposite in sign to RIS for both Cr and Ni. 
Importantly, non-monotonic segregation is found to occur when both thermodynamic and RIS mechanisms are considered, providing a novel physical interpretation of experimental observations.
The model is expected to serve as a key tool in accelerated qualification of irradiated materials.

\end{abstract}





\end{frontmatter}

\section{Introduction}

Austenitic stainless steels (SSs) serve as structural materials in the core and cladding of nuclear fission reactors.
On exposure to radiation, these alloys undergo atomistic damage that ultimately leads to mesoscale and macroscale effects such as radiation-induced segregation (RIS), radiation-enhanced precipitation, and void swelling.
RIS is the local enhancement or depletion of certain elements as a result of irradiation and manifests at extended defects such as grain boundaries (GBs) and dislocations.  RIS adversely affects mechanical properties and resistance to irradiation-assisted stress corrosion cracking  \cite{zinkle2009structural,zinkle2013materials}.
In addition to RIS, thermodynamic segregation (TS) can occur at extended defects under accelerated kinetics due to radiation-enhanced diffusion~\cite{nastar2005segregation}---we will refer to this as radiation-enhanced TS (RETS). TS can occur in isothermal conditions (as opposed to being driven by thermal gradients).
TS at GBs can potentially lead to deleterious effects (embrittlement and sensitization) or advantageous effects (coarsening resistance) depending on the segregating elements and their effect on GB properties such as cohesion, GB energy and mobility~\cite{raabe2014grain}.
Thus, it is crucial to understand the underlying mechanisms that lead to segregation as a function of temperature and irradiation conditions. 

With significant progress being made in additive manufacturing technologies, additively manufactured (AM) austenitic SSs are being considered as prime candidates for future applications in advanced reactors \cite{li2022advanced}. Additive manufacturing can enable new technologies to be deployed and reduce supply-chain burdens on reactor components \cite{li2022advanced}. While AM austenitic SSs have demonstrated promising strength and ductility \cite{byun2021mechanical}, their responses under irradiation conditions are still not well characterized \cite{byun2021mechanical_rep}.
 The microstructure of AM austenitic SSs is heavily dependent upon the build method,  processing parameters, specimen geometry, and post-build heat treatments.  However, general features of AM austenitic SSs include columnar or otherwise non-equiaxed grains, a substructure of dislocation cells that may be equiaxed or columnar, chemical microsegregation at the dislocation cell walls (CWs) and GBs, and chemical segregation along melt pool boundaries \cite{bertsch2020origin}.  The dislocation cell structure consists of closely packed cells on the order of a micron or less in diameter within the grains; the CWs consist of densely tangled dislocations and nanoscale oxides \cite{bertsch2020origin}.  These unique microstructural features of AM austenitic SSs  give rise to a distinct radiation damage response in comparison to conventionally manufactured austenitic SSs \cite{shang2021heavy,chen2024situ}. For example, ion irradiation of AM 316 SS can result in heterogeneous void formation, with voids forming near CWs and not within the center of the cell \cite{chen2024situ}.  Radiation-induced segregation is also observed, with stronger segregation occurring on GBs and lesser, but evident, segregation occurring at CWs \cite{chen2024situ}.  
 
The mechanisms behind RIS in conventional alloys have received considerable experimental and theoretical attention \cite{was2007fundamentals,nastar20121,ardell2016radiation}.
Irradiation displacement cascades produce supersaturated concentrations of vacancies and self-interstitial atoms (SIAs).
While a number of these point defects annihilate or form clusters, the remaining mobile point defects are transported and absorbed (in either a biased or unbiased manner) at dislocations, voids, and GBs.
Owing to atom-vacancy exchange, solute-vacancy drag, and dumbbell migration, absorption of point defects at sinks is accompanied by a flux of atoms.
RIS consequently manifests from the unequal fluxes of different elements to the sinks.
In austenitic SSs, the faster diffusivity of Cr relative to Fe and Ni via the atom-vacancy exchange mechanism has been widely recognized as the reason for Cr depletion at GBs~\cite{was2011assessment}.
However, other potential contributing factors such as vacancy production bias and SIA absorption bias of dislocations have received considerably less attention.
While RIS manifests under non-equilibrium driving forces induced by radiation, TS can be observed under thermal conditions because it is driven by more favorable energetics of certain elements at the GB versus within the bulk~\cite{seah1973grain,raabe2014grain}.
A variety of repulsive and attractive interactions between alloying elements at the GB and within the bulk can lead to TS in multicomponent alloys~\cite{xing2018solute}.
In addition, radiation-enhanced diffusion due to supersaturated point defect concentrations accelerates the kinetics of TS under irradiation, resulting in RETS.
Pre-irradiation segregation (presumably TS) with Cr enrichment has been reported in austenitic SSs~\cite{kenik1998origin,busby1998influence,cole2002influence,li2013atomic}.
Under irradiation, ``W"-shaped Cr segregation profiles have been observed with Cr enrichment at the GB and Cr depletion adjacent to the GB~\cite{kenik1998origin,busby1998influence,was2002emulation,barr2018observation}.
While the latter is attributed to RIS, the former has been speculated to be due to TS, possibly resulting from complex interactions of Cr with other alloying elements at the GB. 
Thus, the exact mechanism causing non-monotonic segregation profiles under irradiation are still under debate.

Classical sharp-interface models typically employ a Dirichlet boundary (fixed at the thermal equilibrium concentration) to describe an ideal sink behavior of a high-angle GB---or more realistically, of a dislocation core.
These simulations are also performed in a 1D setting and are thus restricted to ideal GBs and systems with simple spatial variations in the sink density.
More sophisticated models are therefore needed to describe microstructural effects and realistic GBs, especially relevant for AM austenitic SSs.
Several sharp-interface models have been developed for RIS in austenitic SSs and austenitic Ni-Cr and ferritic Fe-Cr steels.
Allen and Was \cite{allen1998modeling} described composition-dependent vacancy migration energies for FCC Fe-Cr-Ni, and compared the temperature and dose predictions against proton irradiation data.
The effect of Zr and Hf on mitigating RIS in austenitic SS via solute-vacancy trapping was described by Hackett et al.~\cite{hackett2008mechanism}.
Duh et al.~\cite{duh2001numerical} and Field et al.~\cite{field2015defect} developed flux boundary conditions to simulate non-ideal sink behavior of symmetric tilt GBs in 304 austenitic SS.
Yang et al.~\cite{yang2016roles} and Nastar et al.~\cite{nastar1997role} used experimental tracer diffusivity data to incorporate composition-dependent activation energies and frequency factors.
They also incorporated non-ideal thermodynamic factors and preferential solute-SIA binding to more accurately capture the contributions to RIS.
To describe RIS in Ni-Cr, Barnard et al.~\cite{barnard2012modeling} employed ab initio parameters and compared the individual contributions of vacancy and SIA mechanisms.
For the same system, Ozturk et al.~\cite{ozturk2021surface} demonstrated that the effect of production bias can lead to non-monotonic ``W"-shaped RIS profiles. 
For ferritic Fe-Cr alloys, Wharry et al.~\cite{wharry2014mechanism} showed that Cr is depleted near GBs at high temperatures as a result of Cr-vacancy exchange transport, but is enriched at low temperatures due to dominant Cr-SIA binding (the crossover in segregation occurs anywhere between 400\textdegree C to 800 \textdegree C depending on alloy composition and point defect energetics).
The above models, though implemented in 1D, provide the essential framework needed to describe RIS.

In contrast to the sharp-interface models, phase-field (PF) models employ a diffuse interface description and enable spatially resolved microstructure modeling in 2D and 3D, in addition to 1D.
They also offer potential for simulating dynamic spatial evolution of the sink distribution in the microstructure.
While PF models have been developed to study irradiation effects such as void/gas-bubble evolution, precipitation, or phase separation, as well as point defect interaction with dislocations, RIS models have remained limited \cite{tonks2018apply,li2017review,thuinet2018multiscale}.
Piochaud~\cite{piochaud2016atomic} formulated a 1D BCC Fe-Cr model that relies on Onsager transport coefficients calculated from atomistic methods.
Recently, Rezwan et al.~\cite{rezwan2022effect} coupled the rate theory diffusion model of Allen and Was~\cite{allen1998modeling} to a polycrystal PF model in order to investigate the effects of concurrent grain growth and RIS in 2D.
These models also assumed the GB to be ideal, neglected the thermodynamic interactions of alloying elements with GBs and omitted the effect of dislocation bias on RIS.
In addition to the aforementioned limitations, most RIS models assume ideal alloy thermodynamics and kinetics. 
For a more accurate description of RIS mechanisms, formulations employing the complete Onsager transport relations ~\cite{barnard2012modeling,piochaud2016atomic,yang2016roles} are preferred, since they can describe solute-vacancy drag and enable parameterization from analytic and atomistic calculations of Onsager coefficients \cite{barbe2006self,trinkle2017automatic,schuler2020kineclue,dai2023radiation}.
Furthermore, to incorporate realistic alloy thermodynamics from Calculation of Phase Diagrams (CALPHAD) databases, transport relations must employ driving forces based on chemical or diffusion potential gradients~\cite{barnard2012modeling,piochaud2016atomic,yang2016roles}. 
This is also necessary to simulate TS at GBs. 
While the PF method naturally permits such descriptions and PF models for TS in polycrystals have been developed~\cite{cha2002phase,abdeljawad2015stabilization}, a combined description of radiation and thermodynamic effects on segregation has been lacking~\cite{nastar2005segregation}.

In this paper, we develop a multi-order-parameter, multicomponent PF model for GB segregation (RIS and RETS) in polycrystals under irradiation.
To model RIS at the dislocation CWs of the AM microstructure, we leverage the multicomponent RIS model developed for GB segregation, but reduce it (i.e., neglect dynamic evolution of the PF order parameters) to a simple diffusion rate theory model with spatial variation in the mean dislocation density describing the AM cell structure.
The model formulations are presented in Sec.~\ref{sec:model}, with detailed derivations of the equations being provided in the Supplementary Material.
In Sec.~\ref{sec:param}, we parameterize the models for a ternary component FCC Fe-Cr-Ni alloy as a representative system for austenitic SS.
Sec.~\ref{sec:impl:num} details the numerical implementations of the PF model for GB segregation (RIS and RETS), the sharp-interface model for GB RIS and the spatially resolved RIS model for CW segregation. 
In \ref{appendix_A} and \ref{appendix_B}, details and simulations of the sharp-interface RIS models are presented for verification and parameterization of the PF model.
In Sec.~\ref{sec:results}, the contributions of RETS and RIS to GB segregation are assessed. Following this, the results of GB RIS from 1D and 2D simulations are presented as a function of temperature, grain size, and dislocation density. 
Finally, 1D and 2D simulations of RIS to dislocation CWs are presented, then compared with those of GBs.
In Sec.~\ref{sec:discussion}, we discuss the relevance, limitations and future scope of the model formulation and the simulation results.
The work is summarized in Sec.~\ref{sec:conclusions}.

\section{Modeling framework} \label{sec:model}

We treat the polycrystal and dislocation cell structures separately. Within the PF approach, a polycrystal (or dislocation cell structure) of $N$ grains (or dislocation cells) is represented by a set of order parameters $\boldsymbol{\eta}:=\left\{\eta_{1},\eta_{2},\dots,\eta_{N}\right\}$. 
An individual grain $i$ is defined by $\eta_{i} = 1$ and $\eta_{j}=0$ ($\forall ~j\neq i$), while the defected region (GB or dislocation CW) between two adjacent grain regions $i$ and $j$ is given by a smooth variation in $\eta_i$ from $0$ (in grain $j$) to $1$ (in grain $i$), with $\eta_j=1-\eta_i$ due to symmetry.
For the polycrystal, the model is rigorously developed in Sec.~\ref{sec:model:pf_gb} by using the PF approach and will be applicable to GB movement and grain growth.
For the latter, in Sec.~\ref{sec:model:pf_am} we assume the dislocation cell structure to be static in terms of dislocation density and spatial distribution, thus enabling development of a reduced model that does not require dynamic evolution of the PF order parameters.

To describe the flux coupling between atoms ($\phi=1,2,\dots,K$) and point defects (${{\upsilon}}=V,I$), we follow linear irreversible thermodynamics.
The partial fluxes of alloy elements are given in terms of the chemical potential gradients of all components, as \cite{wolfer1983drift}:
\begin{flalign}
    \boldsymbol{J}^{{\upsilon}}_{\phi} &= - \sum^K_{k = 1} L^{{\upsilon}}_{\phi k} \nabla \left(\mu_k + \text{sign}({{\upsilon}}) \mu_{{\upsilon}} \right), && 
\label{eq:partial_flux}
\end{flalign}
where $K$ is the total number of atomic components, $\text{sign}(V)=-1$ for the vacancy ($V$) and $\text{sign}(I)=+1$ for the SIA ($I$, which can be any atom $\phi$). Due to conservation, the total fluxes are given by $\boldsymbol{J}_\phi = \sum_{{\upsilon}} J^{{\upsilon}}_\phi$ and $\boldsymbol{J}_{{\upsilon}} = \text{sign}({{\upsilon}}) \sum^K_{k=1} \boldsymbol{J}^{{\upsilon}}_k$.
From the Onsager reciprocal relations, the rest of the transport coefficients are given by $L_{\phi {{\upsilon}}} = \text{sign}({{\upsilon}}) \sum^K_{k=1} L^{{\upsilon}}_{\phi k}$ and $L_{{{\upsilon}}{{\upsilon}}} = \sum^K_{k=1}\sum^K_{l=1} L^{{\upsilon}}_{kl}$. 
Similar to Piochaud et al.'s treatment for a binary alloy~\cite{piochaud2016atomic}, we reformulate the total fluxes in terms of the diffusion potentials ($\mu_{k1} = \mu_k - \mu_1$) by invoking the Gibbs-Duhem relation $\sum^K_{k=1} c_k \nabla \mu_k + \sum_{{\upsilon}} c_{{\upsilon}} \nabla \mu_{{\upsilon}} = 0$ and considering the point defect concentrations in their dilute limits.
Here, the lattice site concentrations are related as $c_1=1-\sum^K_{k=2} c_k + \sum_{{\upsilon}} \text{sign}({{\upsilon}}) c_{{\upsilon}} \approx 1-\sum^K_{k=2} c_k$, where $1$ refers to the solvent atom, with the rest being solute atoms (here, $c_k$ counts both the on-lattice atoms and the off-lattice SIA atoms).
Thus (see Sec.~{\ref{sm:fluxes}} in the Supplementary Material):
\begin{subequations}
\begin{flalign}
    \boldsymbol{J}_{\phi} &= - \sum^K_{k = 2} L^1_{\phi k} \nabla \mu_{k 1} - \sum_{{{\upsilon}}=V,I} L_{\phi {{\upsilon}}} \nabla \mu_{{\upsilon}}, \\
    \boldsymbol{J}_{{{\upsilon}}} &= - \sum^K_{k = 2} L^1_{{{\upsilon}} k} \nabla \mu_{k 1} - L_{{{\upsilon}} {{\upsilon}}} \nabla \mu_{{\upsilon}}, &&
\end{flalign}
\label{eq:total_flux}
\end{subequations}
where the relative transport coefficients with reference to the solvent $1$ are given by:
\begin{subequations}
\begin{flalign}
    &L^1_{{{\upsilon}} k} = \text{sign}({{\upsilon}}) \left( \sum^K_{j=1} L^{{\upsilon}}_{k j} - c_k L_{{{\upsilon}}{{\upsilon}}}\right), \\ 
    &L^1_{\phi k} = \sum_{{{\upsilon}}} L^{{\upsilon}}_{\phi k} - \, c_k \sum_{{{\upsilon}}} \text{sign}({{\upsilon}}) L_{\phi {{\upsilon}}}. &&
\end{flalign}
\end{subequations}
With Eq.~2, we reduce the number of species to be tracked from $K+2$ to $K+1$. 
The time evolution of the concentrations under irradiation can be written in a general form as:
\begin{subequations} \label{eq:conc-flux}
\begin{flalign} \label{eq:conc-flux_atom}
    &\frac{\partial c_\phi}{\partial t} = -\nabla \cdot \boldsymbol{J}_\phi, \\ \label{eq:conc-flux_defect}
    &\frac{\partial c_{{\upsilon}}}{\partial t} = -\nabla \cdot \boldsymbol{J}_{{\upsilon}} + P_{{\upsilon}} - R_{VI}c_Vc_I - k^2_{{\upsilon}} D_{{\upsilon}}(c_{{\upsilon}}-c^e_{{\upsilon}}), && 
\end{flalign}
\end{subequations}
where $D_{{\upsilon}} = L_{{{\upsilon}}{{\upsilon}}}\theta_{{{\upsilon}}{{\upsilon}}}$ represent the point defect diffusivities{, with $\theta_{\upsilon \upsilon}$ being the thermodynamic factor}. {$c^e_\upsilon=\exp{\left(\frac{-E^f_\upsilon+S^f_\upsilon T}{k_BT}\right)}$ is the point defect concentration at thermal equilibrium, with $E^f_\upsilon$ being the formation energy and $S^f_\upsilon$ being the formation entropy.}
In Eq.~\ref{eq:conc-flux_defect}, the non-conserved mean-field rate terms are the point defect production rate $P_{{\upsilon}}$, recombination reaction rate $R_{VI}$, and sink strength $k^2_{{\upsilon}}$. 
Coupling of the above with the PF microstructure for polycrystals and dislocation cells is presented in the subsequent sections.

\subsection{PF model for GB segregation} \label{sec:model:pf_gb}

\begin{figure*}[htp!] 
\centering
    \begin{subfigure}[t]{0.9\textwidth}
        \includegraphics[width=1\textwidth]{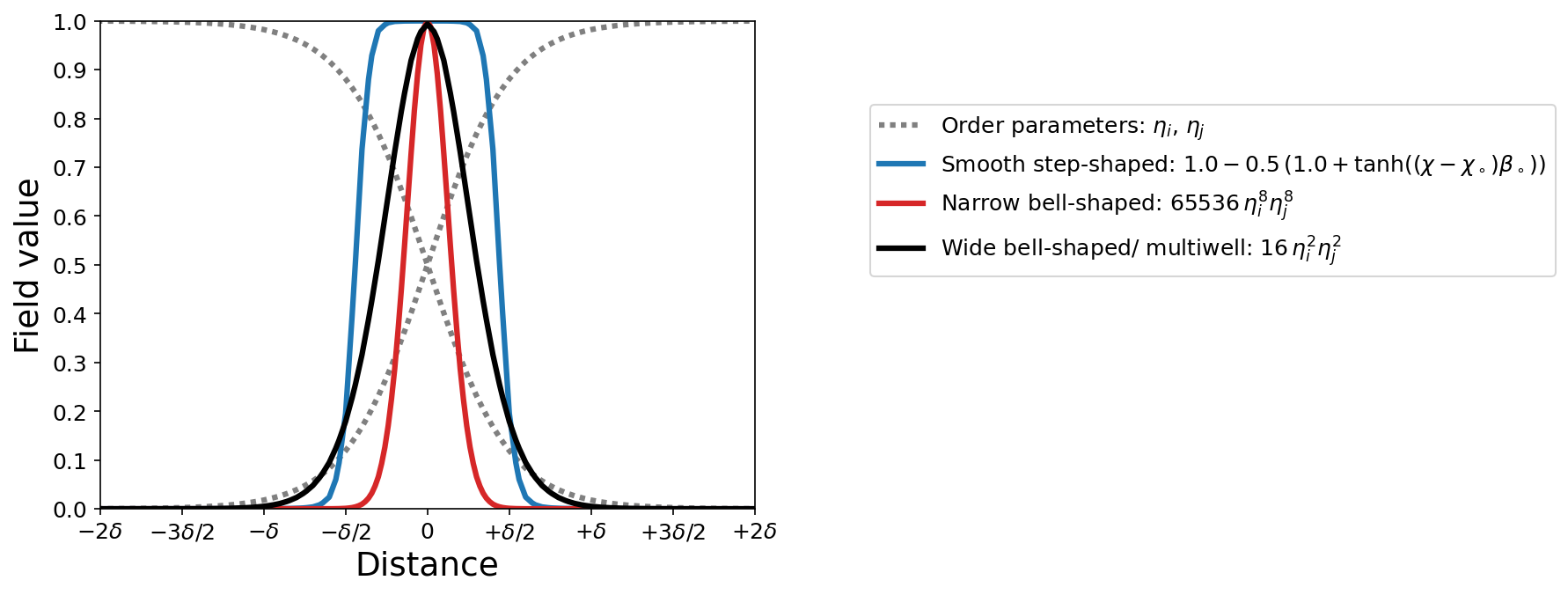}
    \end{subfigure}
    \caption{PF order parameters and functions across the interface (GB in a polycrystal and CW in an AM cell structure). All the three functions are tested as GB sink functions $g_\text{sink}$ in \ref{appendix_A} and \ref{appendix_B}. The narrow bell-shaped (red) function is selected as $g_\text{sink}$ for all results of GB RIS. The smooth step-shaped (blue) function describes the dislocation density at CWs in the AM microstructure. The wide bell-shaped or normalized multiwell function $\bar{g}_\text{mw}$ describes the thermodynamic properties of the GB.}
    \label{fig:pf_functions}
\end{figure*}

We adapt the grand potential approach originally proposed for binary alloy solidification by Plapp~\cite{plapp2011unified}, along with the extension to multicomponent and {multiphase} systems, as proposed by Aagesen et al.~\cite{aagesen2018grand}.
The isothermal grand potential functional for the polycrystalline microstructure is written as:
\begin{flalign} \label{eq:GP_functional}
    {\Omega}
    = \int_V
    \left[\omega^b + (m_0 + \omega^g-\omega^b)\bar{g}_{\text{mw}} + \sum^N_{n=1} \frac{\kappa}{2} \left|\nabla \eta_n \right|^2 \right]
 d V, &&
\end{flalign}
\noindent where the grand potential densities for the bulk ($b$) and GB ($g$) are given by $\omega^\psi = f^\psi - \sum^K_{k=2} c^\psi_k \mu_{k 1} - \sum_{{\upsilon}} c_{{\upsilon}} \mu_{{\upsilon}}$ ($\psi= b$ or $g$). $\omega^g-\omega^b$ is the excess grand potential density in the GB region, {$\bar{g}_\text{mw}$ is a normalized multiwell potential,} $m_0$ is an external barrier height for the GB, and $\kappa$ is the gradient energy coefficient for the order parameter fields.
For simplicity, we only consider distinct energetics (at the GB relative to the bulk) for the alloy component, and ignore distinct energetics of the point defects.
While the second term in the functional penalizes large GB widths, the last term penalizes small GB widths.
We only consider gradient energy contributions for the order parameter fields, and ignore their contributions for the concentration fields~\cite{aagesen2018grand}.
{$\bar{g}_\text{mw}$ is given by $8g_\text{mw}$} (see Fig.~\ref{fig:pf_functions}), where $g_\text{mw}$ is the standard multiwell potential employed in existing PF models of grain growth \cite{moelans2008aniso-grain,aagesen2018grand}:
\begin{flalign} \label{eq:std_gmw}
g_\text{mw} = \frac{1}{4} + \sum^{N}_{i=1}\left(\frac{\eta^4_{i}}{4} - \frac{\eta^2_{i}}{2} \right) + 1.5 \sum^{N}_{i=1}\sum_{j\neq i} \eta^2_{i}\eta^2_{j}. &&
\end{flalign}
We approximate the free energy density of the alloy by using a multivariate Taylor expansion truncated to the second order \cite{choudhury2015method}.
The alloy and (dilute) point defect free energy density are together given by:
\begin{flalign} \label{eq:fe_taylor} \nonumber
    f^\psi =& f^{\psi,\circ} + \sum^{N}_{k=2} \mu^{\psi,\circ}_{k 1}(c^\psi_k - c^{\psi,\circ}_k) + \sum^K_{k=2} \frac{\theta^{\psi,\circ}_{kk}}{2}(c^\psi_k - c^{\psi,\circ}_k)^2 \\
    &+ \sum^K_{k=2}\sum_{l \neq k} \frac{\theta^{\psi,\circ}_{k l}}{2}(c^\psi_k - c^{\psi,\circ}_k)(c^\psi_l - c^{\psi,\circ}_l) + \frac{RT}{V_m} \sum_{{\upsilon}} c_{{\upsilon}} \ln{\left(\frac{c_{{\upsilon}}}{c^e_{{\upsilon}}}\right)}. &&
\end{flalign} 
\noindent Here, the terms denoted by $\circ$ are obtained from the CALPHAD free energy $f_\text{C}$ (see Sec.{~\ref{sm:gb_fe}} in the Supplementary Material) at the equilibrium (between $b$ and $g$) composition or the nominal composition (when $b$ and $g$ are identical in free energy).
The diffusion potentials and thermodynamic factors specific to $\psi$ (superscript is omitted in the following for clarity) are given by $\mu^\circ_{k1}= \left.{\partial f_\text{C}}/{\partial c_k} \right|_{\mathbf{c}^\circ}$ and $\theta^\circ_{kl} = \left. {\partial^2 f_\text{C}}/{\partial c_{k} \partial c_{l}} \right|_{\mathbf{c}^\circ}$, respectively.
$V_m$ is the molar volume of the lattice site, which is assumed constant across the system.
The local concentrations at any point in the system are obtained as:
\begin{subequations} \label{eq:tot_conc}
\begin{flalign}
    c_{k} &= -\frac{\delta \Omega}{\delta \mu_{k}} = -\frac{\partial \omega^b}{\partial \mu_{k1}} + \left(-\frac{\partial \omega^g}{\partial \mu_{k1}} + \frac{\partial \omega^b}{\partial \mu_{k1}} \right)\bar{g}_{\text{mw}} = c^b_{k} + \left(c^g_{k} - c^b_{k} \right)\bar{g}_{\text{mw}}, \label{eq:tot_conc_atomic} \\ 
    c_{{{\upsilon}}} &= -\frac{\delta \Omega}{\delta \mu_{{{\upsilon}}}} = -\frac{\partial \omega^{b}}{\partial \mu_{{\upsilon}}} = -\frac{\partial \omega^{g}}{\partial \mu_{{\upsilon}}}. \label{eq:tot_conc_defect} &&
\end{flalign} 
\end{subequations}
By inverting the expressions $\mu_{\phi1}=\partial f / \partial c_\phi$ and $\mu_{{{\upsilon}}}=\partial f / \partial c_{{\upsilon}}$ obtained from Eq.~\ref{eq:fe_taylor}, the phase concentrations can be expressed in terms of the diffusion potentials as:
\begin{subequations}
\begin{flalign}
    c^\psi_k &= c^{\psi,\circ}_k + \sum^K_{j=2} \chi^\psi_{k j} (\mu_{j1} - \mu^{\psi,\circ}_{j1}), \\
    c^\psi_{{{\upsilon}}} &= c_{{\upsilon}} = c^{e}_{{{\upsilon}}}\exp{\left(\frac{V_m}{RT}\mu_{{\upsilon}} \right)}, &&
\label{eq:phase_conc}
\end{flalign}
\end{subequations}
\noindent where $\boldsymbol{\chi}={\boldsymbol{\theta}}^{-1}$ is the chemical susceptibility matrix for the alloy.

The time evolution equations in terms of the PF concentration variables, diffusion potentials, and order parameters can be expressed after substituting Eq.~\ref{eq:total_flux} in Eq.~\ref{eq:conc-flux} as:
\begin{subequations}
\begin{flalign}
    &\frac{\partial c_\phi}{\partial t} = \nabla \cdot \left(\sum^K_{k = 2} L^1_{\phi k} \nabla \mu_{k1} + \sum_{{{\upsilon}}=V,I} L_{\phi {{\upsilon}}} \nabla \mu_{{\upsilon}} \right), \\
    &\frac{\partial c_{{\upsilon}}}{\partial t} = \nabla \cdot \left(\sum^K_{k = 2} L^1_{{{\upsilon}} k} \nabla \mu_{k1} + L_{{{\upsilon}} {{\upsilon}}} \nabla \mu_{{\upsilon}} \right), \\
    &\hspace{1cm}+ P_{{\upsilon}} - R_{VI}c_Vc_I - k^2_{{{\upsilon}},b} D_{{\upsilon}} (c_{{\upsilon}} - c^e_{{\upsilon}}) - k^2_g D_{{\upsilon}} (c_{{\upsilon}} - c^e_{{\upsilon}}) g_\text{sink}(\boldsymbol{\eta}), &&
\end{flalign}
\label{eq:conc_time_evol}
\end{subequations}
where $k^2_{{{\upsilon}},b}$ is the mean-field sink strength of dislocations in the bulk and is distinct for vacancies and SIAs. 
$k^2_g$ is the local sink strength of the GB acting in the GB region that is identified by $g_\text{sink}$. 
Thus, $k^2_g$ must be determined in conjunction with $g_\text{sink}$.
In \ref{appendix_A} and \ref{appendix_B}, we test three different functions (see Fig.~\ref{fig:pf_functions}) for $g_\text{sink}$ and compare the results of point defect and atomic concentrations from the PF implementation against that of the sharp-interface implementations of GB sinks.
Based on these results, we propose $g_\text{sink}$ as:
\begin{flalign} \label{eq:gsink_narrow_bell}
    g_\text{sink} = 65536 \sum^N_{i=1} \sum_{j>i} \eta^8_i \eta^8_j, && 
\end{flalign}
where the prefactor ensures that $g_\text{sink}=1$ at the GB center (see Fig.~\ref{fig:pf_functions}).
Eqs.~\ref{eq:conc_time_evol} entail both concentration and diffusion potential variables and constitute a mixed formulation~\cite{boutin2022grand} of the grand potential model that can be solved with the constraints in Eq.~\ref{eq:tot_conc}.
With $c:=c(\boldsymbol{\mu},\boldsymbol{\eta})$, the chain rule for differentiation can be employed to reformulate $\partial c(\boldsymbol{\mu},\boldsymbol{\eta})/ \partial t$ in terms of $\partial {\mu}/ \partial t$, or $\nabla \mu$ in terms of $\nabla c$. 
In the former approach, which was employed in the original grand potential formulations ~\cite{plapp2011unified,aagesen2018grand}, the concentration variables are eliminated and Eq.~\ref{eq:conc_time_evol} is only solved for the diffusion potentials.
This approach has been found to not adhere strictly to mass conservation if time integration is performed over large time steps.
This issue is avoided by directly solving the mixed formulation presented above, or by adopting the latter approach ~\cite{chatterjee2021grand,boutin2022grand}, in which the diffusion potential gradients are expanded (see Sec.~{\ref{sm:pf_gb}} in the Supplementary Material) to yield:
\begin{subequations}
\begin{flalign} \label{eq:alt_time_evol_atom} \nonumber
    \frac{\partial c_\phi}{\partial t} =& \nabla \cdot \left(\sum^K_{k = 2} \sum^K_{j=2} L^1_{\phi j} \theta_{j k} \nabla c_k +  \sum_{{{\upsilon}} = V,I} L_{\phi {{\upsilon}}} \theta_{{{\upsilon}} {{\upsilon}}} \nabla c_{{\upsilon}} \right. \\
    &\hspace{1cm}- \left. \sum^K_{k = 2} \sum^K_{j=2} \sum^N_{n=1} L^1_{\phi j} \theta_{j k} (c^g_k - c^b_k) \frac{d\bar{g}_\text{mw}}{d\eta_n} \nabla \eta_n \right), \\ \label{eq:alt_time_evol_defect} \nonumber
    \frac{\partial c_{{\upsilon}}}{\partial t} =& \nabla \cdot \left(\sum^K_{k = 2} \sum^K_{j=2} L^1_{{{\upsilon}} j} \theta_{j k} \nabla c_k + L_{{{\upsilon}} {{\upsilon}}} \theta_{{{\upsilon}}{{\upsilon}}} \nabla c_{{\upsilon}} \right) \\
    &+ P_{{\upsilon}} - R_{VI}c_Vc_I - k^2_{{{\upsilon}},b} D_{{\upsilon}} (c_{{\upsilon}} - c^e_{{\upsilon}}) - k^2_g D_{{\upsilon}} (c_{{\upsilon}} - c^e_{{\upsilon}}) g_\text{sink}(\boldsymbol{\eta}). &&
\end{flalign}
\label{eq:alt_time_evol}
\end{subequations}
\noindent 
Finally, the time evolution equations for the order parameters are given by:
\begin{flalign} \label{eq:eta_evol}
    \frac{\partial \eta_n}{\partial t} = -L_\eta\, \frac{\delta \Omega}{\delta \eta_n} = -L_\eta\, \left[(m_0 + \omega^g-\omega^b)\frac{d\bar{g}_\text{mw}}{d\eta_n} - \sum^N_{n=1} \kappa\nabla^2\eta_n \right], &&
\end{flalign}
\noindent where $L_\eta$ is the GB mobility.

\subsection{Spatially resolved model for CW segregation} \label{sec:model:pf_am}

Dislocation CWs comprised of dislocation tangles in the AM microstructure cannot be treated as interfaces because they do not accommodate relative misorientation between the adjacent crystals. 
Therefore, conventional PF order parameter dynamics (as given by the Allen-Cahn equation in Eq.~\ref{eq:eta_evol}) are not expected to apply to the evolution of the dislocation cell structure.
Rather, the recovery of CW dislocations under thermal and irradiation conditions is expected to occur via annihilation of dislocation dipoles, as facilitated by point defect absorption/dislocation climb.
Although dislocation recovery is important to consider, it is beyond the scope of the present paper. 
In Ref.~\cite{jokisaari2024defect}, we report a preliminary formulation of a model that considers dislocation recovery.
In the present work, we only employ static PF order parameters to initialize and represent the dislocation CWs.

For the present task of describing RIS at the CWs, we propose a simple description with spatially varied but temporally static dislocation density and sink strength.
We also ignore thermodynamic interactions between the solutes and the dislocations at the CW. 
Therefore, considering only the bulk thermodynamics, Eq.~\ref{eq:alt_time_evol} reduces to the following diffusion and rate theory form:
\begin{subequations}
\begin{flalign} \label{eq:am_time_evol_atom}
    \frac{\partial c_\phi}{\partial t} =& \nabla \cdot \left(\sum^K_{k = 2} \sum^K_{j=2} L^1_{\phi j} \theta_{j k} \nabla c_k +  \sum_{{{\upsilon}} = V,I} L_{\phi {{\upsilon}}} \theta_{{{\upsilon}} {{\upsilon}}} \nabla c_{{\upsilon}} \right), \\ \nonumber \label{eq:am_time_evol_defect}
    \frac{\partial c_{{\upsilon}}}{\partial t} =& \nabla \cdot \left(\sum^K_{k = 2} \sum^K_{j=2} L^1_{{{\upsilon}} j} \theta_{j k} \nabla c_k + L_{{{\upsilon}} {{\upsilon}}} \theta_{{{\upsilon}}{{\upsilon}}} \nabla c_{{\upsilon}} \right) \\
    &+ P_{{\upsilon}} - R_{VI}c_Vc_I - k^2_{{{\upsilon}},b}(\rho) D_{{\upsilon}} (c_{{\upsilon}} - c^e_{{\upsilon}}), &&
\end{flalign}
\label{eq:am_time_evol}
\end{subequations}
where $k^2_{{{\upsilon}},b}(\rho)$ is the spatially varying sink strength.
The distribution in dislocation density and initial concentration (AM microsegregation) are defined as:
\begin{flalign} \label{eq:rho_spatial_dep}
    &\rho_b = \rho_{b,c} + \left(\rho_{b,w}-\rho_{b,c} \right) g_\text{cw}, \\
    &c = c_{b,c} + \left(c_{b,w}-c_{b,c}\right) g_\text{cw}, &&
\end{flalign}
where $g_\text{cw}$ is a function of auxiliary PF order parameters given by:
\begin{flalign} 
    &g_\text{cw} = 1 - \frac{1}{2} \left[1 + \tanh \left(\left(\chi_\text{bnds} - \chi^\circ_\text{bnds} \right) \beta_\circ \right) \right], &&
\end{flalign}
where $\chi_\text{bnds} = \sum_i \eta^2_i$, $\chi^\circ_\text{bnds} = 0.75$, and $\beta_\circ = \frac{2 \tanh^{-1}(0.8)}{0.11}$ \cite{simon2022mechanistic}.
Note that with this approach, we simply use the PF order parameters as auxiliary variables to initialize the required microstructure and CW widths.
To ensure the required CW width is realized, Eq.~\ref{eq:eta_evol} is solved for a few time steps to evolve the auxiliary PF order parameters so that the accurate width is achieved.
The functional forms of $g_\text{sink}$ and $g_\text{cw}$ are described in Sec.~\ref{sec:impl:num} and are plotted in Figs.~\ref{fig:pf_functions} and~\ref{fig:2D_micro}.

\section{Parameterization} \label{sec:param}

The models presented in Sec.~\ref{sec:model} are parameterized for the ternary FCC Fe-17Cr-12Ni system.
Because Fe is the solvent, it is taken as the reference component 1; the solutes Cr and Ni are taken as components 2 and 3.
The key parameters are summarized in Table~\ref{tab:myfirstlongtable}.

\centerline{Table 1. Default model parameters.} 
\footnotesize{
\begin{longtable}{l l l l} 
\toprule
{\textbf{Symbol} }  &  {\textbf{Description} }  & { \textbf{Value} }  & {\textbf{Ref.}} \\ 
\midrule
{$c^\circ_{Cr}$} & {Nominal Cr concentration} & {0.17} & {}  \\
{$c^\circ_{Ni}$} & {Nominal Ni concentration} & {0.12} & {}  \\
{$T$} & {Temperature} & {773 K (500\,$^\circ$C)} & {}  \\
{$P_{{\upsilon}}$} & {Point defect production rate} & {$2\times10^{-6}$ dpa/s} & {}  \\
{$E^{f}_V$} & {Vacancy formation energy} & {1.8 eV} & {\cite{yang2016roles}}  \\
{$S^f_V, {S^f_I}$} & {Vacancy{, SIA} formation entropy} & { 2$k_B${, 0}} & {\cite{was2007fundamentals}}  \\
{{$Q_{Fe}$}} & {{Activation energy for vacancy diffusion of Fe}} & {{2.89 eV}} & {\cite{yang2016roles}}  \\
{{$Q_{Cr}$}} & {{Activation energy for vacancy diffusion of Cr}} & {{2.88 eV}} & {\cite{yang2016roles}}  \\
{{$Q_{Ni}$}} & {{Activation energy for vacancy diffusion of Ni}} & {{2.86 eV}} & {\cite{yang2016roles}}  \\
{{$P_{Fe}$}} & {{Frequency factor for vacancy diffusion of Fe}} & {{$3.8\times10^{13}$ nm$^2$/s}} & {\cite{yang2016roles}}  \\
{{$P_{Cr}$}} & {{Frequency factor for vacancy diffusion of Cr}} & {{$5.5\times10^{13}$ nm$^2$/s}} & {\cite{yang2016roles}}  \\
{{$P_{Ni}$}} & {{Frequency factor for vacancy diffusion of Ni}} & {{$1.5\times10^{13}$ nm$^2$/s}} & {\cite{yang2016roles}}  \\
{$E^f_I$} & {SIA formation energy} & {3.6 eV} & {\cite{allen1998modeling}}  \\
{$E^m_I$} & {SIA migration energy} & {0.6 eV} & {Based on \cite{yang2016roles}}  \\
{$E^b_{CrI}-E^b_{FeI}$} & {SIA binding energy for Cr} & {0.025 eV}  & {Based on \cite{yang2016roles}}  \\
{$E^b_{NiI}-E^b_{FeI}$} & {SIA binding energy for Ni} & {$-0.0075$ eV} & {Based on \cite{yang2016roles}}  \\
{$\omega^\circ_{kI}$} & {SIA jump frequency prefactor} & {$1.5\times10^{12}$ s$^{-1}$} & {\cite{allen1998modeling}}   \\
{$f_I$} & {Atom-SIA correlation factor} & {0.44} & {\cite{allen1998modeling}}\\
{$f_\circ$} & {FCC correlation factor} & {0.78} & {\cite{manning1971correlation}}  \\
{$a_\circ$} & {Lattice parameter}  & {$0.35$ nm} & {\cite{hackett2008mechanism}}   \\
{$b$} & {Burgers vector} & {0.25 nm} & {\cite{jourdan2015influence}}   \\ %
{$\lambda_I$} & {SIA jump distance}  & {$a_\circ/2$ nm} & {\cite{was2007fundamentals}}  \\
{$V_a$} & {Atomic volume} & {$a^3_\circ/4$} & {\cite{was2007fundamentals}}  \\
{$r_\circ$} & {Recombination radius} & {$2a_\circ$} & {\cite{was2007fundamentals}}  \\
{$Z_V$} & {Dislocation sink efficiency for vacancy} & {1} & {\cite{chang2013dislocation}}  \\
{$Z_I$} & {Dislocation sink efficiency for SIA} & {1.2} & {\cite{chang2013dislocation}}  \\
{$\rho_b$, $\rho_{b,c}$} & {Dislocation density in bulk and cell} & {$10^{14}$ m$^{-2}$} & {\cite{bertsch2020origin}}  \\
{$\rho_{b,w}$} & {Dislocation density at CW} & {$10^{15}$ m$^{-2}$} & {}  \\
{$\sigma$} & {Relative GB density} & {0.8, 0.9} & {\cite{wang2021density}}  \\
{$\kappa$} & {Gradient energy coefficient for GB} & {$7.5\times10^{-10}$ J/m} & {This work}  \\
{} & {Gradient energy coefficient for CW} & {$7.5\times10^{-8}$ J/m} & {This work}  \\
{$m_\circ$} & {External barrier height for GB} & {$7.5\times10^8$ J/m$^3$} & {This work}  \\
{} & {External barrier energy for CW} & {$7.5\times10^6$ J/m$^3$} & {This work}  \\
{$\Delta \omega^e_g$} & {Intrinsic barrier energy height of GB and CW} & {$0$ J/m$^3$} & {This work}  \\
{$k^2_g$} & {Local sink strength for GB} & {34.6 nm$^{-2}$} & {This work}  \\
{$\delta$} & {GB width} & {1 nm} & {}  \\
{} & {CW width} & {100 nm} & {\cite{shang2021heavy}}  \\
{$d_{1D}$} & {Grain size for 1D simulations} & {1 $\mu$m} & {}  \\
\hline
\label{tab:myfirstlongtable}
\end{longtable}
}
\normalsize

\subsection{Transport properties} \label{sec:param_transport}

Following Yang et al.~\cite{yang2016roles}, the Onsager coefficients are calculated from the partial diffusivities $d_{k{{\upsilon}}}$ by using Manning's relations for concentrated multicomponent alloys~\cite{lidiard1986note}:
\begin{flalign}
    L^{{\upsilon}}_{k l} = \left(\frac{c_k c_{{\upsilon}} d_{k {{\upsilon}}}}{RT/V_m} \right)\left(\delta_{k l} + \frac{2 c_l d_{l{{\upsilon}}}}{M_o \sum_{j} c_j d_{j{{\upsilon}}}} \right), &&
\end{flalign}
where $\delta_{kl}$ is the Kronecker delta and $M_\circ=2f_\circ/(1-f_\circ)$, with $f_\circ$ being the geometric correlation factor for the FCC lattice. 
The partial diffusivities for vacancy and SIA-mediated transport are given by~\cite{yang2016roles}:
\begin{subequations}
\begin{flalign} \label{eq:partial_dV}
    &d_{k V} = P_k \exp{\left(\frac{-Q_k + E^f_V}{k_B T}\right)}, \\ \label{eq:partial_dI}
    &d_{k I} = \frac{1}{6} \lambda^2_k z f_I \omega^\circ_{k I} \exp{\left(-\frac{E^m_I}{k_B T}\right)\beta_k}, &&
\end{flalign}
\end{subequations}
where the activation energy $Q_k$ and frequency factor $P_k$ for the vacancy transport are temperature- and composition-dependent expressions of the Redlich-Kister form that have been optimized using tracer diffusivity data provided by Yang et al.~\cite{yang2016roles}.
The SIA diffusivities are given by the atom-SIA binding model of Wiedersich et al.~\cite{wiedersich1979theory}, where $\lambda_I$ is the jump distance of $I$, $z_I$ is the site coordination number, $\omega^\circ_{kI}$ is the jump frequency prefactor, $E^m_I$ is the SIA migration energy (assumed identical for all elements), and ${\beta}_{k}=\exp{\left((E_{kI}-E_{FeI})/k_BT)\right)}/\left\{c_{Fe}+\sum_k c_{k}\exp{\left((E_{kI}-E_{FeI})/k_BT\right)}\right\}$ is the SIA binding factor.
These parameters are provided in Table~\ref{tab:myfirstlongtable}.

Since the vacancy activation energies $Q_k$ are derived from experimental tracer diffusivity data,
the vacancy migration energy for diffusion is given by $E^m_{Vk} = Q_k-E^f_V$. 
$Q_k$ is found to lie in the {2.86--2.89}~eV range, and $E^f_V=1.8$~eV is chosen based on a first-principles study on dilute FCC Fe-Cr-Ni~\cite{klaver2012defect}. 
This yields an effective migration energy in the range of 1.05 to 1.1 eV.
With this approach, $E^f_V$ is an uncertain parameter that must be more accurately estimated for a concentrated austenitic SS composition. 
The vacancy diffusivity at 500\,$^\circ$C is $D_V\approx5.5\times10^5$~nm$^2$/s and the partial diffusion coefficient ratios are $d_{CrV}/d_{FeV}=1.62$ and $d_{NiV}/d_{FeV}=0.67$.
These ratios reflect the faster diffusivity of Cr and the slower diffusivity of Ni in FCC Fe. 
The SIA parameters in the model are quite uncertain, since they are not experimentally accessible and are challenging to determine via atomistic techniques.
$E^m_I$ = 0.9 eV, as chosen by Yang et al.~\cite{yang2016roles}, resulted in a low diffusivity for SIA in comparison to the vacancy.
Therefore, we chose a lower value of $E^m_I=0.6$ eV~\cite{yang2016roles}, yielding $D_I\approx 1.3\times10^7$ nm$^2$/s at 500\,$^\circ$C---an order of magnitude higher than $D_V$. 
With regards to SIA binding, first-principles studies of SIA in dilute \cite{klaver2012defect} and concentrated \cite{piochaud2014first} FCC Fe--Ni--Cr indicate an attractive interaction between SIAs and Cr, a repulsive interaction with Ni, and an ideal interaction with Fe. 
Thus, the positive value for $E^b_{CrI}-E^b_{FeI}$ models favorable transport of Cr via SIAs, whereas the negative value for $E^b_{NiI}-E^b_{FeI}$ models unfavorable transport of Ni via SIAs. 
The resulting binding factors for the SIA partial diffusivities are ${\beta}_{CrI}=1.37$, ${\beta}_{NiI}=0.84$, and ${\beta}_{FeI}=0.94$. 
The ratios of SIA partial diffusion coefficients are $d_{CrI}/d_{FeI}=1.63$ and $d_{NiI}/d_{FeI}=0.89$. 
At lower temperatures ($<500^\circ$\,C), the increased importance of Cr-SIA binding means that Cr transport via SIA is expected to become more important (i.e. $d_{CrI}/d_{FeI}>d_{CrV}/d_{FeV}$). 
All simulation of RIS performed in this paper employ the preferential solute-SIA binding discussed above. A comparison of these results with the case without solute-SIA binding is performed in Sec.{~\ref{sm:sec:solute-SIA}} of the Supplementary Material.

\subsection{{Phase free energy}} \label{sec:impl:eqseg}
The free energy densities for the bulk and GB phases were fit to the Taylor form in Eq.~\ref{eq:fe_taylor} by using the CALPHAD free energy in Eq.~\ref{eq:calphad}. 
Following the method proposed by Kamachali et al.~\cite{kamachali2020model,wang2021density}, we describe the GB free energy by modifying the bulk CALPHAD free energy via the relative atomic density $\sigma$.  
$\sigma$ is defined as the ratio of atomic volume at the GB over that in the bulk.
Physically, $\sigma$ relates to the free volume at a GB due to atomic disorder or the misorientation angle of a symmetric tilt GB.
Thus, $\sigma=1$ indicates a GB structure identical to the bulk whereas $\sigma<1$ indicates a GB structure distinct from the bulk.
The relative density is used to modify the bulk free energy, providing a thermodynamic driving force for the interaction of diffusing species with the GB:
\begin{flalign} \label{eq:calphad}
f_C = c_{Fe}G^\circ_{Fe}(\sigma) + c_{Cr}G^\circ_{Cr}(\sigma) + c_{Ni}G^\circ_{Ni}(\sigma) + \sigma^2\Delta H_{\text{mix}} - T\Delta S_{\text{mix}}, &&
\end{flalign}
where $G^\circ_k(\sigma) = \sigma^2 H^\circ_k - \sigma T S^\circ_k$. 
The modification to the pure component free energies $G^\circ_{k}(\sigma)$ by $\sigma$ describes the higher free energy of the pure metal's GB relative to the pure metal's bulk.
Thus, in the alloy, the element with lower GB energy in its pure metal state would be energetically favored to segregate to reduce the alloy's GB energy.
The modification to the enthalpy of mixing (fourth term on the right side of Eq.~\ref{eq:calphad}) captures the reduction in interaction strengths between the alloying elements at the GB.
These modifications can be imagined to arise from broken bonds at the GB.
The pure element enthalpies $H^\circ_k$ and entropies $S^\circ_k$ are obtained from Dinsdale~\cite{dinsdale1991sgte}, whereas the optimized parameters of the Redlich-Kister mixing enthalpy $\Delta H_{\text{mix}}$ are obtained from Miettinen~\cite{miettinen1999thermodynamic}. 
Temperature-dependent expressions for these quantities are provided in Sec.{~\ref{sm:gb_fe}} in the Supplementary Material.
The pressure and magnetic contributions are expected to be negligible and are thus ignored.
We assume the far field bulk composition to be the nominal alloy composition (i.e., $c^{b,\circ}_{Cr} = 0.17$ and $c^{b,\circ}_{Ni} = 0.12$), and evaluated the bulk parameters of Eq.~\ref{eq:fe_taylor} by using Eq.~\ref{eq:calphad}. 
By solving the equilibrium conditions $\partial f^b_C/\partial c^b_k = \partial f^g_C/\partial c^g_k$ at a given $\sigma$ and $T$, we determine the GB parameters $c^{g,\circ}_k$, $f^{g,\circ}$, $\mu^\circ_{kFe}=\mu^{b,\circ}_{kFe}=\mu^{g,\circ}_{kFe}$, and $\theta^{g,\circ}_{kk}$ of Eq.~\ref{eq:fe_taylor}.

\subsection{{Point defect reaction rates}} \label{sec:impl:rate}

The reaction rate coefficients in Eq.~\ref{eq:alt_time_evol_defect} and~\ref{eq:am_time_evol_defect} are defined as follows.
We set $P_{V}=P_I=2\times 10^{-6}$ dpa/s representing conditions of neutron or proton irradiation~\cite{allen1998modeling}.
For the present work, we do not consider the formation and effects of clusters in the collision cascade and thus neglect the resulting point defect production bias.
The recombination reaction coefficient is given by $R_{VI} = 4\pi r_\circ(D_V+D_I)/ V_a$, with $r_\circ$ being the recombination radius. 
The point defect diffusivities are derived from the partial diffusivities in Sec.~\ref{sec:param_transport} as $D_{{\upsilon}} = (c_{Ni}d_{Ni{{\upsilon}}}+c_{Cr}d_{Cr{{\upsilon}}}+c_{Fe}d_{Fe{{\upsilon}}})/f_\circ$. 

The sink strength of dislocation is given by $k^2_{{{\upsilon}},b}=\rho_b Z_{{\upsilon}}$, where $Z_{{\upsilon}}$ is the sink efficiency of dislocations for absorption of point defects.
Realistically, $Z_{{\upsilon}}$ is a function of the dislocation density and character~\cite{kohnert2019sink}.
Relations for $Z_{{\upsilon}}$ based on calculations of the elastic interaction between dislocation configurations and the point defect have revealed a bias for SIA absorption (i.e., $Z_I>Z_V$), and this bias increases with dislocation density.
For the sake of simplicity, we choose dislocation-density-independent values of $Z_I=1.2$ and $Z_V=1$, amounting to a bias factor $(Z_I-Z_V)/Z_V$ of 20\%~\cite{chang2013dislocation}.
We note that estimates of bias factors ranging from 2 to 30\% have been reported using different modeling techniques and the discrepancy between these have been discussed in Ref.~\cite{golubov20121}.
The choice of $Z_I = 1.2$ therefore represents a strong bias.

Based on assessments in \ref{appendix_A} and \ref{appendix_B}, the narrow bell-shaped sink function (Eq.~\ref{eq:gsink_narrow_bell}) is chosen and the local sink strength of the diffuse GB for this choice of sink function, corresponding to $C_\text{sink}=0.32$, is identified from Eq.~\ref{aeq:rbc_pf_relation} as $k^2_g=34.6$ nm$^{-2}$.
This sink strength can be related to a symmetric tilt GB of $7^\circ-15^\circ$ misorientation angle~\cite{duh2001numerical,gu2017point}.

\subsection{Equilibrium PF properties} \label{sec:impl:equil}

We assume the excess GB energy to be $\gamma = 1$ J/m$^2$ and the planar GB width at equilibrium to be $\delta=1$~nm. 
These properties are realized by setting the model parameter values as $m_0=7.5\times10^8$ J/m$^3$ and $\kappa=7.5\times10^{-10}$ J/m.
The relationship between the equilibrium GB properties and the model parameters are described in Sec.{~\ref{sm:eq_gb}} of the Supplementary Material.
For the sake of simplicity in parameterization and interpretation of results, we omitted the grand potential dependence (i.e. $\omega_g - \omega_b$ term) from the PF evolution in Eq.~\ref{eq:eta_evol}. 
This assumption is acceptable for the purposes of the current study since we are only interested in capturing the magnitude of TS (via Eq.~\ref{eq:alt_time_evol_atom}) and not in assessing the effects of TS on grain coarsening dynamics.
The complexity that would arise from a more rigorous approach is discussed in Sec.~{\ref{appendix_C}} of the Supplementary Material.

For the AM microstructure, per the literature~\cite{shang2021heavy,chen2024situ}, a CW width of $\delta = 100$ nm is considered.
We choose a dislocation density of $\rho_{b,c}=10^{14}$ m$^{-2}$ in the cell, and set $\rho_{b,w}$ in the CW to be 5--15 times higher. 
For $\rho_{b,w} = 10^{15}$ m$^{-2}$ distributed uniformly across the CW width, the dislocation energy of the CW can be estimated as $\frac{1}{2}Gb^2\rho_{c,w} \delta \approx 1$ J/m$^2$, where $G=77$ GPa is the shear modulus. 
To realize a stationary CW of $\delta = 100$ nm, we set the PF model parameters as $m_0=7.5\times10^6$ J/m$^3$ and $\kappa=7.5\times10^{-8}$ J/m from Sec.~\ref{appendix_C} of the Supplementary Material.
The interpolation scheme to determine the total atomic thermodynamic factors in Eq.~\ref{eq:alt_time_evol} and~\ref{eq:am_time_evol} from the phase-specific atomic thermodynamic factors obtained above is provided in Sec.~{\ref{sm:pf_gb}} of the Supplementary Material.
The thermodynamic factors for the point defect species are simply given by ${\theta}_{{{\upsilon}}{{\upsilon}}}=(RT/V_m)/c_{{\upsilon}}$.

\section{{Implementation}} \label{sec:impl:num}

The model is implemented using the open-source MOOSE (Multiphysics Object Oriented Simulation Environment) framework.
Weak forms of the partial differential equations are spatially discretized via the finite element method.
For time integration, the implicit second-order backward differentiation method is used to obtain the system of nonlinear equations at each time step. These are then solved using Newton's method.
Mesh elements with linear Lagrange shape functions are used for the nonlinear and auxiliary variables.
2D domains are meshed with four-node quadrilateral elements.
For both 1D and 2D simulations, a uniform mesh with adaptive time stepping {(see Sec.~\ref{sm_time_stepper_results} for details on the {IterationAdaptiveDT} scheme and the results of testing the parameters)}, a nonlinear relative tolerance of $10^{-8}$, and a nonlinear absolute tolerance of $10^{-10}$ are employed.
The order parameters for the 2D polycrystal and 2D CW microstructures (Fig.~\ref{fig:2D_micro}) are generated using the Voronoi tessellation and grain growth algorithm implemented in MOOSE \cite{permann2020}.

\subsection{GB segregation}
\begin{figure}[ht]
\centering
    \begin{subfigure}[t]{0.4\textwidth}
        \includegraphics[width=1\textwidth]{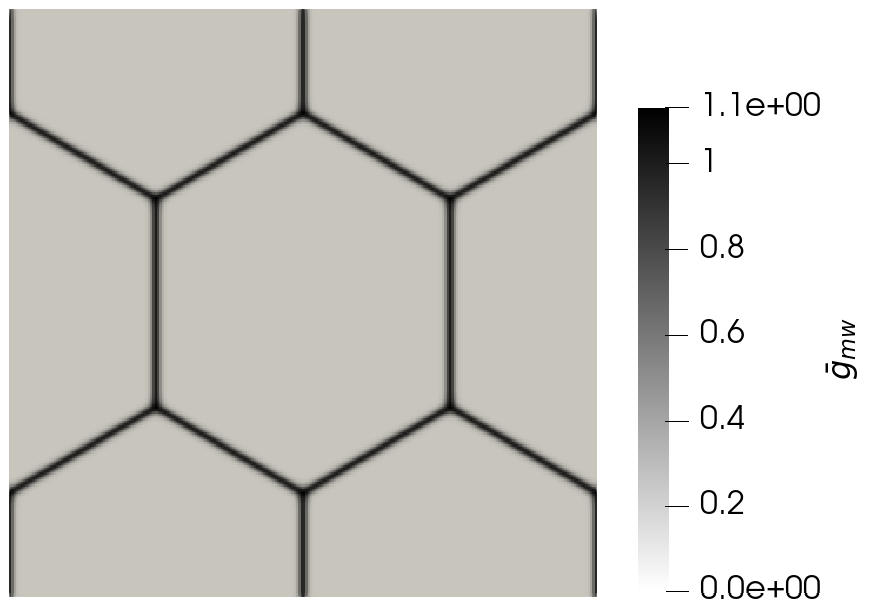}
    \caption{}
    \label{subfig:2D_micro:pc}
    \end{subfigure}
    ~ 
    \begin{subfigure}[t]{0.39\textwidth}
        \includegraphics[width=1\textwidth]{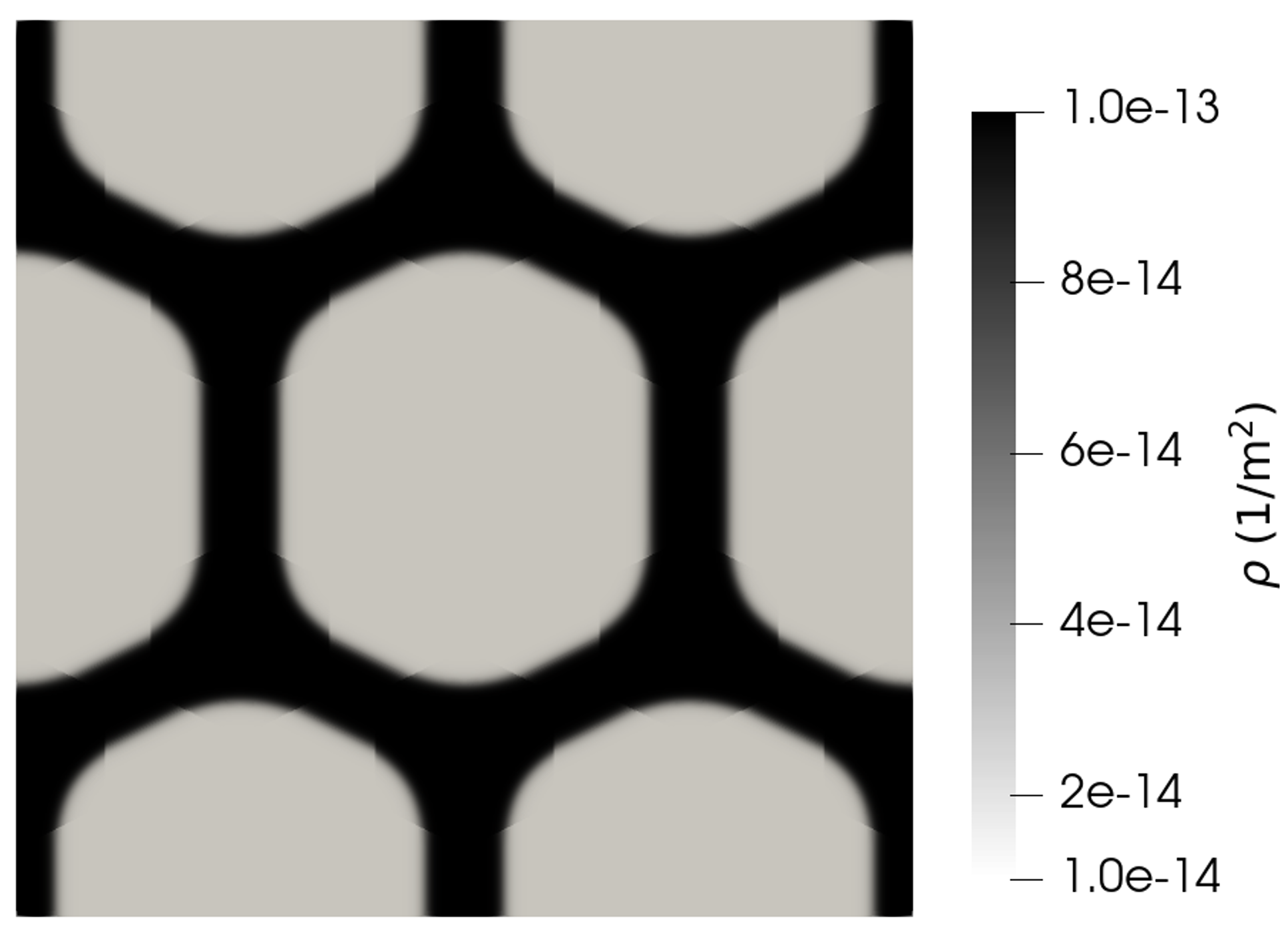}
    \caption{}
    \label{subfig:2D_micro:am}
    \end{subfigure}
    \caption{2D PF microstructures for (a) polycrystal and (b) AM dislocation cells. (a) Normalized multiwell potential $\bar{g}_\text{mw}$. (b) Dislocation density distribution and the sink indicator function $g_\text{sink}$.}
    \label{fig:2D_micro}
\end{figure}

Simulations of GB segregation are performed in 1D and 2D with periodic boundary conditions.
For the PF implementations, Eqs.~\ref{eq:alt_time_evol_atom},~\ref{eq:alt_time_evol_defect}, and~\ref{eq:eta_evol} are solved. 
For a GB width of $\delta=1$ nm, a mesh element size of $0.1$ nm is chosen, based on a mesh convergence test {(i.e. separate simulations with different mesh element sizes were performed to identify the mesh element size at which the numerical solutions converge)}.
The different PF simulation methods adopted in this work are as follows:
\begin{itemize}
    \item 1D PF: The domain constitutes a bicrystal with a full GB width $\delta$ at the center and two GBs---each of half GB width $\delta/2$---at the two ends of the domain.
    \item 2D (hexagonal) PF (Fig.~\ref{subfig:2D_micro:pc}): A square domain with hexagonal grains and GB width $\delta$ is employed. Due to periodic boundary conditions, the system contains four equiaxed grains.
\end{itemize} 
In addition to the above PF simulations, we performed sharp interface simulations (see \ref{appendix_A} and \ref{appendix_B}) for RIS at GBs by employing Eq.~\ref{eq:alt_time_evol} with boundary conditions.
For this, we simply omit the PF sink term $-k^2_gD_{{\upsilon}}(c_{{\upsilon}}-c^e_{{\upsilon}})g_\text{sink}$ from Eq.~\ref{eq:alt_time_evol_defect} and impose a Dirichlet boundary condition (DBC) or Robin boundary condition (RBC) for the point defect concentration variables at the edges of the domain defining the GB.
The DBC sets $c_{{\upsilon}} = c^e_{{\upsilon}}$ at the GB, thus describing an ideal sink.
The RBC is applied as a flux normal to the GB given by $J_{{\upsilon}}.\hat{n} = D_{{\upsilon}} \alpha^{-1} (c_{{\upsilon}}-c^e_{{\upsilon}})$, where $\alpha$ is a characteristic length for point defect absorption ~\cite{gu2017point}.
The effect of this parameter on the non-ideal GB sink behavior is studied in ~\ref{appendix_B}.
The different sharp-interface simulations of RIS at GBs performed in this work are:
\begin{itemize}
    \item 1D DBC: The domain constitutes a single grain with a GB at each end as specified by the DBC.
    \item 1D RBC: Similar to the above, but with the RBC specified at the ends of the domain.
    \item 2D (square) DBC: A square domain is used, with a DBC imposed on the edges (GBs).
\end{itemize}
For those methods that employ a DBC, sharp gradients in the point defect concentrations arise, making mesh convergence difficult to achieve. 
Thus, the results for mesh sizes ranging from a fraction of nanometer to 1 nm are reported; these are the resolutions achieved by high resolution experimental characterization techniques.
The results of the RBC method converged at a mesh size of 0.1 nm.
The parameterization for the RBC method are presented in~\ref{appendix_A}, and \ref{appendix_B} compares and verifies the results from 1D PF RIS simulations against 1D DBC and RBC simulations.

\subsection{CW segregation}
Simulations of RIS at CWs are performed in 1D and 2D by using the spatially resolved model in Eq.~\ref{eq:am_time_evol}.
Auxiliary PF order parameters are employed to initialize the microstructure with dislocation cells and CWs.
Periodic boundary conditions are imposed on the concentration variables.
A CW of width $\delta=100$ nm is employed, and Eq.~\ref{eq:rho_spatial_dep} is used to initialize a spatial variation in the dislocation density.
The following simulation methods are adopted:
\begin{itemize}
    \item 1D cells: The domain constitutes two subgrains with a CW at the center of the domain.
    \item 2D hexagonal cells: A square domain consisting of hexagonal subgrains and CWs.
    \item 2D cells with DBC: A square grain consisting of one hexagonal subgrain and six one-half subgrains; the DBC is applied to the edges/GBs of the square grain.
\end{itemize}

\section{Results} \label{sec:results}

We first present the results of GB segregation in 1D and 2D systems as a function of irradiation dose and temperature.  We examine four different GB segregation mechanisms: RIS without TS, TS without RIS, radiation-enhanced TS (RETS) without RIS, and RIS in combination with RETS (RIS+RETS).   
The effects of grain size and dislocation bias on RIS are then presented.
The results obtained through verification and comparison of the 1D PF RIS simulations against sharp-interface simulations are provided in \ref{appendix_A} and \ref{appendix_B}.
Next, we present the results of RIS at dislocation CWs in 1D and 2D cells. 
The effects of cell size, CW density, initial microsegregation, and GB sink are examined.
{Results of point defect concentrations are provided in Sec.~\ref{sm_results} of the Supplementary Material.}

\subsection{{{1D simulations of GB segregation}}} \label{results:GB_mechanisms}

A periodic bicrystal with a grain size $d_{1D}$ of 1 $\mu$m is considered.
To simulate the RIS or TS mechanism individually, the other mechanism is turned off.
Since TS is impossible when the energetics at the GB are identical to that in the bulk, the
TS mechanism is turned off by simply setting the relative GB atomic density $\sigma=1$, resulting in the free energy of the GB being identical to that of the bulk.
Since RIS does not occur when the partial diffusivities of the atomic components are equal, RIS is turned off by setting the vacancy activation energies of the solutes equal to that of Fe (i.e., $Q_{Cr}=Q_{Ni}=Q_{Fe}$) and setting the SIA binding energies to be zero (i.e., the binding factors are $\beta_k=1$).
With this approach, radiation-enhanced diffusivities are still retained due to point defect supersaturation, thus yielding RETS without RIS.
Unless stated otherwise, all results correspond to the default parameters of $500\,^\circ$C, $\sigma=0.8$, $\rho_b = 10^{14}$ m$^{-2}$, and $Z_I=1$ (see Table~\ref{tab:myfirstlongtable}).

\subsubsection{Segregation profiles} \label{results:GB_mechanisms:1D_profiles}

\begin{figure}[htp!] 
\centering
    \begin{subfigure}[t]{0.48\textwidth}
        \includegraphics[width=1\textwidth]{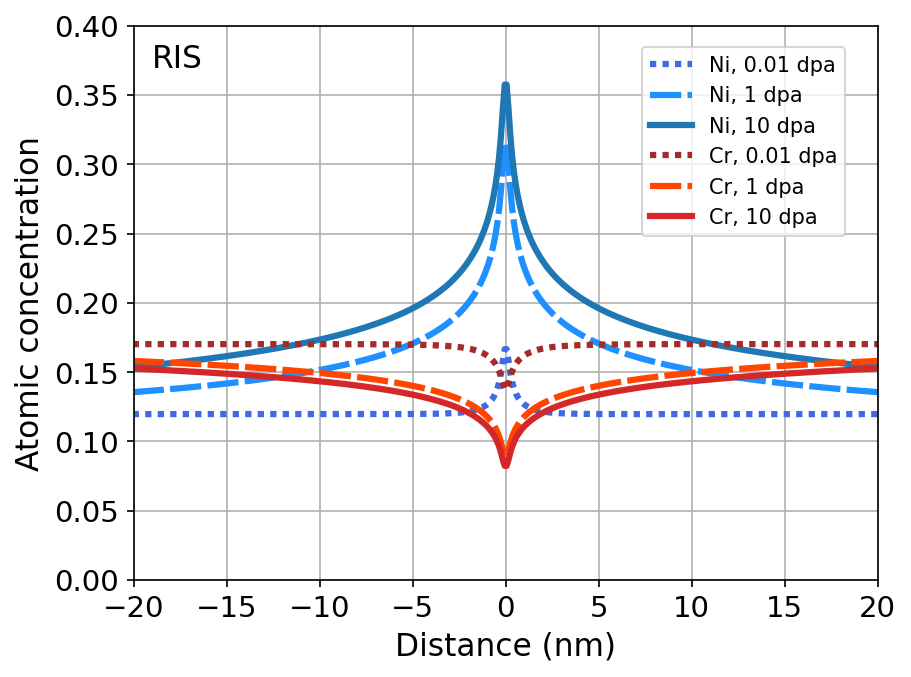}
    \caption{}
    \label{subfig:1D_ris_zoom}
    \end{subfigure}
    \begin{subfigure}[t]{0.48\textwidth} 
        \includegraphics[width=1\textwidth]{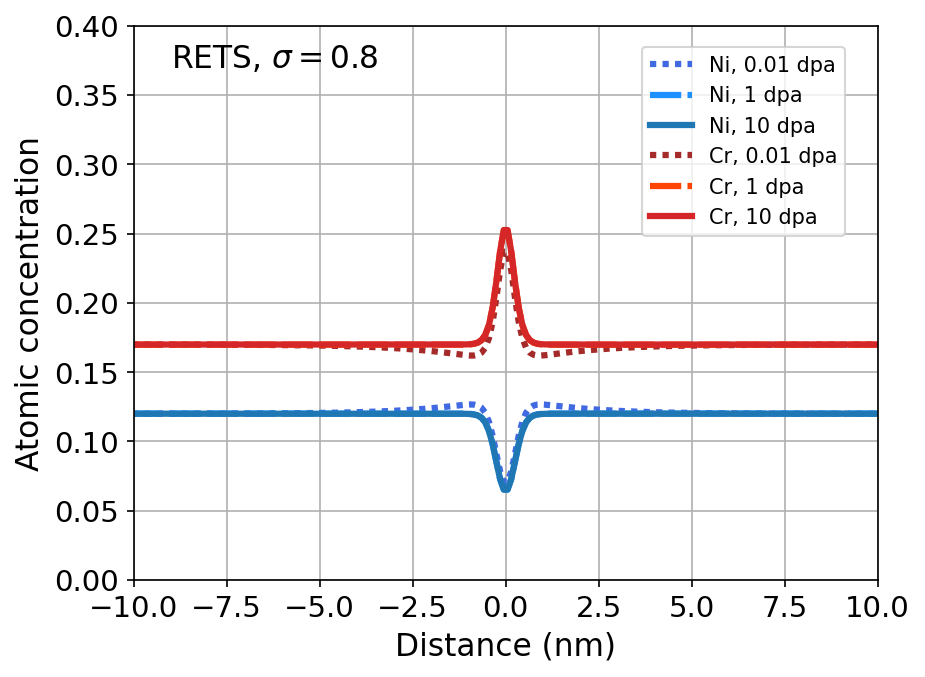}
    \caption{}
    \label{subfig:1D_rets_zoom}
    \end{subfigure}
    \caption{Segregation at the GB in a 1D system of 1 $\mu$m grain size when under irradiation at $2\times10^{-6}$ dpa/s and 500\,$^\circ$C. 
    (a) The RIS mechanism. (b) The RETS mechanism.}
    \label{fig:1D_ris_rets}
\end{figure}

\begin{figure}[htp!] 
\centering
    \begin{subfigure}[t]{0.48\textwidth}
        \includegraphics[width=1\textwidth]{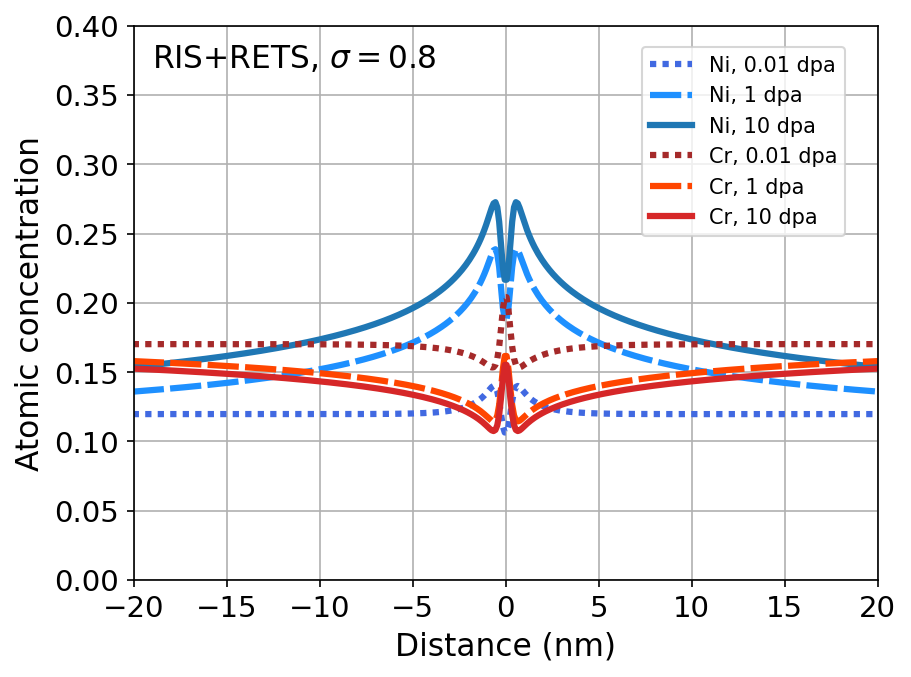}
    \caption{}
    \label{subfig:1D_ris_and_rets_strong_zoom}
    \end{subfigure}
    \begin{subfigure}[t]{0.48\textwidth}
        \includegraphics[width=1\textwidth]{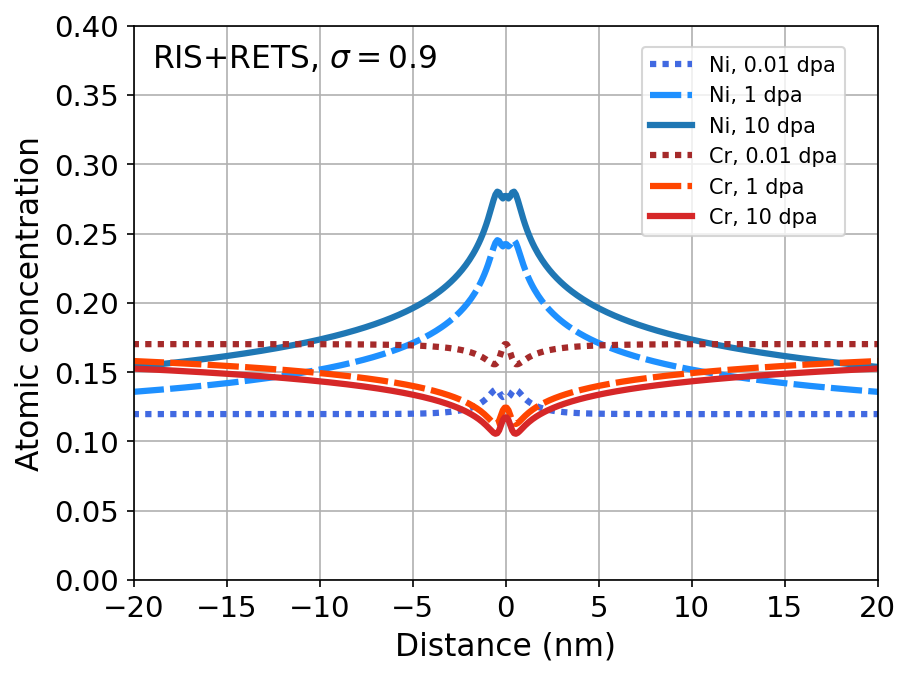}
    \caption{}
    \label{subfig:1D_ris_and_rets_weak_zoom}
    \end{subfigure}
    \caption{Segregation due to RIS and TS mechanisms at the GB in a 1D system of 1 $\mu$m grain size when under irradiation at $2\times10^{-6}$ dpa/s and 500\,$^\circ$C. 
    (a) Strong TS corresponding to a relative GB density of $\sigma=0.8$. (b) Weak TS corresponding to $\sigma=0.9$.}
    \label{fig:1D_ris_and_rets}
\end{figure}

Concentration profiles for the RIS mechanism are plotted for different doses in Fig.~\ref{subfig:1D_ris_zoom}.
Ni enrichment (inverted ``V"-shaped profile) and Cr depletion (``V"-shaped profile) are observed, with wide segregation widths of over 100 nm on either side of the GB center.
Significant changes to RIS were found to occur between 0.01 dpa (1 hour) and 10 dpa (60 days).
Concentration profiles for the RETS mechanism are plotted in Fig.~\ref{subfig:1D_rets_zoom}.
Unlike RIS, RETS is seen to evolve rapidly and reach equilibrium at just after 0.01 dpa (1 hour).
Moreover, Ni depletion and Cr enrichment, which are opposite to their behaviors during RIS, are observed.
In further contrast, RETS shows sharp profiles that only extend to about 0.5 nm on either side of the GB center.
Next, Fig.~\ref{fig:1D_ris_and_rets} shows profiles from the combined RIS+RETS simulation. 
Here, two different cases are shown, with TS being strong ($\sigma=0.8$) in one case and moderate ($\sigma=0.9$) in the other.
While the low-magnification plots of the profiles ({see Sec.~\ref{sm_results} in the Supplementary Material}) resemble the monotonic RIS profiles, the high-magnification plots (Figs.~\ref{subfig:1D_ris_and_rets_strong_zoom} and~\ref{subfig:1D_ris_and_rets_weak_zoom}) clearly show non-monotonic variation close to the GB.
For the case with strong TS, a slight Cr enrichment persists at the GB center, even at 10 dpa.
The profiles resemble a ``W" shape for Cr and an ``M" shape for Ni.
The non-monotonic profiles are less pronounced for the case of weak TS. 
Note that for the sake of simplicity, the local GB sink strength $k^2_g$ was kept the same and only $\sigma$ was varied; however, in a more realistic scenario, both $k^2_g$ and $\sigma$ would vary as a function of the GB structure.

\subsubsection{Dose and temperature dependence} \label{results:GB_mechanisms:1D_temp}

\begin{figure}[htp!] 
\centering
    \begin{subfigure}[t]{0.475\textwidth}
        \includegraphics[width=1\textwidth]{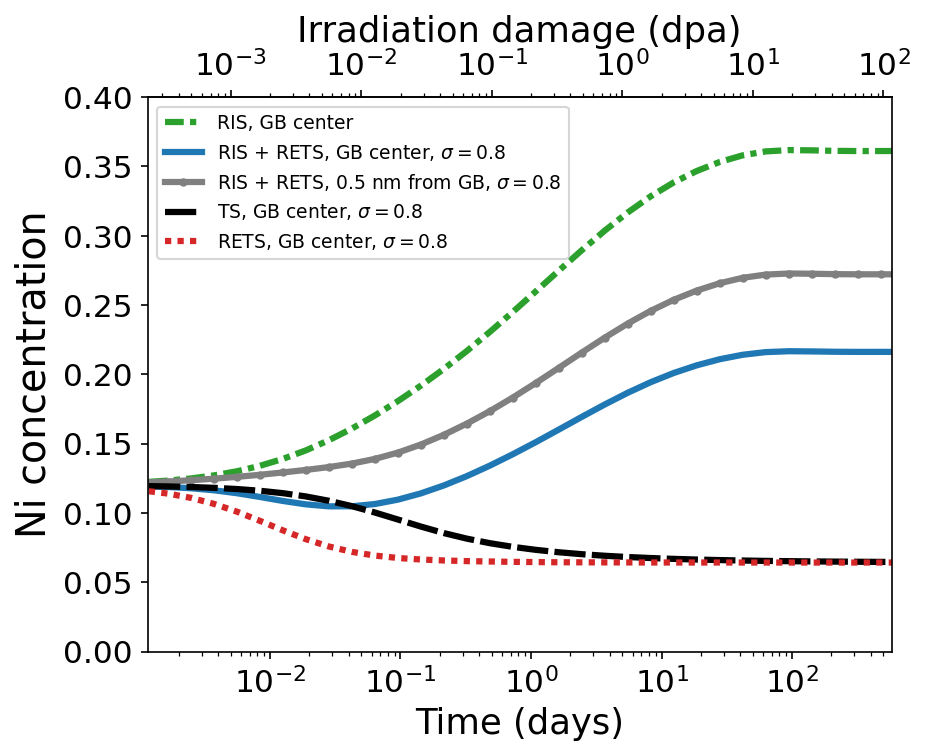}
    \caption{}
    \label{subfig:1D_Ni_vs_time}
    \end{subfigure}
    ~
    \hspace{0.1cm}
    \begin{subfigure}[t]{0.475\textwidth}
        \includegraphics[width=1\textwidth]{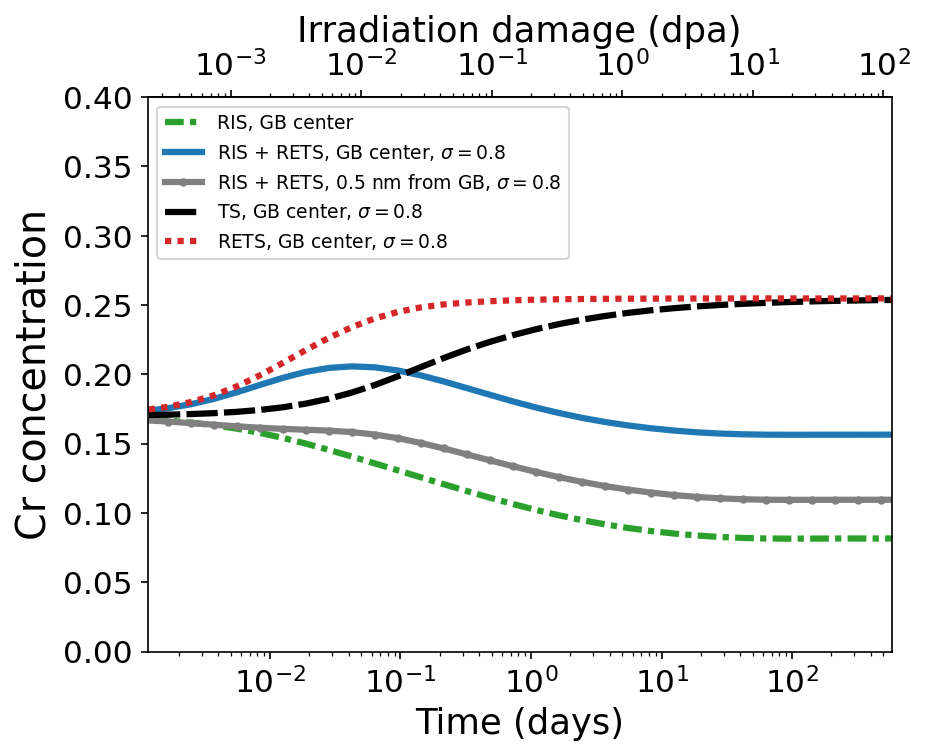}
    \caption{}
    \label{subfig:1D_Cr_vs_time}
    \end{subfigure}
    \caption{Evolution of GB concentrations of (a) Ni and (b) Cr in a 1D system of 1 $\mu$m grain size when under irradiation at $2\times10^{-6}$ dpa/s and 500\,$^\circ$C. Different segregation mechanisms are separately plotted. For the combined case of RIS+RETS (with $\sigma=0.8$), the concentration at 0.5 nm from the GB center is also shown.}
    \label{fig:1D_conc_vs_time}
\end{figure}

The irradiation time and dose evolution of GB concentrations are plotted in Figs.~\ref{subfig:1D_Ni_vs_time} and~\ref{subfig:1D_Cr_vs_time}.
Concentrations 0.5~nm away from the GB center are also plotted for the RIS+RETS simulations.
In addition to RETS, we show the evolution of TS in the absence of irradiation.
We observe RETS to evolve rapidly and reach equilibrium faster (in a matter of hours or just over 0.01 dpa) in comparison to TS without irradiation (a few days), but the steady-state compositions are the same in both cases.
In contrast, steady states for RIS and RIS+RETS are reached in a few months, and the steady-state compositions differ.
The steady-state enrichment of Ni and depletion of Cr are significantly lower for the combined RIS+RETS case than for RIS alone.
In the initial period of evolution for RIS+RETS, the GB concentrations somewhat follow that of RETS, with Ni showing a slight depletion and Cr showing significant enrichment.
This indicates that TS dominates in the early stages of irradiation.
However, at 0.5~nm from the GB center, a monotonic evolution in segregation as governed by RIS is seen, with no influence from TS.

\begin{figure}[htp!] 
\centering
    \begin{subfigure}[t]{0.485\textwidth}
        \includegraphics[width=1\textwidth]{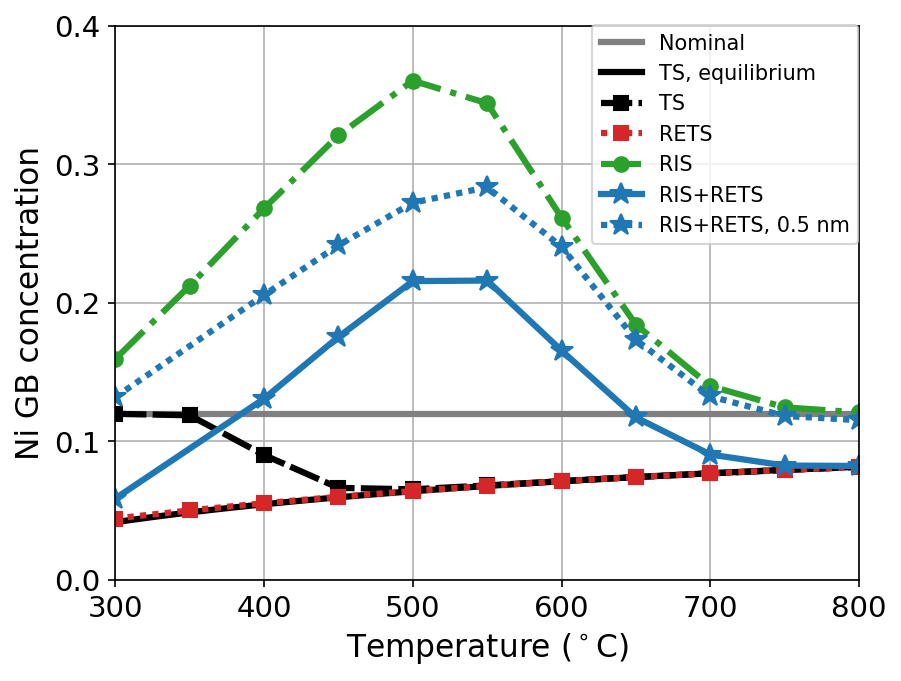}
    \caption{}
    \label{subfig:1D_Ni_vs_temp}
    \end{subfigure}
    ~
    \begin{subfigure}[t]{0.485\textwidth}
        \includegraphics[width=1\textwidth]{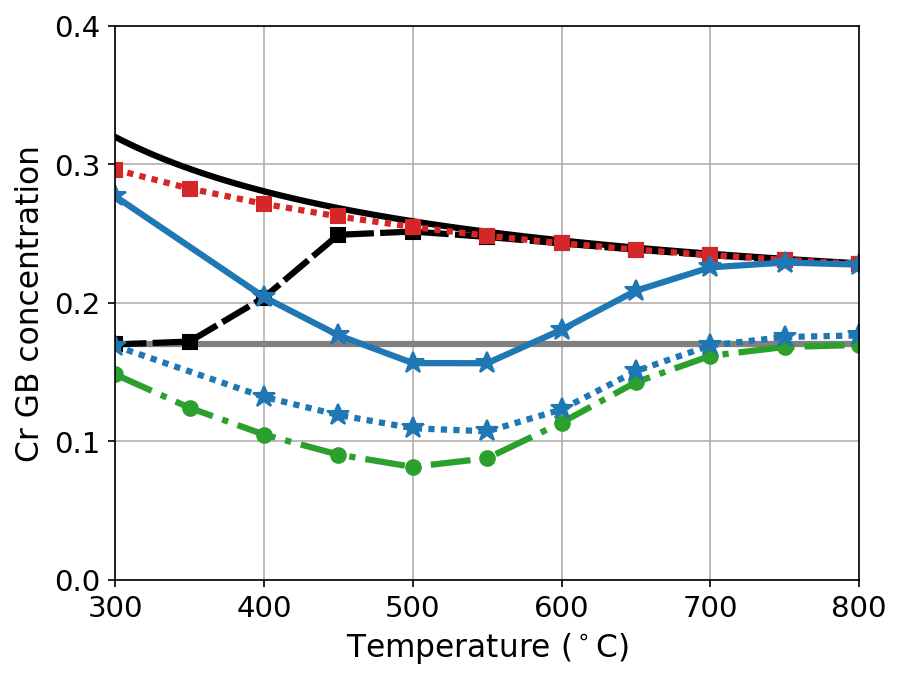}
    \caption{}
    \end{subfigure}
    \label{subfig:1D_Cr_vs_temp}
    \caption{GB segregation of (a) Ni and (b) Cr as a function of temperature, as calculated from a 1D system of 1 $\mu$m grain size. The irradiation mechanism corresponds to 10 dpa or 60 days, and the TS mechanism ($\sigma=0.8$) corresponds to 60 days of thermal aging. For RIS+RETS, the concentrations are taken at the GB center as well as at 0.5~nm from the GB center.}
    \label{fig:1D_seg_vs_temp}
\end{figure}

For the different mechanisms, GB concentrations at 60 days (10 dpa) are plotted as a function of temperature in Fig.~\ref{fig:1D_seg_vs_temp}.  The equilibrium segregation concentrations (labeled as: TS, equilibrium) are analytically calculated from the CALPHAD free energies (Sec.~\ref{sec:impl:eqseg}) for comparison.  
Excellent agreement is found for the steady-state RETS composition with the equilibrium analytical calculation; Cr enrichment and Ni depletion occur at all temperatures as a result of the more favorable energetics of Cr at the GB. RETS is found to decrease steadily with temperature, although some effect of RETS is seen even at 800\,$^\circ$C.  

However, TS without irradiation shows lower segregation at the end of the simulated 60 days than that observed with RETS below 450\,$^\circ$C, indicating that equilibrium TS is not achieved during that time frame due to the extremely slow diffusivity at lower temperatures.
In contrast, RIS varies strongly with temperature and peaks at around 500\,$^\circ$C. 
Negligible RIS is observed above 750\,$^\circ$C due to thermal back diffusion.
Conversely, below 500\,$^\circ$C, increasing contributions from recombination and preferential SIA diffusivity (see Sec.~\ref{sm:sec:solute-SIA} in the Supplementary Material) lead to decreasing RIS (here, SIAs are mobile but vacancies have reduced mobility; thus, recombination is promoted, resulting in a reduced loss of vacancies to sinks).
Compared to RIS alone, RIS+RETS for $\sigma=0.8$ (strong TS) shows lower Ni enrichment at intermediate temperatures, while a transition to Ni depletion is observed below 400\,$^\circ$C and above 650\,$^\circ$C. 
For RIS+RETS of Cr, RETS dominates across the temperature range, resulting in an effective Cr enrichment at the GB center.
However, at 0.5~nm from the GB center, there is negligible influence from RETS; therefore, the temperature trend follows that of RIS.
Finally, we note that for $\sigma=0.95$ (weak TS, see Sec.~\ref{sm_results} of the Supplementary Material), the RIS+RETS results are very close to that of RIS, even at the GB center.

\subsubsection{Dislocation sink density and bias effect} \label{results:GB_RIS:disl_effect}

\begin{figure*}[htp!] 
\centering
    \begin{subfigure}[t]{0.485\textwidth}
        \includegraphics[width=1\textwidth]{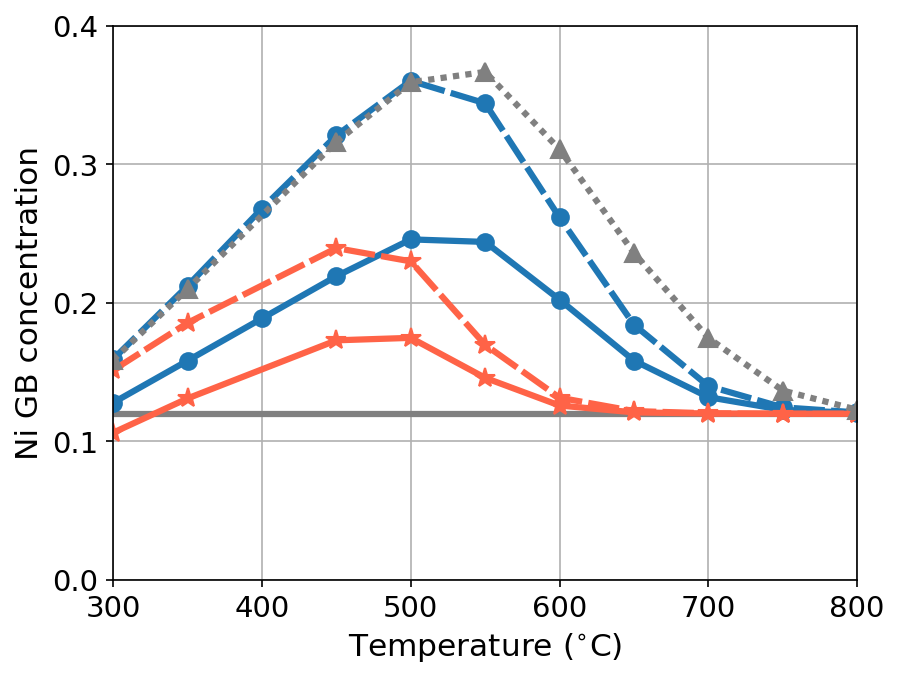}
    \caption*{\hspace{0.5cm}(a)}
    \end{subfigure}
    ~
    \begin{subfigure}[t]{0.485\textwidth}
        \includegraphics[width=1\textwidth]{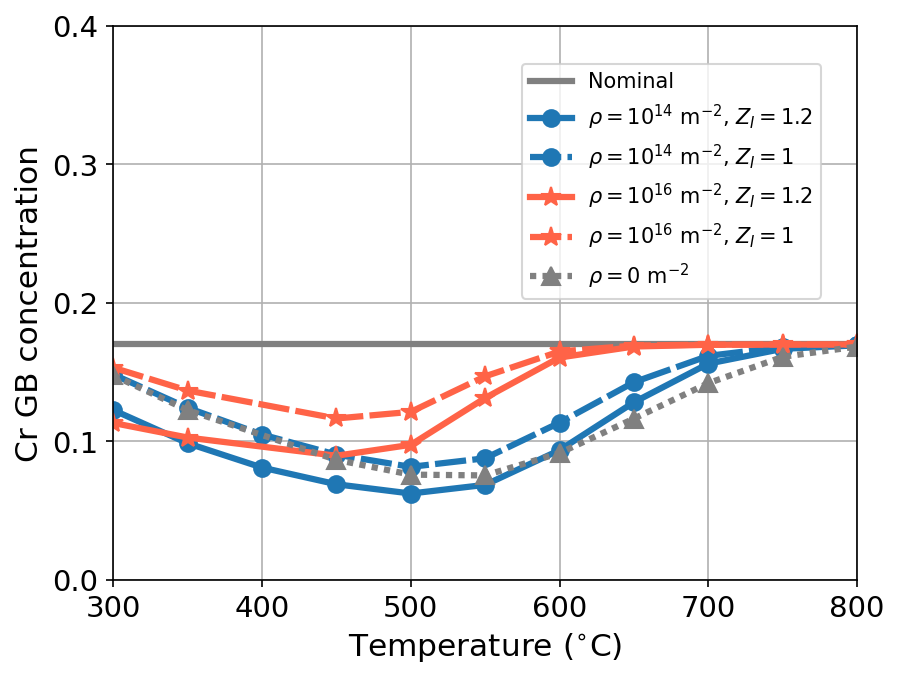}
    \caption*{\hspace{0.5cm}(b)}
    \end{subfigure}
    \caption{GB RIS of (a) Ni and (b) Cr as a function of temperature for different bulk dislocation densities $\rho_b$ and absorption efficiencies $Z_I$ for SIA. The calculations are from a 1D system of 1 $\mu$m grain size irradiated to 10 dpa or 60 days. The dashed lines show results from the DBC implementation of 1 nm mesh size.}
    \label{fig:1D_conc_vs_temp_disl}
\end{figure*}

In Fig.~\ref{fig:1D_conc_vs_temp_disl}, we compare the temperature dependence of RIS for different bulk dislocation densities $\rho_b$ and absorption efficiencies $Z_I$.
Increasing $\rho_b$ beyond $10^{14}$ m$^{-2}$ leads to a significant reduction in Ni enrichment and Cr depletion.
A more significant reduction in RIS is noted at higher temperatures, with the peak in RIS shifting to lower temperatures.
Introducing a bias for SIA absorption $Z_I=1.2$ results in biased effects on the RIS of Ni and Cr.
While Ni enrichment is significantly suppressed, Cr depletion is seen to be slightly enhanced.
In contrast to the effect of $\rho_b$, peak RIS is not significantly affected by $Z_I$.

\subsection{{{2D simulations of GB segregation}}} \label{results:GB_mechanisms:2D_poly}

\begin{figure}[htp!] 
\centering
    \begin{subfigure}[t]{0.4\textwidth}
        \includegraphics[width=1\textwidth]{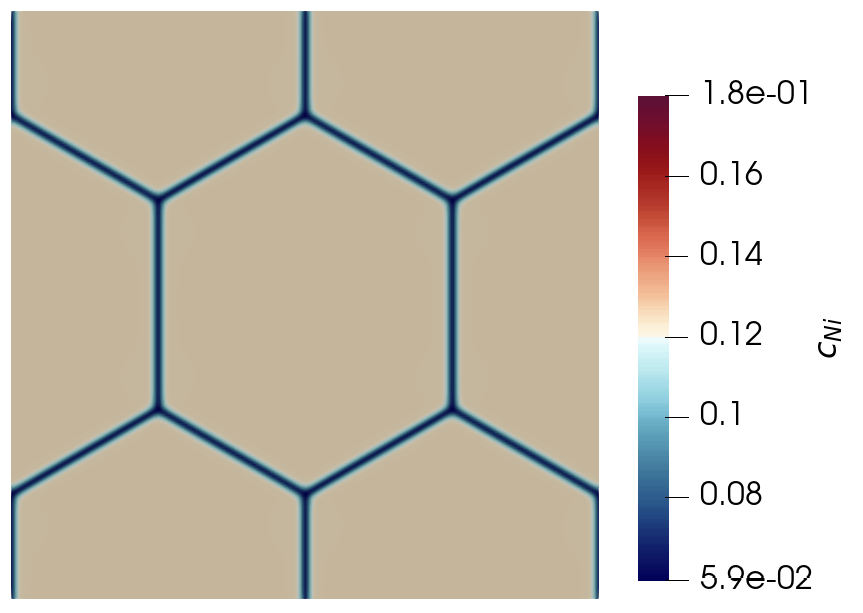}
    \caption{}
    \label{subfig:2D_rets_Ni}
    \end{subfigure}
    ~~
    \begin{subfigure}[t]{0.4\textwidth}
        \includegraphics[width=1\textwidth]{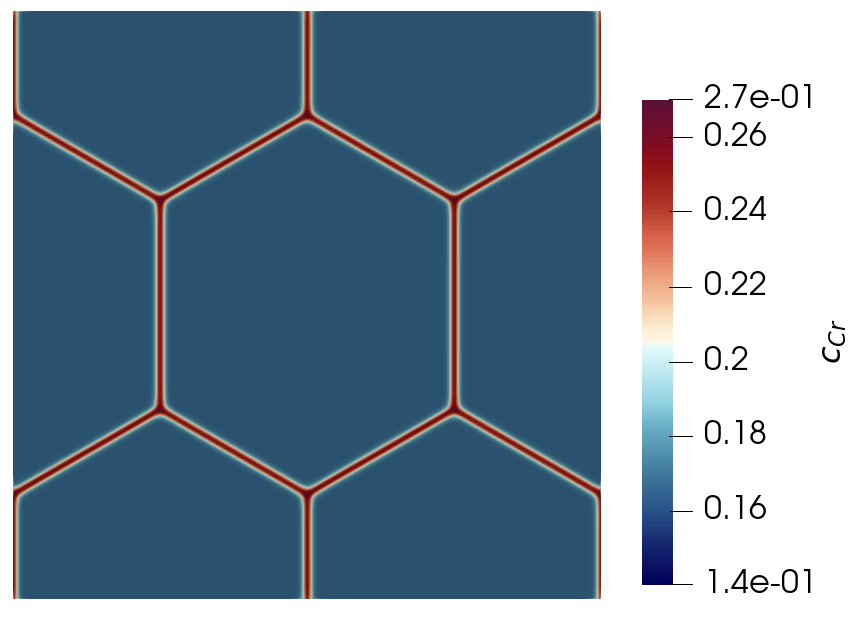}
    \caption{}
    \label{subfig:2D_rets_Cr}
    \end{subfigure}
    ~~~
    \begin{subfigure}[t]{0.4\textwidth}
        \includegraphics[width=1\textwidth]{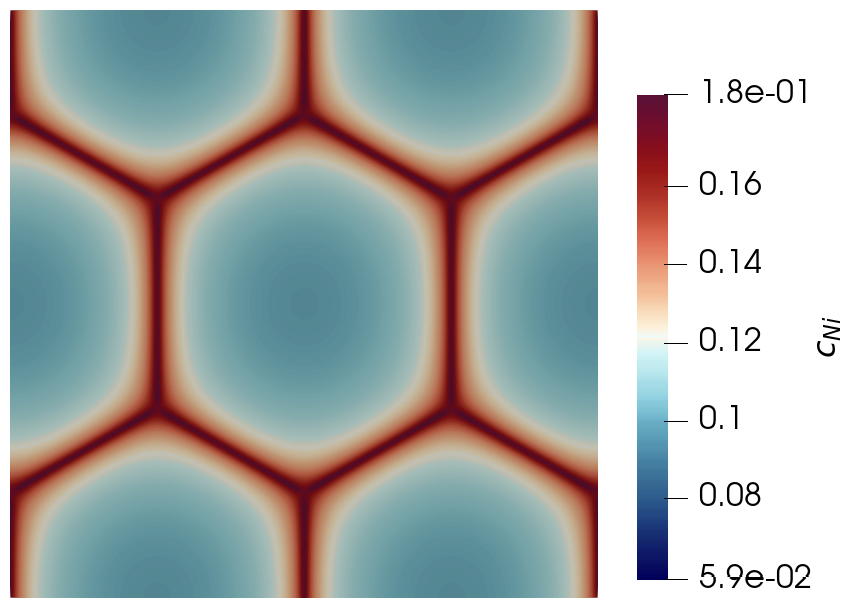}
    \caption{}
    \label{subfig:2D_ris_Ni}
    \end{subfigure}
    ~~
    \begin{subfigure}[t]{0.4\textwidth}
        \includegraphics[width=1\textwidth]{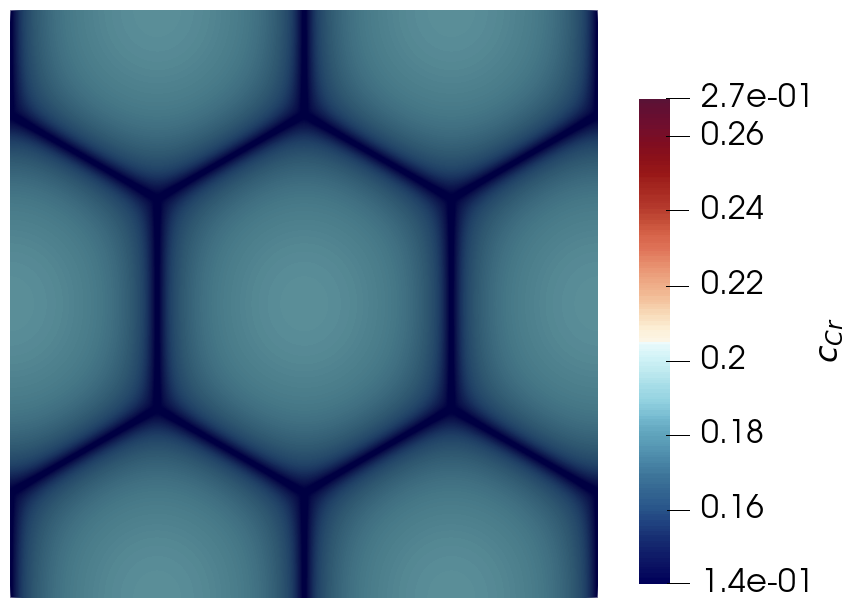}
    \caption{}
    \label{subfig:2D_ris_Cr}
    \end{subfigure}
    ~~~
    \begin{subfigure}[t]{0.41\textwidth}
        \includegraphics[width=1\textwidth]{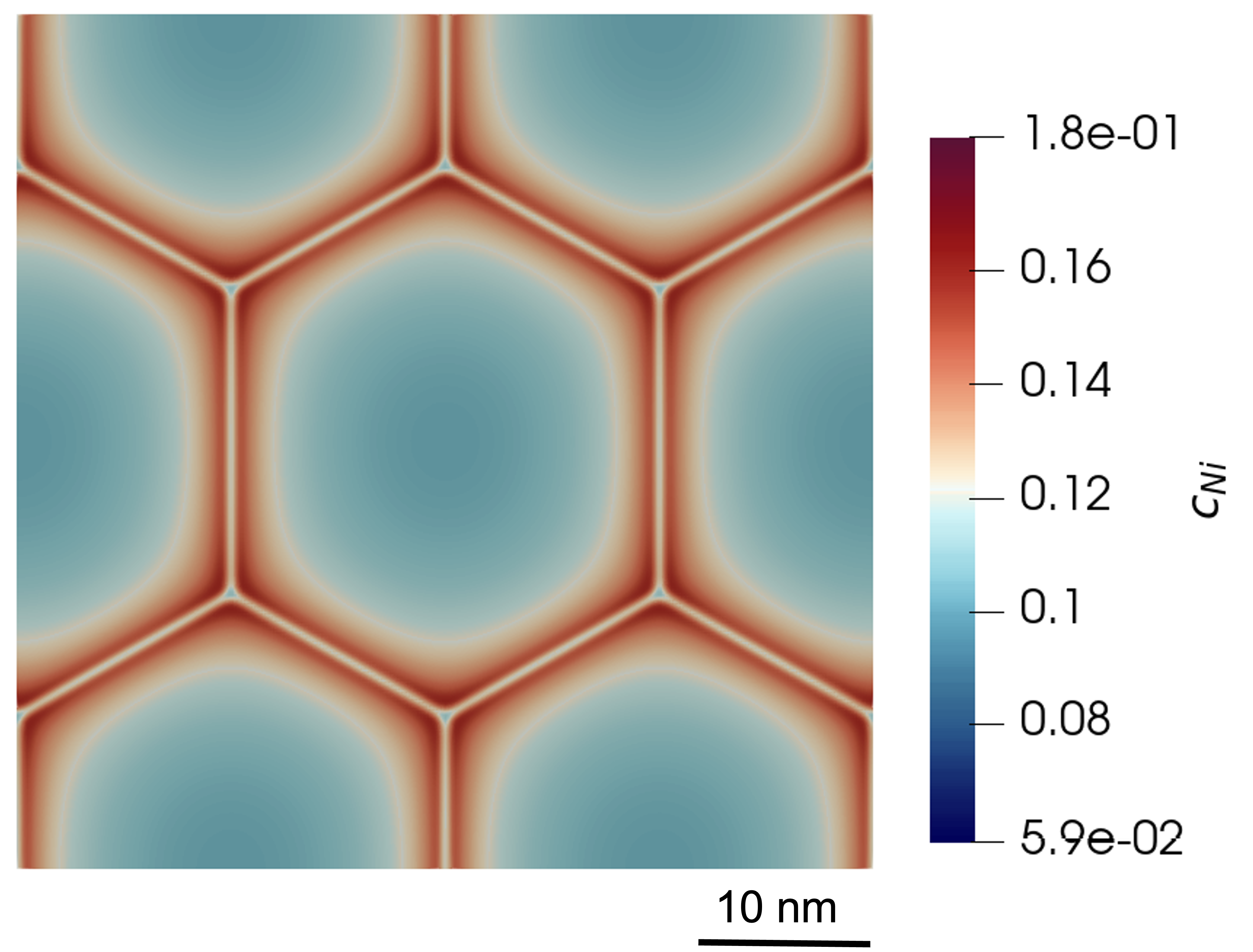}
    \caption{}
    \label{subfig:2D_ris_rets_Ni}
    \end{subfigure}
    ~~
    \begin{subfigure}[t]{0.41\textwidth}
        \includegraphics[width=1\textwidth]{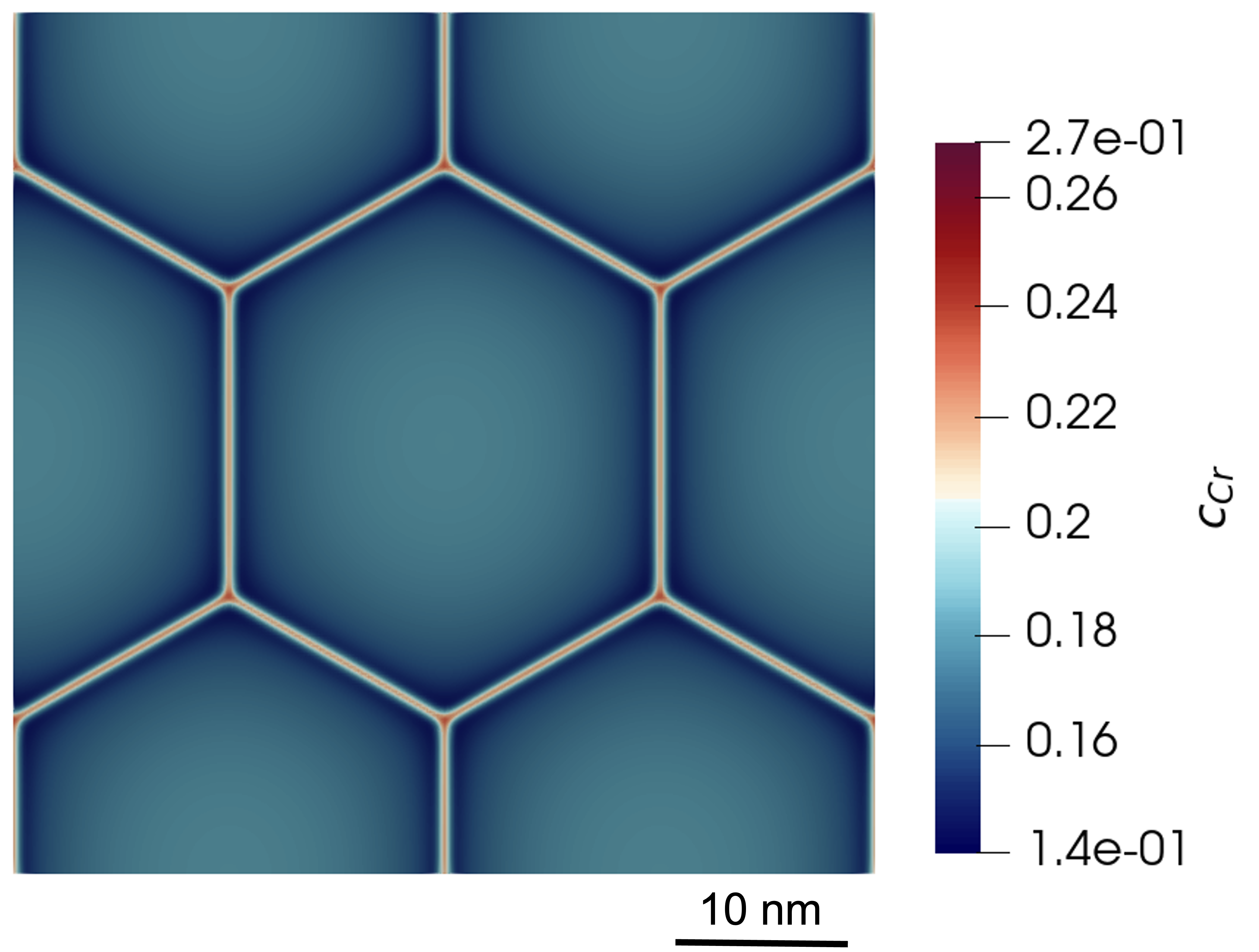}
    \caption{}
    \label{subfig:2D_ris_rets_Cr}
    \end{subfigure}
    \caption{Segregation at GBs in a 2D hexagonal polycrystal system of $50\times50$ nm$^2$ size irradiated to 10 dpa at 500\,$^\circ$C. (a,b) RETS mechanism for $\sigma=0.8$. (c,d) RIS mechanism. (e,f) RIS+RETS mechanisms.}
    \label{fig:compare_seg_vs_time}
\end{figure}

RETS, RIS, and RIS+RETS simulations are performed on a hexagonal polycrystal microstructure within a square domain of 50 $\times$ 50~nm$^2$ and 100 $\times$ 100~nm$^2$.
Since 2D PF simulations with a 1 nm GB width are computationally expensive, they are only performed for nanocrystalline grain sizes of $\approx$ 28~nm and 56~nm (diameters of the area equivalent circular grains).
(To assess RIS for larger grain sizes, in Sec.~\ref{results:grain_size}, we perform 2D simulations using the DBC method---which is computationally less expensive than PF---and demonstrate the equivalence between the PF and DBC methods for select choices of GB width and mesh size, respectively.)
As with the 1D simulations, a uniform (nominal) composition was employed as the initial condition.
Concentration maps from the RETS simulation (for $\sigma=0.8$) are shown in Figs.~\ref{subfig:2D_rets_Ni} and~\ref{subfig:2D_rets_Cr}.
At 10 dpa (60 days), the GB center (far from the triple junction) is depleted in Ni (concentration of $\approx0.06$) and enriched in Cr (concentration of $\approx0.27$). These concentrations are very close to those observed in 1D simulations of large grains.
RETS is observed to be slightly higher at the triple points. This is due to a slightly higher value of the normalized multiwell potential $\bar{g}_\text{mw}=1.1$ at triple points as compared to $\bar{g}_\text{mw}=1$ at the GB center. 
Concentration maps from the RIS simulation are shown in Figs.~\ref{subfig:2D_ris_Ni} and~\ref{subfig:2D_ris_Cr}.
At 10 dpa (60 days), monotonically varying Cr depletion and Ni enrichment profiles are observed, with a GB center concentration of {0.18} for Ni and {0.14} for Cr.
RIS is found to be identical in all the grains, due to the same grain size, but is expected to vary over certain grain sizes, as discussed in Sec.~\ref{results:grain_size}.
However, due to the smaller grain size in the 2D simulations, RIS is significantly lower than the 1D result from the 1 $\mu$m grain size.
Here, the effect of dimensionality (1D vs. 2D) also contributes to the variation in RIS.
These effects are compared in the next section.
Finally, concentration maps from the combined RIS+RETS simulation are shown in Figs.~\ref{subfig:2D_ris_rets_Ni} and~\ref{subfig:2D_ris_rets_Cr}.
Non-monotonic ``W"- and ``M"-shaped variation in Ni and Cr concentrations, respectively, are observed across the GB.
While the regions away from the GB center are depleted in Cr, the GB center itself is enriched relative to the nominal concentration.
On the other hand, Ni enrichment is observed at all locations across the GB. 
However, the extent of enrichment at the GB center is lower than that caused by RIS alone.

\subsection{{{Effect of grain size and dimensionality on GB RIS}}} \label{results:grain_size}

\begin{figure*}[htp!] 
\centering
    \begin{subfigure}[t]{0.8\textwidth}
        {\includegraphics[width=1\textwidth]{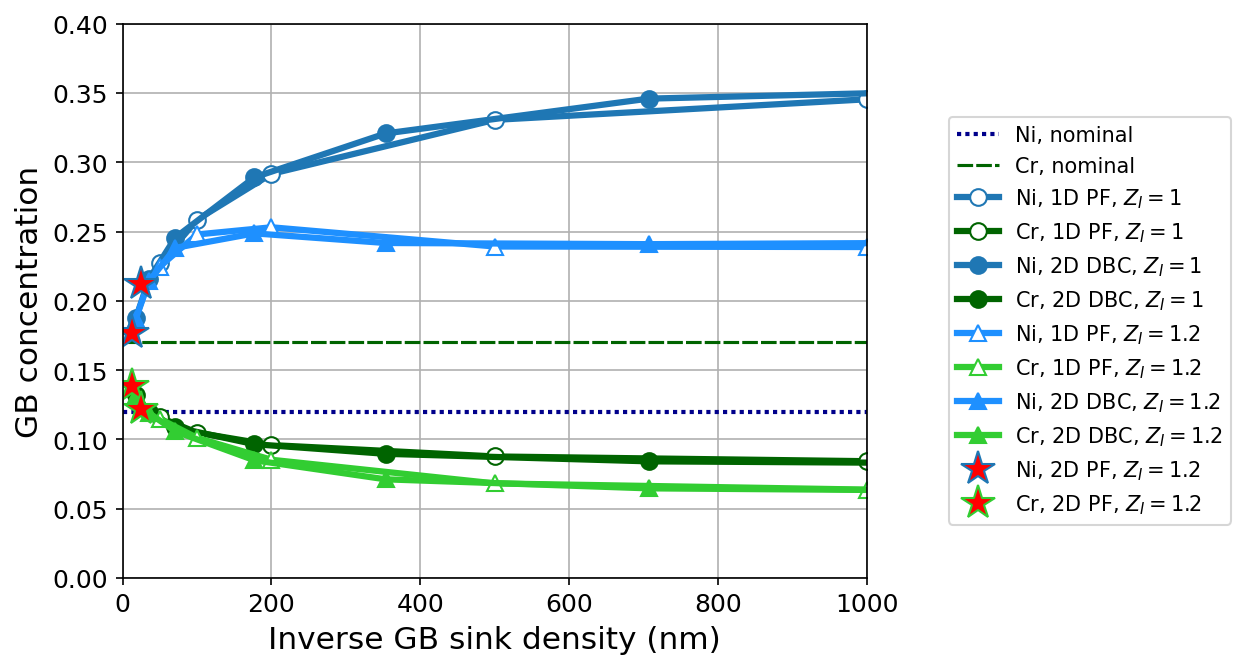}}
    \end{subfigure}
    \caption{GB segregation of as a function of the inverse of GB density ($d_\text{1D}$) for a bulk dislocation density of $\rho=10^{14}$ m$^{-2}$ and dislocation absorption efficiencies of $Z_I=1$ and $Z_I=1.2$ for SIAs. The results correspond to different simulation domains---1D PF, 1D DBC, 2D (hexagonal) PF, and 2D (square) DBC---all irradiated to 10 dpa at 500\,$^\circ$C. }
    \label{fig:1D_2D_RIS_vs_grain}
\end{figure*}

To enable comparison of RIS for the different grain sizes, geometries, and dimensions employed in this work, we define the GB sink density as the ratio of the GB length per grain to the grain area.
Thus, for the 1D system of grain length $d_\text{1D}$, the GB sink density is $1/d_\text{1D}$.
For the 2D square grain (implemented using the DBC method in Sec.~\ref{sec:impl:num}) with edge length $d_\text{2D,sq}$, the GB sink density is $2/d_\text{2D,sq}$.
Finally, for the 2D hexagonal grain whose short diagonal is $d_\text{2D,hex}$, the GB density is $2/d_\text{2D,hex}$.
In Fig.~\ref{fig:1D_2D_RIS_vs_grain}, we plot RIS from these different systems as a function of the inverse of GB sink density. 
The results are shown for two different dislocation sink bias cases: $Z_I=1$ and $Z_I=1.2$.
We note that the RIS concentrations from the 2D simulations are taken from the middle of those GB edges farthest from the corners/triple points.
At low GB sink densities or large grain sizes ($d_\text{1D}>500$ nm, right side of the plot), RIS is nearly constant, with no change in Ni and Cr concentrations.
With increased GB sink density or decreased grain size ($d_\text{1D}<500$ nm), RIS behavior reduces and the GB concentrations tend toward the nominal values.
No difference in RIS is observed between $Z_I=1$ and $Z_I=1.2$ at very high GB sink densities or very small grain sizes ($d_\text{1D}<100$ nm, left side of the plot).
However, for lower GB sink densities or larger grain sizes ($d_\text{1D}>100$ nm), the effect due to dislocation sink strength dominates, resulting in significant RIS differences between $Z_I=1$ and $Z_I=1.2$.
The biased absorption of SIAs when $Z_I=1.2$ leads to a significant suppression of Ni enrichment and a slight enhancement of Cr depletion.
Overall, for a given $Z_I$, excellent agreement is observed among the 1D DBC (1 nm mesh), 1D PF, and 2D (square) DBC (1 nm mesh) systems of equivalent GB sink density.
And while the 2D (hexagonal) PF simulation was only performed for two grain sizes ($d_\text{2D,hex}$ of 25 and 50 nm), the concentrations (``star'' markers) show excellent agreement with the other systems of equivalent GB sink density.

\subsection{{{RIS in AM microstructure}}} \label{results:ris_am}

Simulations of RIS at 500\,$^\circ$C were performed in 1D and 2D dislocation cells representative of the AM microstructure. 
The effects of dislocation cell size and dislocation density are studied to capture the variations observed in AM alloys depending on the AM technique, build geometry and process parameters.

\subsubsection{1D simulations of CW segregation} \label{results:ris_am:1D}

\begin{figure}[htp!]
\centering
    \begin{subfigure}[t]{0.475\textwidth}
        \includegraphics[width=1\textwidth]{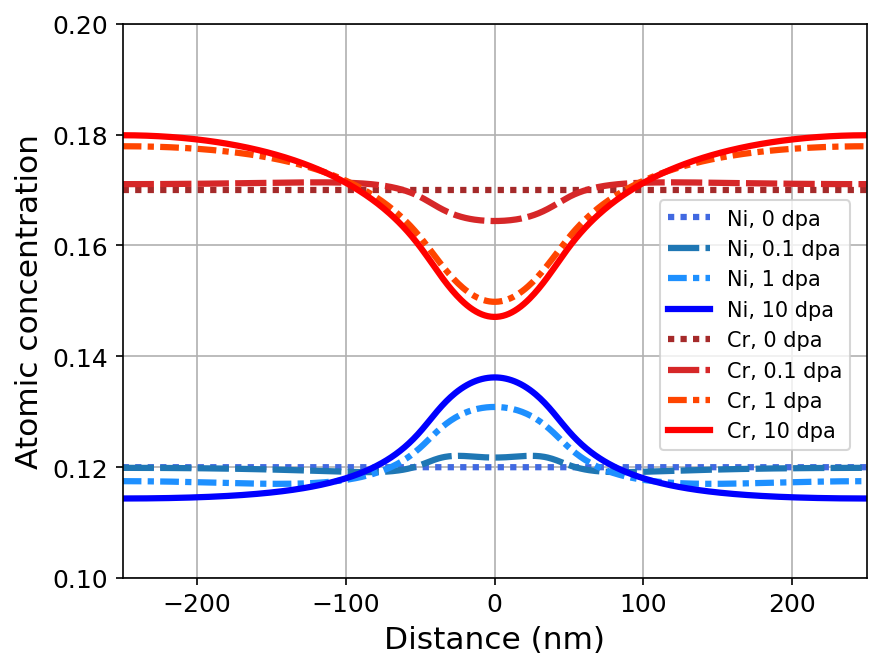}
    \caption{}
    \label{subfig:1D_am_noic_profile}
    \end{subfigure}
    ~
    \begin{subfigure}[t]{0.475\textwidth}
        \includegraphics[width=1\textwidth]{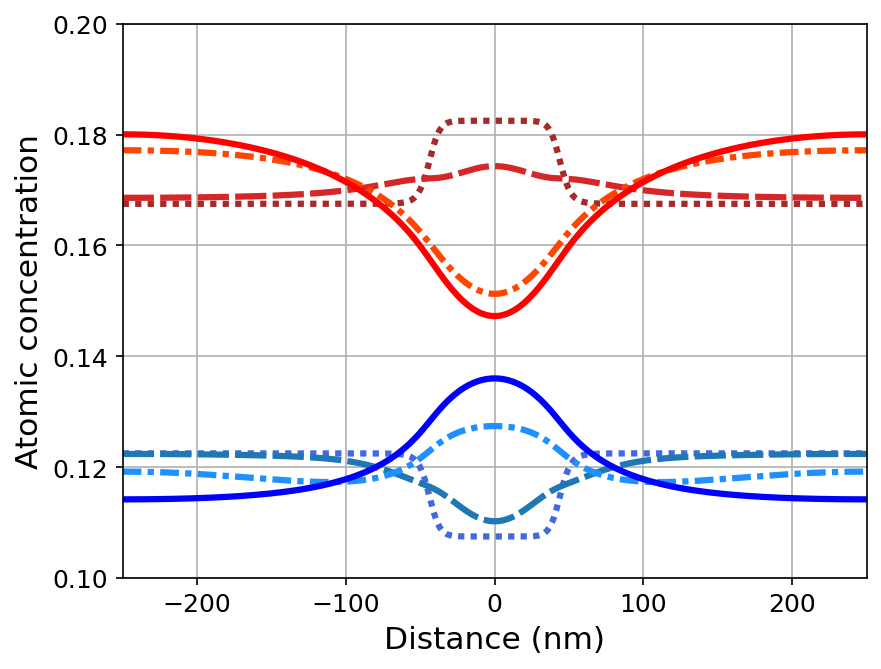}
    \caption{}
    \label{subfig:1D_am_ic_profile}
    \end{subfigure}
    \caption{RIS at a dislocation CW in a 1D AM microstructure of 500 nm dislocation cell size irradiated at 500\,$^\circ$C. The AM system corresponds to a cell interior dislocation density of $\rho_{b,c}=10^{14}$ m$^{-2}$ and a 100-nm-wide CW of dislocation density $\rho_{b,w}=10^{15}$ m$^{-2}$, with a dislocation absorption efficiency of $Z_I=1.2$ for SIA. (a) Initial condition with a homogeneous composition. (b) Initial condition consisting of pre-irradiation segregation at the CW.}
    \label{fig:1D_am_profile}
\end{figure}

1D simulations were performed on dislocation cell lengths ranging from 0.15 to 2 $\mu$m, and on a 100 nm wide dislocation CW~\cite{shang2021heavy,chen2024situ} at the center of the cell.
Concentration profiles from simulations with a CW dislocation density $\rho_{b,w}=10^{15}$ m$^{-2}$ and a dislocation absorption bias $Z_I=1.2$ for SIAs are shown in Fig.~\ref{fig:1D_am_profile}.
In Fig.~\ref{subfig:1D_am_noic_profile}, the initial condition was a uniform nominal composition, whereas in~\ref{subfig:1D_am_ic_profile}, a pre-irradiation microsegregation of Cr enrichment and Ni depletion was assumed following general observations in Ref.~\cite{shang2021heavy}.
Similar to RIS at GBs, RIS at CWs is characterized by Cr depletion and Ni enrichment at 1 dpa. 
For the case with pre-irradiation microsegregation, the effect of RIS is seen relative to the starting concentrations.
While the RIS profiles are distinct at low doses due to their different starting conditions, they become nearly identical at above 1~dpa as they approach steady state.
No persistent ``W" or ``M" shape is observed as the thermodynamic interaction between the solutes and dislocations at the CW were ignored.
Due to the biased absorption of SIAs by dislocations, RIS of Cr is greater in magnitude than RIS of Ni.

\begin{figure*}[htp!] 
\centering
    \begin{subfigure}[t]{0.48\textwidth}
        \includegraphics[width=1\textwidth]{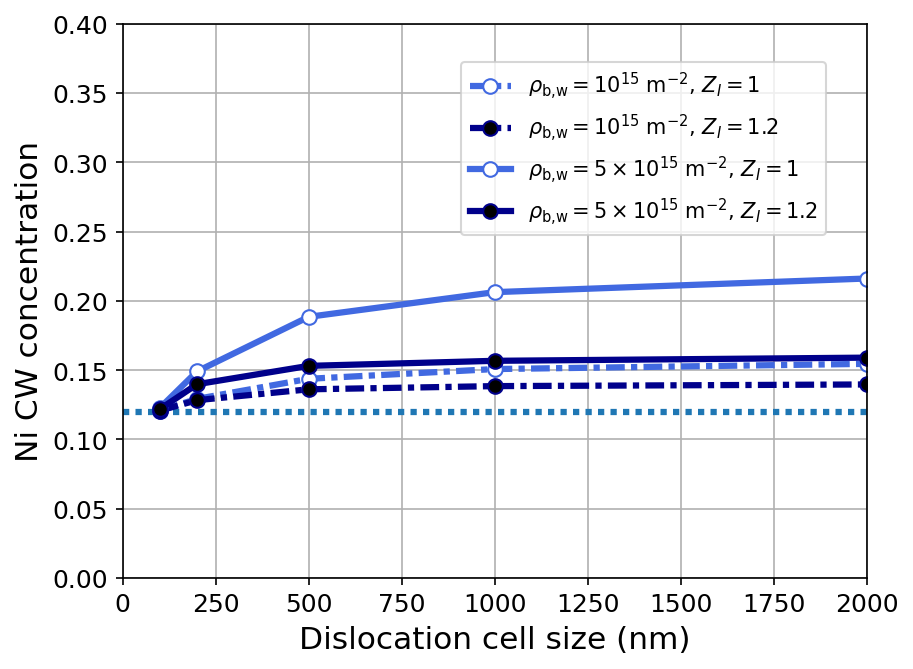}
    \caption{}
    \label{subfig:1D_CW_Ni_vs_size}
    \end{subfigure}
    ~~
    \begin{subfigure}[t]{0.48\textwidth}
        \includegraphics[width=1\textwidth]{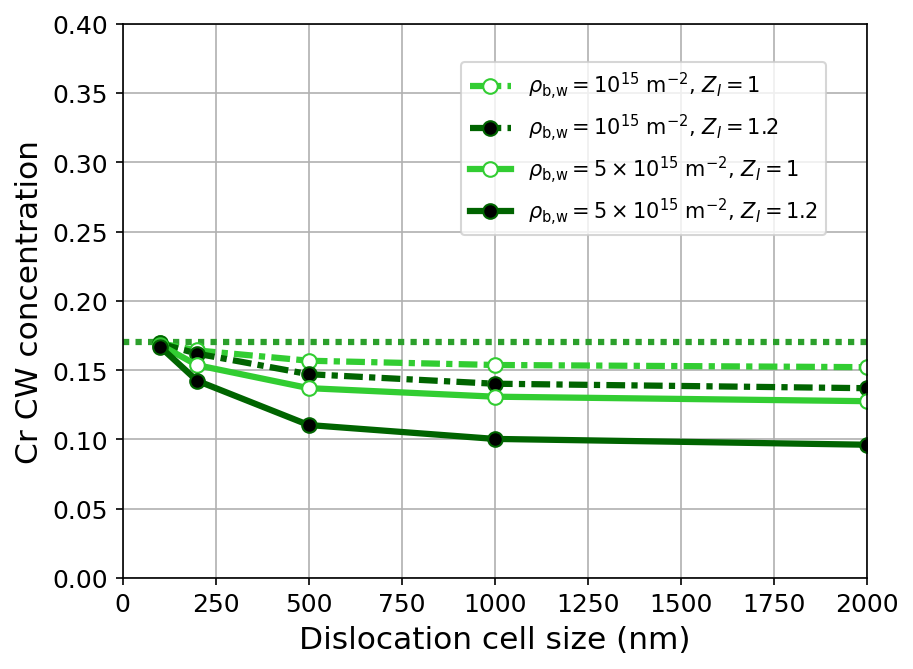}
    \caption{}
    \label{subfig:1D_CW_Cr_vs_size}
    \end{subfigure}
    \caption{RIS at a dislocation CW in a 1D AM microstructure as a function of dislocation cell length. The results correspond to different dislocation absorption efficiencies (i.e., $Z_I=1$ and $Z_I=1.2$) for SIAs. The AM system with a cell interior dislocation density of $\rho_{b,c}=10^{14}$ m$^{-2}$ and a CW density of $\rho_{b,w}=10^{15}$ m$^{-2}$ was irradiated to 1 dpa at 500\,$^\circ$C.}
    \label{fig:1D_CW_seg_vs_size}
\end{figure*}

In Fig.~\ref{fig:1D_CW_seg_vs_size}, concentrations of Ni and Cr from the center of the CW at 1 dpa are plotted for different cell sizes with the same CW width.
As with the grain size dependence of GB RIS (Fig.~\ref{fig:1D_2D_RIS_vs_grain}), RIS at CWs demonstrates a cell size dependence that decreases with diminishing cell size, but reaches a constant value beyond a size of 500 nm.
For comparison, results from different CW dislocation densities $\rho_{b,w}$ and sink efficiencies $Z_I$ are plotted.
Due to the increasing sink strength of CW with the increase in dislocation density $\rho_{b,w}$, greater RIS is observed.
However, for a given $\rho_{b,w}$, the absorption bias $Z_I=1.2$ leads to greater Cr depletion and lower Ni enrichment.
This effect of bias on RIS at CWs is similar to that observed for GBs. 

\subsubsection{2D simulations of CW segregation} \label{results:ris_am:2D}

\begin{figure}[htp!]
\centering
    \begin{subfigure}[t]{0.4\textwidth}
        \includegraphics[width=1\textwidth]{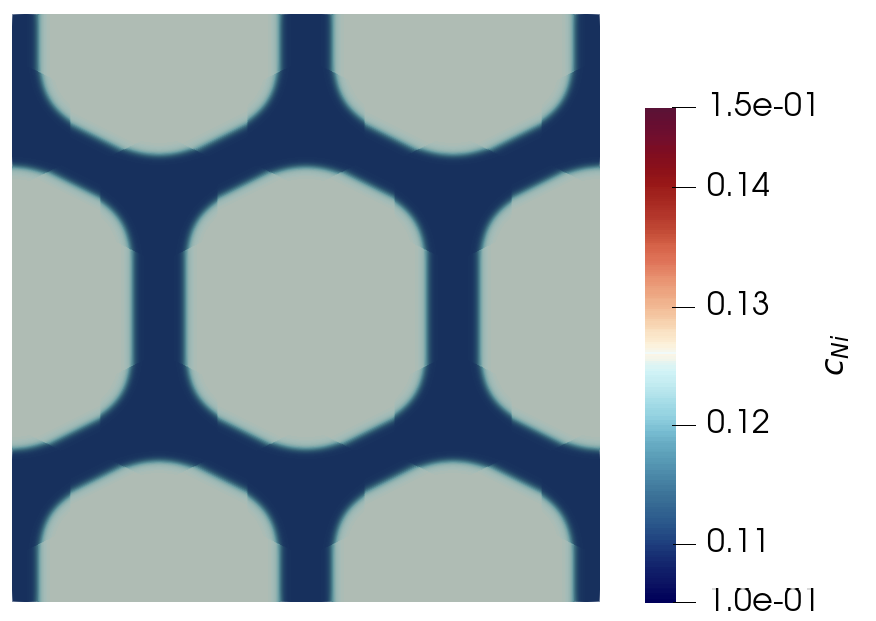}
    \caption{}
    \label{subfig:2D_am:ic_Ni}
    \end{subfigure}
    ~
    \begin{subfigure}[t]{0.4\textwidth}
        \includegraphics[width=1\textwidth]{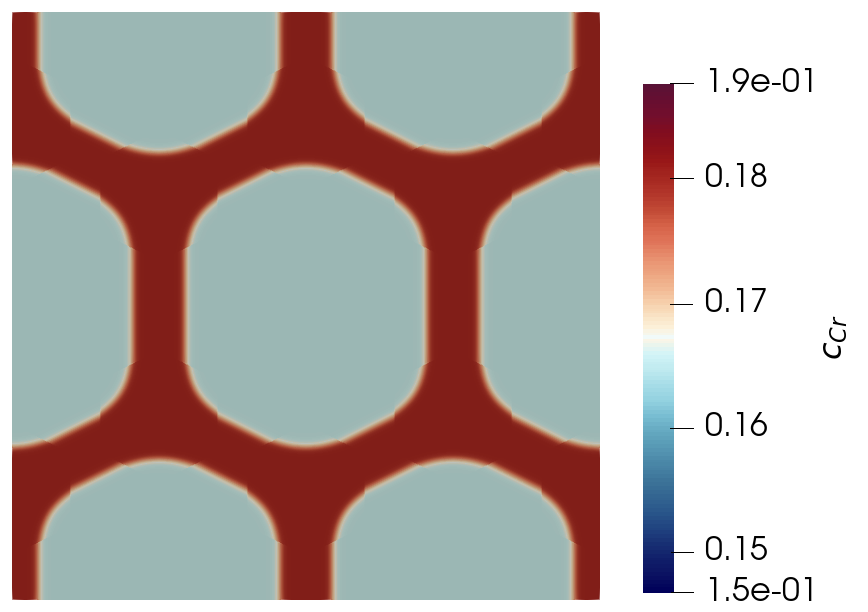}
    \caption{}
    \label{subfig:2D_am:ic_Cr}
    \end{subfigure}
    ~~~
    \begin{subfigure}[t]{0.4\textwidth}
        \includegraphics[width=1\textwidth]{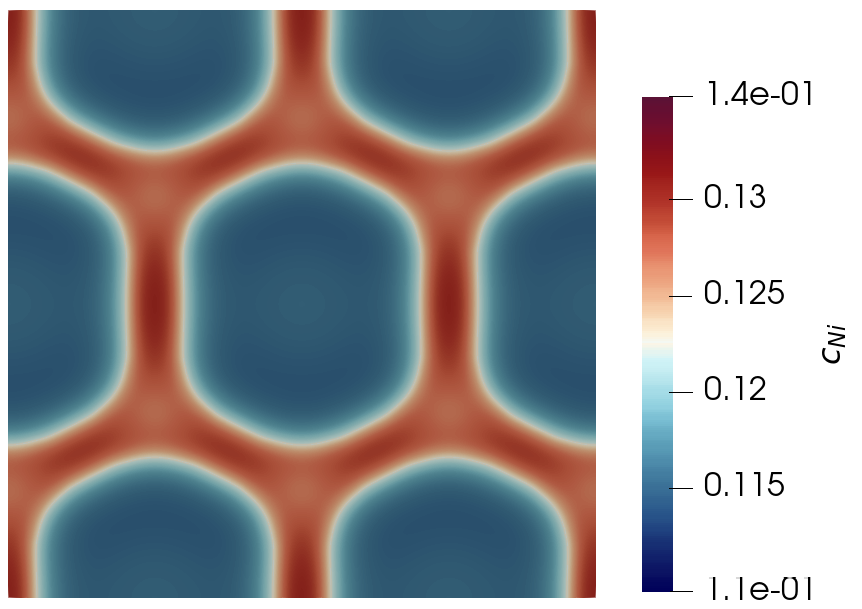}
    \caption{}
    \label{subfig:2D_am:no_dbc_Ni}
    \end{subfigure}
    ~
    \begin{subfigure}[t]{0.4\textwidth}
        \includegraphics[width=1\textwidth]{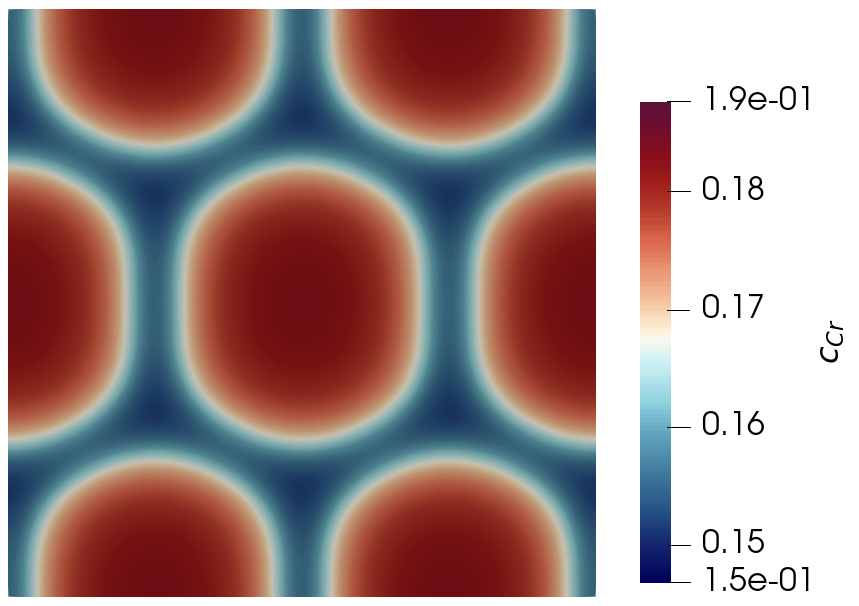}
    \caption{}
    \label{subfig:2D_am:no_dbc_Cr}
    \end{subfigure}
    ~~~
    \begin{subfigure}[t]{0.4\textwidth}
        \includegraphics[width=1\textwidth]{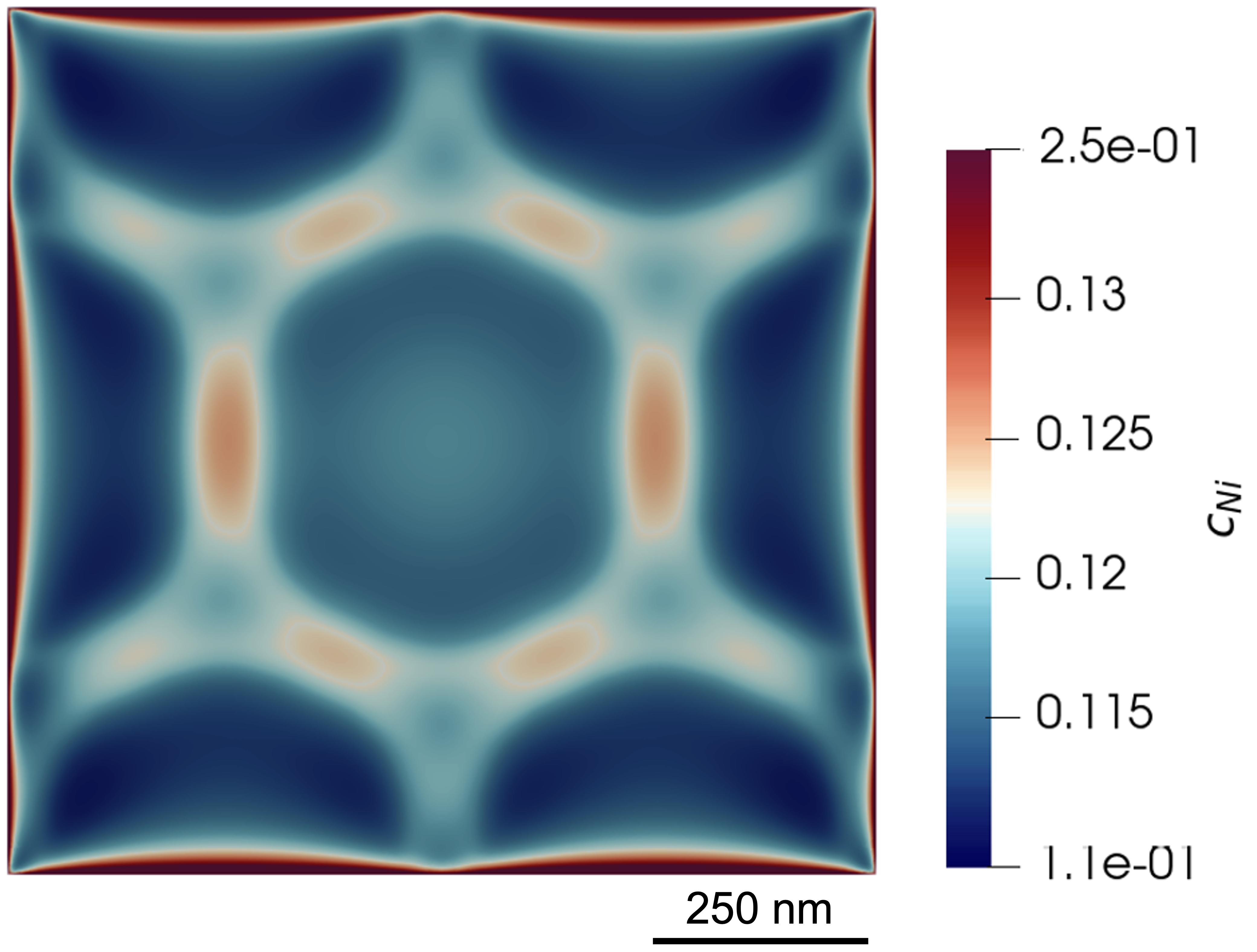}
    \caption{}
    \label{subfig:2D_am:dbc_Ni}
    \end{subfigure}
    ~
    \begin{subfigure}[t]{0.4\textwidth}
        \includegraphics[width=1\textwidth]{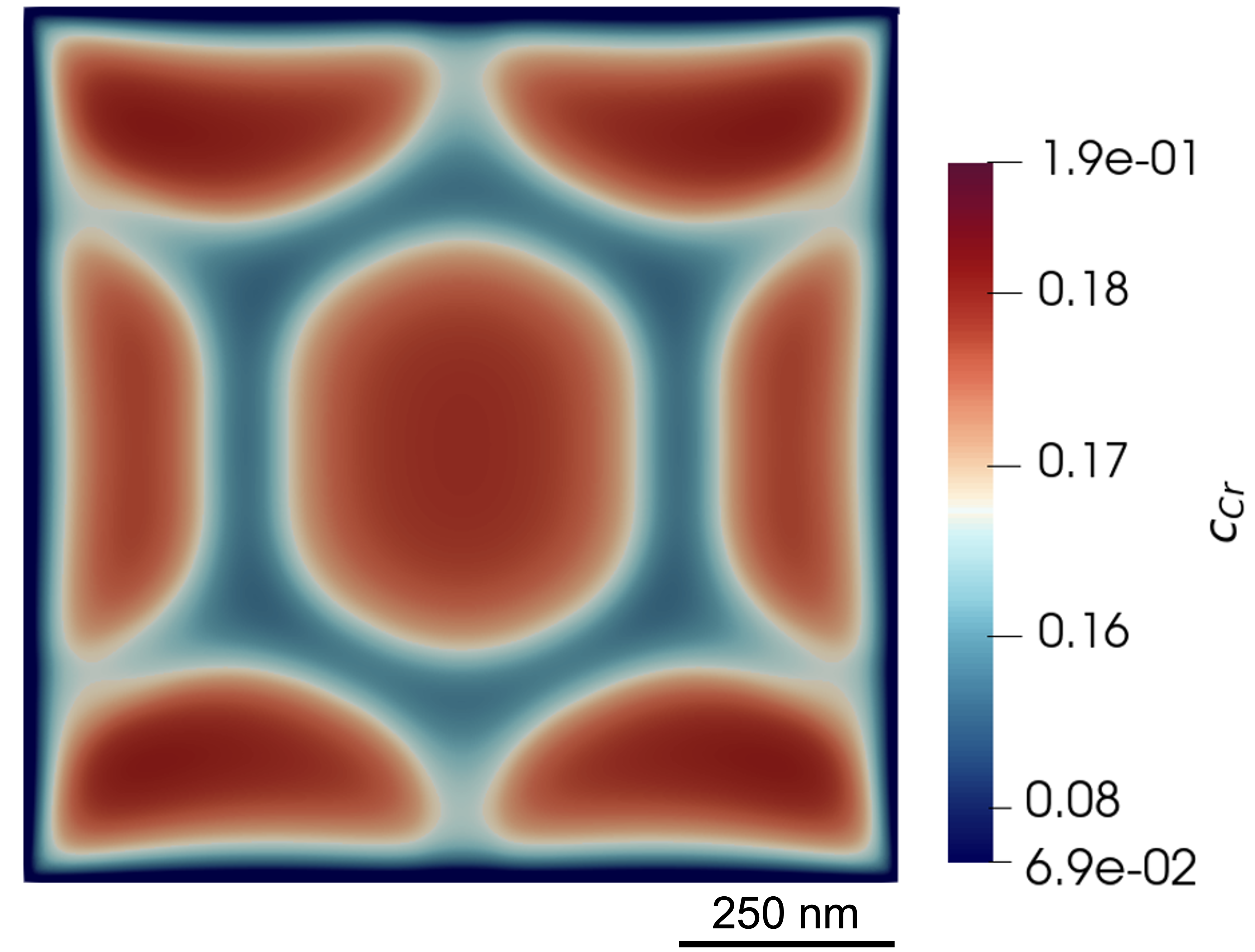}
    \caption{}
    \label{subfig:2D_am:dbc_Cr}
    \end{subfigure}
    \caption{RIS in a 2D hexagonal AM cell structure $500\times500$ nm$^2$ in size and irradiated to 1 dpa at 500\,$^\circ$C. (a,b) Initial condition consisting of pre-irradiation segregation at the CW. (c,d) Without DBC representing GB. (e,f) With DBC for point defect concentration applied to the edges of the square domain. The RIS width near the GB is exaggerated for illustration purposes by setting the same color level from 0.14 to 0.25 for $c_{Ni}$ in (e), and from 0.069 to 0.15 for $c_{Cr}$ in (f).}
    \label{fig:2D_am}
\end{figure}

2D simulations of RIS in dislocation cells are performed within a square domain of {1} $\times$ {1} $\mu$m$^2$ (Fig.~\ref{subfig:2D_micro:am}).
The pre-irradiation microsegregation at CWs with Cr enrichment and Ni depletion are shown in Figs.~\ref{subfig:2D_am:ic_Ni} and~\ref{subfig:2D_am:ic_Cr}, respectively.
The dislocation densities are set to $\rho_{b,c} = 10^{14}$ m$^{-2}$ and $\rho_{b,w}=10^{15}$ m$^{-2}$, and an SIA absorption bias given by $Z_I=1.2$ is used.
Two separate simulations are performed to test the influence of GBs on RIS in dislocation cells.
In the first, hexagonal cells $\approx0.5$ $\mu$m in width are initialized within a square domain.
This effectively models dislocation cells in large grains or those far from the influence of GBs.
The concentration maps in Figs.~\ref{subfig:2D_am:no_dbc_Ni} and~\ref{subfig:2D_am:no_dbc_Cr} 
correspond to 1~dpa and demonstrate Ni enrichment and Cr depletion at the CWs. 
As with the 1D simulations, a flip in segregation from the pre-irradiation microsegregation is observed under irradiation.
In the second simulation, the DBC is applied to all boundaries of the square domain to simulate RIS in dislocation cells near GBs or within small grains.
The composition maps in Figs.~\ref{subfig:2D_am:dbc_Ni} and~\ref{subfig:2D_am:dbc_Cr} show significant RIS at the GBs (edges), while RIS at the CWs is relatively low.
The maximum change in Cr concentration (in site fraction) is $-0.025$ at the CW, whereas it is $-0.1$ at the GB.
Similarly, the maximum change in Ni concentration is just $0.025$ at the CW but 0.125 at the GB.
Overall, reduced RIS at CWs is observed in the presence of the GBs due to the greater sink strength of the GBs.

\section{{Discussion}} \label{sec:discussion}

In the nuclear industry, there is an ongoing effort to increase the operating temperature of reactor designs to improve thermal efficiency, while material costs should remain as low as possible for improved economics.  As a result, iron-based materials with high-temperature mechanical performance (e.g., up to 750 \textdegree C or even greater) are of interest. 
However, such high temperatures result in appreciable thermally-driven kinetics for the diffusion of species that does not occur at lower operation temperatures.  Thus, understanding the contributions of the different physical mechanisms to the observed in-reactor segregation is vital for qualification of materials for advanced reactor technologies. 
To this end, we developed a new microstructure-level model to study different mechanisms of segregation in AM austenitic SSs under irradiation.

\subsection{PF model for GB segregation}

A multi-order-parameter PF model was developed that describes both RIS and RETS in multicomponent polycrystals.
The formulation also preserves the complete set of Onsager coefficients from the Onsager force-flux relations, thus allowing non-ideal kinetics ($L_{kV}/L^{{\upsilon}}_{kk}\neq-1$) to be captured.
Therefore, RIS mechanisms of solute-vacancy exchange or inverse Kirkendall effect ($L_{kV}/L^V_{kk}<0$), solute-vacancy drag ($L_{kV}/L^V_{kk}>0$) or vacancy wind effect, and solute-SIA binding can be incorporated.
While Manning's relations \cite{lidiard1986note} for concentrated multicomponent alloy were used in this work 
to parameterize $L^{{\upsilon}}_{kl}$, one could utilize analytic methods (e.g., the self-consistent mean-field and the Green's function) or atomistic calculations (e.g., kinetic Monte Carlo and molecular dynamics) ~\cite{ardell2016radiation} to accurately parameterize the off-diagonal $L_{ij}$ coefficients for systems in which their contribution or solute-vacancy drag is deemed important~\cite{choudhury2011ab}.

The grand potential formulation for TS locally imposes the equal diffusion potential condition between the bulk and GB phase free energies.
In this regard, the model is similar in scheme to the binary component TS model of Cha et al.~\cite{cha2002phase}. 
By employing carefully constructed Taylor free energies that allow predefined analytic relations for the phase concentrations~\cite{welland2017linearization}, the grand potential approach avoids the computational expense of explicitly solving the equal diffusion potential conditions to determine the phase concentrations.
Furthermore, we employed a mixed formulation \cite{boutin2022grand} of the grand potential model---involving both concentration $c_k$ and diffusion potential $\mu_{k1}$ variables related via Eq.~\ref{eq:tot_conc}---to strictly conserve the mass of the atomic components in the system.
A physically relevant GB width of 1 nm was employed in this work, limiting the overall size of the system that could be simulated in 2D.
Further study is warranted to explore the possibility of employing artificially large GB widths---to preserve quantitative GB energy, segregation, and solute drag effects---and enable mesoscale simulations of grain growth in large systems.

In \ref{appendix_A}, we proposed a method for parameterizing the RIS model for the local sink strength of the diffuse GB, based on the sharp-interface RBC.
As demonstrated in Fig.~\ref{fig:1D_verify_gb_sink}, this approach enables non-ideal GB sink behavior to be captured within the PF framework and provides a direct link to the GB structure.
Several experimental, analytical, and atomistic studies have described the GB misorientation dependence on GB sink absorption and RIS~\cite{duh2001numerical,field2015defect,xia2020radiation,gu2017point,barr2014anisotropic}. 
In the future, the present model can be coupled to a misorientation-based polycrystal PF model to investigate the effects of texture and GB anisotropy.
Further study is needed to examine the effect of the diffuse GB width on the solute excess, and not just on the GB center concentration.

With the development of advanced, high-resolution compositional characterization, complex nanoscale segregation profiles across GBs, such as ``W'' and ``M'' contours, have been revealed, necessitating new models to explain experimental observations.  The model formulation that we have developed enables a systematic investigation for TS without irradiation, RIS without TS, TS with radiation-enhanced diffusion (RETS), and RIS+RETS.  Our results show that there are important characteristic differences in the segregation profiles depending on the active physical mechanism(s), temperature, and radiation damage.  The model and its results enable a new means of interpreting experimental data. It also provides insights on how to structure experimental investigations into segregation mechanisms, given that the simulation results show qualitative differences in evolution with time and accumulated radiation damage. Thus, the interpretation of observed segregation behavior under irradiation is dependent upon the knowledge of the initial composition profile and the expected steady state, including the time and damage level to achieve it.

\subsection{GB segregation in austenitic SS}

Our results indicate that under irradiation, RIS and RETS concentration profiles have different characteristic fingerprints:  TS (driven by GB energetics) with or without radiation-enhanced diffusion in austenitic Fe-Cr-Ni will result in very narrow (approximately 1 nm) composition profiles with Cr enrichment and Ni depletion at the GBs.
The narrow profile widths of equilibrium TS are due to the short-range thermodynamic interactions localized to the GB plane.
Conversely, RIS (mediated by defect energetics and sustained long-range diffusion of point defects to GB sinks) results in wider composition profiles across the GB, with Cr depletion and Ni enrichment. In addition, the time scales at which these two behaviors reach steady state differ greatly, with TS and RETS reaching steady state over a few hours and 0.01 dpa, while RIS and RIS+RETS take much longer.  When RIS+RETS is active, the modeled segregation profiles have characteristics of both the mechanisms, resulting in ``W'' or ``M''-shapes. 

Several reports of thermal segregation observed during heat treatments performed prior to irradiation and RIS exist~\cite{kenik1998origin,busby1998influence}, though to our knowledge, there is a lack of systematic study in the literature on TS at GBs in austenitic SSs.  Only one study by Li et al.~\cite{li2013atomic} specifically probed the thermal segregation of 304 SS annealed at 500 \textdegree C for 30 minutes; with the use of atom probe tomography (APT), they found enrichment of Cr and other minor elements and depletion of Ni at GBs.
This observation is consistent with our simulations of Cr enrichment and Ni depletion via the TS mechanism (Fig.~\ref{fig:1D_ris_rets}). Other studies further investigate the influence of irradiation and RIS on preexisting segregation profiles.
For instance, Kenik et al.~\cite{kenik1998origin} reported pre-irradiation Cr enrichment in austenitic 304 and 316 SSs heat treated at ~1050\textdegree C for 20 minutes; for a similar alloy and conditions, Busby et al.~\cite{busby1998influence} also observed pre-irradiation segregation.  The pre-irradiation Cr enrichments observed in Refs.~\cite{kenik1998origin, busby1998influence} have been attributed to thermal non-equilibrium segregation resulting from the flux of supersaturated vacancies to the GB that arises during the cooling of the alloy from high temperatures~\cite{cole2002influence}.
According to this theory, for Cr to enrich at the GB, it would need to either 1) be the slower-diffusing species, contrary to experimental~\cite{yang2016roles} and atomistic~\cite{piochaud2014first} tracer diffusivity data, or 2) be dragged by vacancies via complex formation, which is also not supported by atomistic data~\cite{was2011assessment}, or 3) have certain thermodynamic interactions with impurities that segregate by the thermal non-equilibrium mechanism~\cite{ardell2016radiation}.
The alternative explanation we have considered is that Cr enriches at the GB due to TS~\cite{li2013atomic,wang2021density} (or, although unexplored in this work, the similar mechanism of thermodynamic co-segregation with impurity elements such as B and C ~\cite{li2013atomic,barr2018observation,yan2023effect}), and that pre-irradiation TS is not necessary for the development of the ``W" shape~\cite{nastar2005segregation}.

Upon irradiation, multiple studies have documented the evolution of GB segregation.  After proton irradiation to 1 dpa at 360\,$^\circ$C, ``W''-shaped Cr profiles were observed in the study by Kenik et al.~\cite{kenik1998origin}. For a similar alloy  irradiated (at 7$\times10^{-6}$ dpa/s) up to 5 dpa, the study by Busby et al.~\cite{busby1998influence} found the ``W''-shaped Cr profile first developed that later transitioned to a ``V''-shaped profile with increasing dose.  A more recent study by Barr et al.~\cite{barr2018observation} reported non-transient ``W''-shaped Cr and ``M''-shaped Ni profiles in 316 SS that underwent neutron irradiation (at ~2$\times 10^{-7}$ dpa/s) to 31 dpa in the temperature range of 390 to 410\,$^\circ$C.
Their APT profiles resemble our simulated profiles of Fig.~\ref{fig:1D_ris_and_rets} except for the slightly wider segregation widths, which might be attributed to trajectory aberration~\cite{li2013atomic}.  Our simulations suggest that such a non-transient ``W'' shape could develop for strong TS, while the transient ``W'' shape could develop for weak TS under RIS+RETS (Fig.~\ref{fig:1D_ris_and_rets}).
In another study employing APT, 304 SS that underwent neutron irradiation to 3.5 dpa was found to show both ``W" and ``V"-shaped Cr depletion~\cite{lach2021correlative}.
Interestingly, 2D concentration maps along the GB plane showed heterogeneous segregation with regions of enrichment and depletion for both Ni and Cr, suggesting the occurrence of an ``M"-shaped profile for Ni at certain locations of the GB. 

The modeling approach for TS under irradiation undertaken in the present work utilized a simple atomic density modification of the Fe-Cr-Ni free energy to demonstrate that ``W''-shaped Cr and ``M''-shaped Ni profiles resulting from RIS+RETS are indeed stable even at large doses. 
Although this approach ignores the possibility of heterogeneous segregation, ballistic mixing, and the evolution of GB structure with irradiation, it still reproduces complex GB segregation from a few fundamental physical principles. 
The effects currently neglected in the model could alter the degree of TS predicted by equilibrium and offer explanations for the observation (or lack thereof) of transient and non-transient ``W"-shaped profiles in the literature.  

In addition, this approach currently makes the simplification of ignoring the formation of secondary phases and the more complex interactions that are likely to occur in the presence of minor alloying elements and impurities.
For instance, the formation of Cr-rich carbides at the GB is known to deplete Cr along the GB during thermal and radiation-induced sensitization~\cite{dong2015microchemical,li2013atomic}.
These precipitates could destabilize TS of Cr in their vicinity, resulting in ``V'' profiles instead of ``W'' profiles~\cite{dong2015microchemical}.  
GB carbide formation might alter or eliminate the central peak that would occur in RIS+RETS, unless the line profile perpendicular to the GB is extracted over the carbide phase itself. In fact, Cr enrichment via TS may be required to produce Cr-rich carbides at GBs during irradiation, as RIS alone leads to depletion of Cr.
Additional impurities such as P and B and minor elements such as Si and Mo complicate accurate modeling and the interpretation of experimental results.  In irradiated austenitic SSs containing Si, Si is known to enrich the GB via SIA transport and Ni-Si clusters are known to form at the sinks~\cite{jiao2011novel,toyama2012grain}; Si is therefore likely to affect the segregation behavior of Ni at GBs. 
Irradiation is also known to alter pre-irradiation thermal segregation of other elements: while C, P, B and Mo enrich the GB under thermal conditions~\cite{cole2002influence,li2013atomic,tomozawa2013solute}, 
P enrichment is reported to be enhanced under irradiation, B enrichment appears to be largely unaffected, while that of C has been found to be complex and correlated with the segregation behavior of Cr~\cite{jiao2011novel}. Such complexities due to multiple segregation mechanisms and thermodynamic interactions between alloying elements are observed in ferritic Fe-Cr alloys as well. For example, in an irradiated Fe-Cr-Al~\cite{field2015evaluation}, Si has been shown to enrich the GB plane while Cr enriches the regions immediately adjacent to Si; Cr-rich clusters are additionally found within the bulk. In the ferritic phase, and possibly in the austenitic phase, segregation of Cr is likely affected by both Cr-carbide precipitation and possible interactions of Cr with Si or other elements segregating via RETS or RIS.
With better atomistic inputs, more sophisticated segregation models can be pursued in the future to describe these complex multinary interactions of Ni and Cr with other elements.
On the other hand, more systematic experimental characterizations of TS in high-purity austenitic Fe-Cr-Ni are needed to validate the predictions of the present model.

Our results of the grain size dependence of RIS (Fig.~\ref{fig:1D_2D_RIS_vs_grain}) indicate that nanocrystalline Fe-Cr-Ni alloys can exhibit superior radiation damage tolerance.
Indeed, Sun et al.~\cite{sun2015superior} found that an ultrafine-grained 304L (100~nm grain size) demonstrates superior resistance to void swelling and precipitation compared to its coarse-grained counterpart (35~$\mu$m grain size).
While radiation-resistant alloys with a high sink density (dislocations and GBs) have been proposed, they tend to suffer from microstructural instability due to the high energetic driving force for reducing these defects in the microstructure.
In this regard, the reduction in GB energy and/or mobility via TS of alloying elements offers opportunities to stabilize the microstructure.
Our PF model provides a preliminary formulation to investigate such concepts for austenitic SS.

The fundamental differences in segregation profiles and the time scales to reach steady state are important for assessing accelerated irradiation qualification efforts. Ion irradiation can induce radiation damage orders of magnitude faster than neutron irradiation, making it attractive as a substitute for neutron irradiation; however, accelerated irradiation damage changes the balance of kinetic factors driving microstructural evolution \cite{taller2024approach}.  Increased irradiation temperature is typically used to compensate for increased damage rates in ion irradiation to more closely match neutron irradiation damage. However, increased temperature will also alter TS or RETS behavior. By using the model developed in this work, the observed segregation behavior in ion-irradiated and neutron-irradiated material can be assessed to determine how well the different physical phenomena are being matched under different irradiation conditions.  

\subsection{RIS at dislocation CWs}

The results of the effects of dislocation density and absorption bias on RIS (Figs.~\ref{fig:1D_conc_vs_temp_disl} and~\ref{fig:1D_CW_seg_vs_size}) are important for cold-worked and AM alloys that have a high density of as-processed dislocations. They are also important for annealed alloys at high levels of radiation damage, because dislocations and subgrain structures are known to develop under irradiation.
A simple rate theory diffusion model with spatial variation in dislocation density and sink strength was utilized to study RIS in representative AM microstructures.
The inhomogeneous distribution of dislocations arising from the additive manufacturing process was found to result in RIS at the dislocation CWs in addition to GBs. 
Our simulation results are generally in qualitative agreement with the preliminary characterization results found in the literature. 
In heavy ion irradiation of 316LN at 450\,$^\circ$C, Ni enrichment and Cr depletion at both high-angle GBs and dislocation CWs were observed~\cite{shang2021heavy}.
As predicted by our simulations, RIS at both CWs and GBs was observed, with peak RIS being lower in magnitude but greater in width for the CWs.
We presented the results of irradiations of up to a dose of only 1 dpa, since significant dislocation recovery is expected in reality at higher dpa levels.
For instance, Chen et al.~\cite{chen2024situ} observed the dislocation cell structures in 316H and 316L to recover and homogenize at doses above a few dpa under ex situ ion irradiations at 300\,$^\circ$C and 600\,$^\circ$C.
Therefore, concurrent evolution of RIS with the dislocation cell structure must to be considered for a complete picture.
In a recent work~\cite{jokisaari2024defect}, we coupled the present model of Sec.~\ref{sec:model:pf_am} with a dislocation evolution model in order to study such concurrent evolution.
In that work, we found pipe diffusion to be important in accurately accounting for the kinetics of dislocation recovery.
With experimental evidence for pre-irradiation microsegregation persisting at the CWs ~\cite{chen2024situ}, it will be important in the future to also account for TS to dislocations, as well as its effects on dislocation recovery.
Our PF model for GB segregation is also expected to be useful for AM austenitic SS, as Cr enrichment at high-angle GBs in AM 316LN SS~\cite{shang2021heavy} has been observed.

\subsection{Limitations of the RIS model}
Although the simulation results of GB segregation profiles show a consistent trend with experimental observation, the model is limited by uncertainties in the SIA energetics of austenitic SS. Due to the lack of comprehensive experimental and simulation data, the model neglects potential differences in the mobility of mixed interstitial dumbbells (e.g., Fe-Cr, Fe-Ni, and Ni-Cr) by using a single SIA migration energy~\cite{yang2016roles}. Additionally, the binding energies of interstitial dumbbells are treated as adjustable parameters based on previous studies~\cite{yang2016roles,wiedersich1979theory,LAM1983106}.
Atomistic calculations~\cite{piochaud2014first} of $\langle 100 \rangle$ dumbbell energetics in Fe-20Cr-10Ni have shown that Fe-Fe is the most stable and Cr is likely to occur as mixed dumbbells, whereas Fe-Ni, Ni-Ni and Cr-Cr are very unlikely to occur.
While the binding energies (i.e. favorable Cr and unfavorable Ni transport via SIA) utilized in Yang et al.~\cite{yang2016roles} and in the present work are qualitatively consistent with the atomistic calculations, more accurate parameterization is required. These energetics are crucial as they influence the preferred formation and mobility of mixed dumbbell configurations. Therefore, these uncertainties may lead to deviations of the predicted Onsager coefficients, RIS tendencies, and RIS profiles. 
The significance of migration barriers and binding energies of mixed interstitial dumbbells has been demonstrated in various BCC Fe-based steels~\cite{Messina_SCMF,Ke_RIS_RPV,Ke_RIS_FM}, and similar chemical coupling mechanisms between solutes and point defects are expected to occur in austenitic SS. The lack of available energy and property data involving SIAs in austenitic SS highlights the need for accurate thermo-kinetic property prediction. Such information is typically difficult to obtain from experiments, and the complex magnetic and chemical interactions caused by concentrated solute elements further challenge accurate descriptions of alloy energetics. 

Our RIS results with absorption bias could be an overestimation due to the assumption of a strong bias factor of 20\%, although the results will only differ quantitatively, not qualitatively. 
An accurate value for $Z_I$ is not well known due to the complexity of elastic interaction between the different SIA configurations and the specifics of dislocation character and distribution~\cite{golubov20121}. Improved understanding of absorption bias is therefore needed for more accurate predictions of its effect on RIS.
Additionally, the RIS model can be improved by considering defect clusters and voids, which are important for microstructure evolution in nuclear materials exposed to high-flux irradiation environments, such as those in advanced reactors~\cite{AITKALIYEVA2017253}. In addition to sinks such as GBs and dislocations, defect clusters and voids act as temporary and permanent traps, respectively, for point defects~\cite{chen2024situ}. 
For example, oversized solutes such as Hf, Zr, Ti and Nb have been suggested to bind with vacancies and reduce Cr depletion under certain irradiation conditions~\cite{hackett2008mechanism,ardell2016radiation}.
Interstitial impurities such as C and N have been suggested to reduce Ni and Si enrichment in alloys similar to 304 and 316 SS due to the increased formation of dislocations loops~\cite{ardell2016radiation}.  Incorporating these factors can improve the quantitative predictions of the model.
Their formation and evolution can significantly impact the dose rate sensitivity
of RIS. 
Developing an integrated model that couples these factors is crucial for understanding microstructure evolution ~\cite{jokisaari2024defect,jokisaari_2023,Ke_flux_effect}. This can help establish the physics-based correlation between ion irradiation and neutron irradiation for accelerating the qualification of modern nuclear structural materials~\cite{taller2024approach}.

\section{{Conclusions}} \label{sec:conclusions}

A multi-order-parameter PF model was developed to describe radiation-induced composition changes arising from multiple mechanisms such as preferential solute-point defect transport (RIS), biased sinks at dislocations, and solute-GB thermodynamic interaction (TS).  The dependence of RIS on temperature and GB sink density was presented using 1D and 2D simulations.
An expression for the local sink strength of the diffuse GB was developed and verified using a sharp-interface RBC.  Although the PF model was implemented for static GBs, it is expected to be further useful in studying composition evolution at moving GBs. 

The mesoscale model was used to simulate FCC Fe-Cr-Ni as a simplified system representative of austenitic SSs.  We find characteristic ``fingerprints'' for segregation profiles depending on the physical mechanism(s) involved. RETS without RIS results in narrow segregation profiles at the GB, while RIS without RETS results in broad segregation profiles. When both RETS and RIS occur, non-monotonic ``W''- and ``M''-shaped profiles for Cr and Ni, respectively, were found to persist to large irradiation doses.
In addition, the time scales to achieve steady-state segregation at GB differ between the RIS and RETS mechanisms and are a function of temperature.  For example, at 500\textdegree C, RETS achieves steady state within hours, TS within days, and RIS and RIS+RETS over several months.  
Furthermore, we observe the effect of biased point defect absorption on RIS.  For large dislocation densities, unbiased point defect absorption by dislocations suppressed RIS of both Cr and Ni at the GB; however, strongly biased absorption of SIAs suppressed Ni enrichment but slightly enhanced Cr depletion relative to the unbiased case.

In addition to studying GB segregation, a reduced model was applied to study RIS in spatially resolved dislocation cell structures, which is representative of AM materials.
Similar to RIS at the GB, Cr depletion and Ni enrichment were predicted to occur at dislocation CWs.  The magnitude of segregation at CWs is less in comparison to that at GBs, but the width of the segregation is greater due to the greater width of the dislocation CWs.  
In the absence of thermodynamic interactions with dislocations, pre-irradiation microsegregation at the CW was found to have little effect on the RIS profile at large damage levels.  In the future, the model is expected to be coupled with dislocation generation and recovery models in order to describe the concurrent evolution of RIS and dislocation cell structure.

Our modeling results provide evidence for a new interpretation of experimentally observed GB segregation profiles of irradiated materials, as well as the first computational study of segregation at dislocation CWs commonly found in AM materials.  Although our results qualitatively align well with experimental results in the literature for austentic SSs, future experimental investigations on high-purity austenitic Fe-Cr-Ni are needed for detailed validation of the model.  In addition to providing new physical interpretations of experimental data, the proposed model can be a key tool in accelerated qualification of irradiated materials. 

\appendix
\setcounter{section}{0}
\setcounter{figure}{0}
\setcounter{equation}{0}
\renewcommand{\theequation}{A\arabic{equation}}
\renewcommand{\thetable}{A\arabic{table}}
\renewcommand{\thefigure}{A\arabic{figure}}

\section{Analytic relations for RIS at GBs} \label{appendix_A}
\subsection{Sink strength of GB}
In the sharp-interface implementation, an RBC for the point defect concentration is imposed as:
\begin{flalign} \label{eq:abs_rate_rbc}
    J_{{\upsilon}}{\cdot}\hat{n} = D_{{\upsilon}} \alpha^{-1} (c_{{\upsilon}}-c^e_{{\upsilon}}), && 
\end{flalign}
where $J_{{\upsilon}}{\cdot}\hat{n}$ is the flux of point defect normal to the GB plane {and $\alpha^{-1}$ is a parameter governing the rate of point defect absorption or emission by a non-ideal GB \cite{gu2017point}}.
In the PF implementation, the total absorption rate across a planar GB per unit area can be written as:
\begin{flalign} \label{eq:abs_rate_pf}
    S_{{\upsilon}} = \int^{+\delta/2}_{-\delta/2} k^2_g D_{{\upsilon}} (c_{{\upsilon}}-c^e_{{\upsilon}}) g_\text{sink} dx.  && 
\end{flalign}
The absorption rates between the sharp-interface model with RBC and the PF model can be related as $2 J_{{\upsilon}}{\cdot}\hat{n} = S_{{\upsilon}}$.
However, since $c_{{\upsilon}}$ is expected to vary spatially within the GB, one cannot accurately evaluate the above integral. 
As an approximation, we evaluated it by assuming that $c_{{\upsilon}}$ is uniform within the diffuse GB, and related the coefficients between the sharp-interface and PF models as:
\begin{flalign} \label{eq:abs_rate_equiv} \nonumber
    2 D_{{\upsilon}} \alpha^{-1} &= \int^{+\delta/2}_{-\delta/2} k^2_g D_{{\upsilon}} g_\text{sink} dx  \\
    &= k^2_g D_{{\upsilon}}  \int^{1}_{0} g_\text{sink} \frac{dx}{d\eta_i} d\eta_i = k^2_g D_{{\upsilon}} \delta \,C_\text{sink},  &&
\end{flalign}
where $C_\text{sink}$ can be obtained analytically or numerically. 
For different choices of the PF sink indicator functions shown in Fig.~\ref{fig:pf_functions}, we get: $C_\text{sink} = 0.67$ for $g_\text{sink} = 16 \eta^2_i \eta^2_j$ (wide bell-shaped), $C_\text{sink}=0.32$ for $g_\text{sink} = 65536 \eta^8_i \eta^8_j$ (narrow bell-shaped), and $C_\text{sink}=0.89$ for $g_\text{sink} = 1-0.5\left[1+\tanh{((\chi-\chi_\circ)\beta_\circ)}\right]$ (smooth step-shaped).
Using Eq.~\ref{eq:abs_rate_equiv}, we parameterize $k^2_g$ in the PF model as:
\begin{flalign} \label{aeq:rbc_pf_relation}
    k^2_g = \frac{2 \alpha^{-1}}{\delta\,C_\text{sink} }. &&
\end{flalign}
Neglecting elastic interactions between point defects and GB dislocations, and recombination in the bulk, expressions of $\alpha$ in terms of misorientation angle or dislocation spacing have been derived by Duh et al.~\cite{duh2001numerical} (for low-angle symmetric tilt and high-angle GB) and Gu et al.~\cite{gu2017point} (for low-angle symmetric tilt GB).

\subsection{Steady-state concentrations}

With the sharp-interface model, the evolution of point defect concentration in 1D follows:
\begin{flalign} \label{aeq:conc_evol_rbc}
    \frac{\partial c_{{\upsilon}}}{\partial t} = -\frac{d J_{x,{{\upsilon}}}}{dx}+ P_{{\upsilon}} - k^2_{b,{{\upsilon}}}D_{{\upsilon}}(c_{{\upsilon}}-c^e_{{\upsilon}}), &&
\end{flalign}
with an RBC (Eq.~\ref{eq:abs_rate_rbc}) applied to model the GB sink.
Here, recombination between vacancies and SIAs, which is only expected to contribute at low temperatures, has been neglected. 
At steady state ($\partial c_{{\upsilon}}/\partial t = 0$), the point defect concentrations at the grain center ($dJ_{x,{{\upsilon}}}/dx = 0$) are given by:
\begin{flalign} \label{aeq:rbc_conc_bulk}
    c_{{\upsilon}} = c^e_{{\upsilon}} + \frac{P_{{\upsilon}}}{k^2_{b,{{\upsilon}}} D_{{\upsilon}}}.  &&
\end{flalign}
Furthermore, neglecting bulk sinks ($k^2_{b,{{\upsilon}}}=0$) and integrating Eq.~\ref{aeq:conc_evol_rbc} from the grain center to the GB plane gives us:
\begin{flalign} 
    \int^{J^b_{x,{{\upsilon}}}}_{J^g_{x,{{\upsilon}}}} dJ_{x,{{\upsilon}}} =  \int^{\pm d/2}_0 P_{{\upsilon}} dx, &&
\end{flalign}
where $d$ is the grain length and $J^g_{x,{{\upsilon}}}(x=0)$ and $J^b_{x,{{\upsilon}}}(x=\pm \frac{d}{2})$ are the fluxes at the GB plane and grain center, respectively.
In the RBC model, $J_{x}$ varies monotonically from the grain center to the GB plane, with the maximum in $J_{x}$ occurring at the GB plane.
Noting that $J^b_x=0$, and $J^g_{x,{{\upsilon}}}(x=0)$ is given by Eq.~\ref{eq:abs_rate_rbc}, we obtain the concentration at the GB, as:
\begin{flalign} \label{aeq:rbc_gb_conc}
    c_{{\upsilon}} = c^e_{{\upsilon}} + \frac{P\,d}{2D_{{\upsilon}}\alpha^{-1}}. &&
\end{flalign}
By substituting Eq.~\ref{aeq:rbc_pf_relation}, we can get an approximate relation for the GB center concentration in the PF model:
\begin{flalign} \label{aeq:pf_gb_conc}
    c_{{\upsilon}} = c^e_{{\upsilon}} + \frac{P\,d}{\delta\,C_\text{sink}k^2_gD_{{\upsilon}}}. &&
\end{flalign}
We note that in the PF model, the point defect absorption at the GB is diffuse, and the flux can exhibit a non-monotonic variation from the grain center to the GB center.

\begin{figure}[htp!]
\centering
    \begin{subfigure}[t]{0.48\textwidth}
        {\includegraphics[width=1\textwidth]{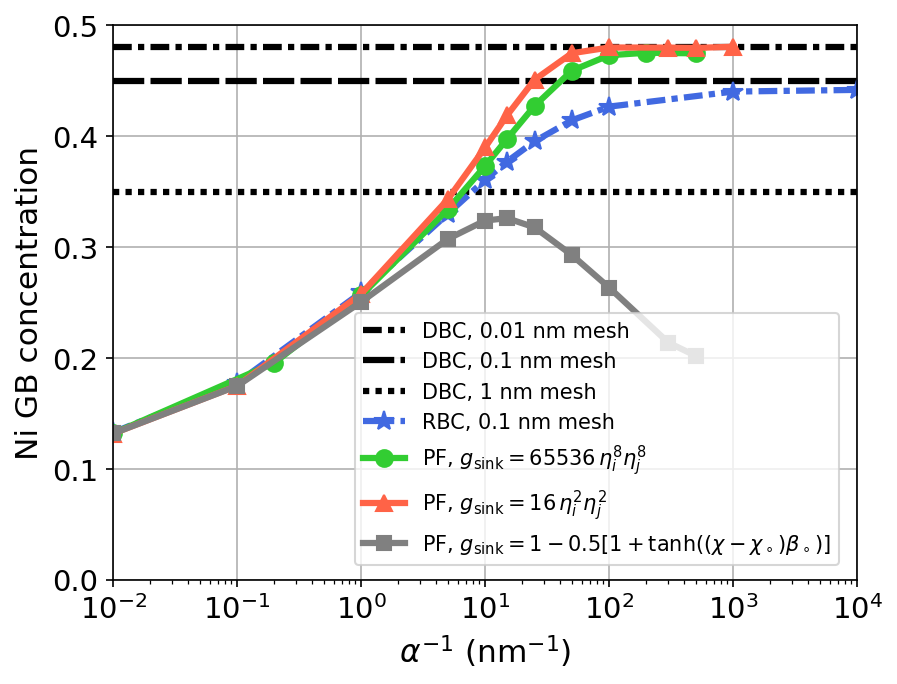}}
    \caption{}
    \label{subfig:1D_verify_gb_sink:Ni}
    \end{subfigure}
    ~
    \begin{subfigure}[t]{0.48\textwidth}
        {\includegraphics[width=1\textwidth]{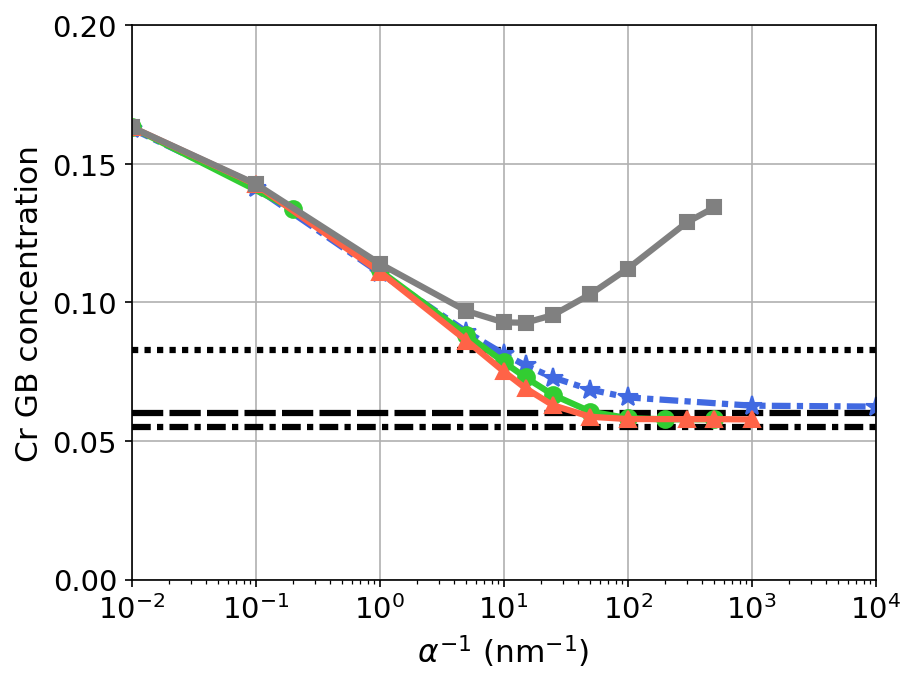}}
    \caption{}
    \label{subfig:1D_verify_gb_sink:Cr}
    \end{subfigure}
    ~~~ 
    \begin{subfigure}[t]{0.48\textwidth}
        {\includegraphics[width=1\textwidth]{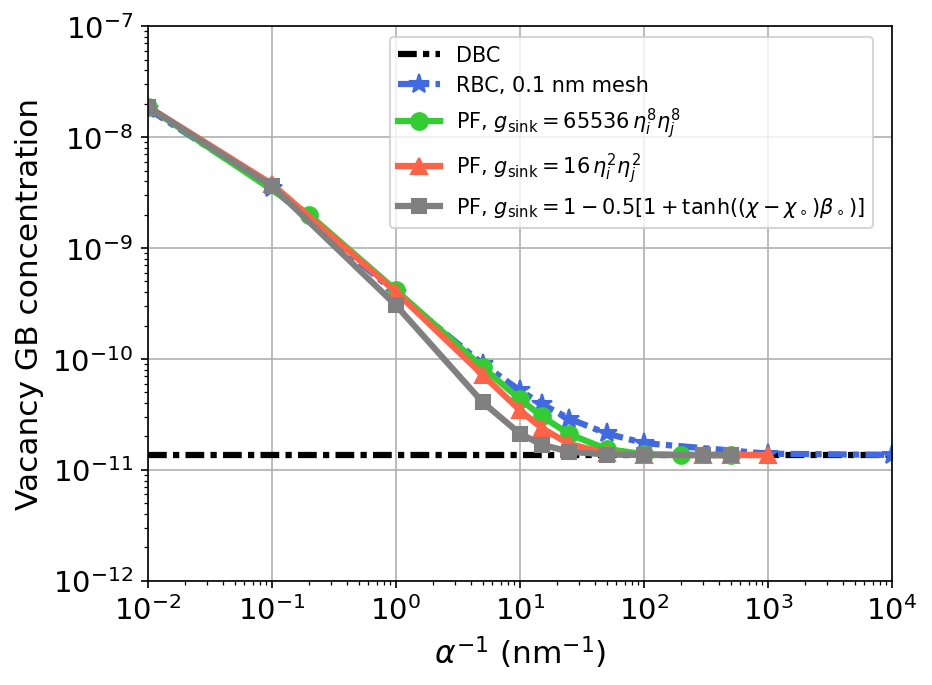}}
    \caption{}
    \label{subfig:1D_verify_gb_sink:vac}
    \end{subfigure}
    ~  
    \begin{subfigure}[t]{0.48\textwidth}
        {\includegraphics[width=1\textwidth]{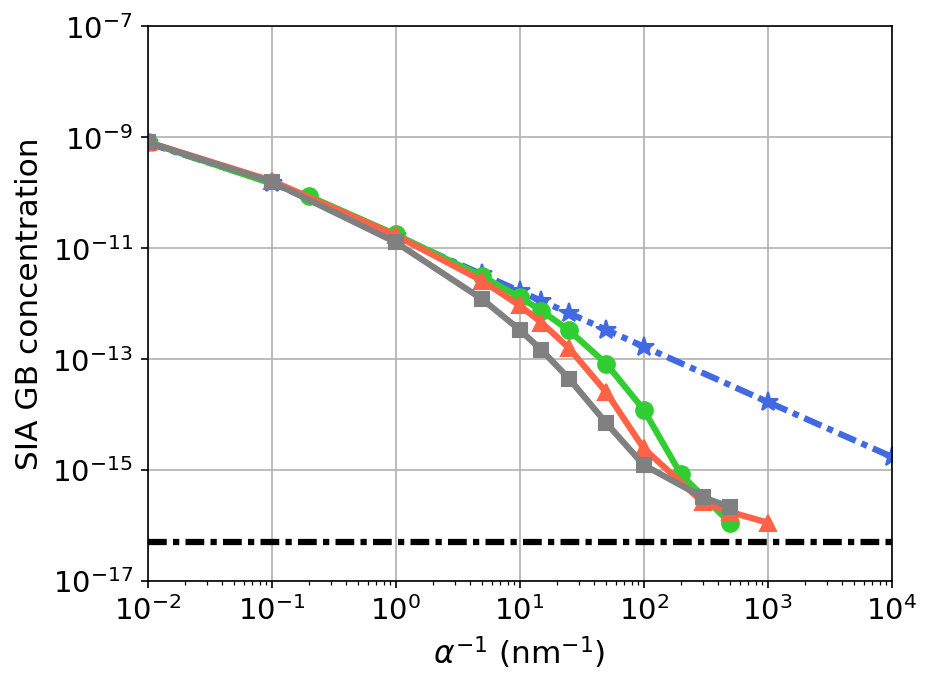}}
    \caption{}
    \label{subfig:1D_verify_gb_sink:sia}
    \end{subfigure}
    \caption{RIS at GB for (a) Ni, (b) Cr, (c) vacancy, and (d) SIA as a function of $\alpha^{-1}$ as calculated from a 1D system of 1 $\mu$m grain size irradiated to 10 dpa. Results are shown for different PF sink functions and compared against sharp-interface simulations with DBC and RBC.}
    \label{fig:1D_verify_gb_sink}
\end{figure}

\section{Verification of PF RIS results} \label{appendix_B}

In this appendix, the steady-state GB center concentrations from the PF implementation are compared against the GB concentrations from DBC and RBC implementations.
All results correspond to a grain size of 1 $\mu m$ and the parameters listed in Table~\ref{tab:myfirstlongtable}.
Details on the different implementations, including mesh sizes, are provided in Sec.~\ref{sec:impl:num}.
In Fig.~\ref{fig:1D_verify_gb_sink}, the results are compared as a function of $\alpha^{-1}$ at 500\,$^\circ$C.
At low values of $\alpha^{-1}<1$, excellent agreement is found between the different methods.
For $\alpha^{-1}>1$, the PF model with the smooth step-shaped sink function shows significant deviation in SIA concentration, as well as differing trends in the atomic concentrations.
On the other hand, for $1<\alpha^{-1}<10$, reasonably good agreement is found between the PF models with bell-shaped sink functions and the RBC model; for $\alpha>10$, some deviations are observed between these methods.
Due to the finite width of the GB in the PF model, point defect concentrations tend to reach ideal (equilibrium or DBC) values at a lower $\alpha^{-1}$.
The DBC method with a mesh size of 1 nm closely aligns with the PF models that employ the bell-shaped sink function.

\begin{figure}[htp!]
\centering
    \begin{subfigure}[t]{0.48\textwidth}
        \includegraphics[width=1\textwidth]{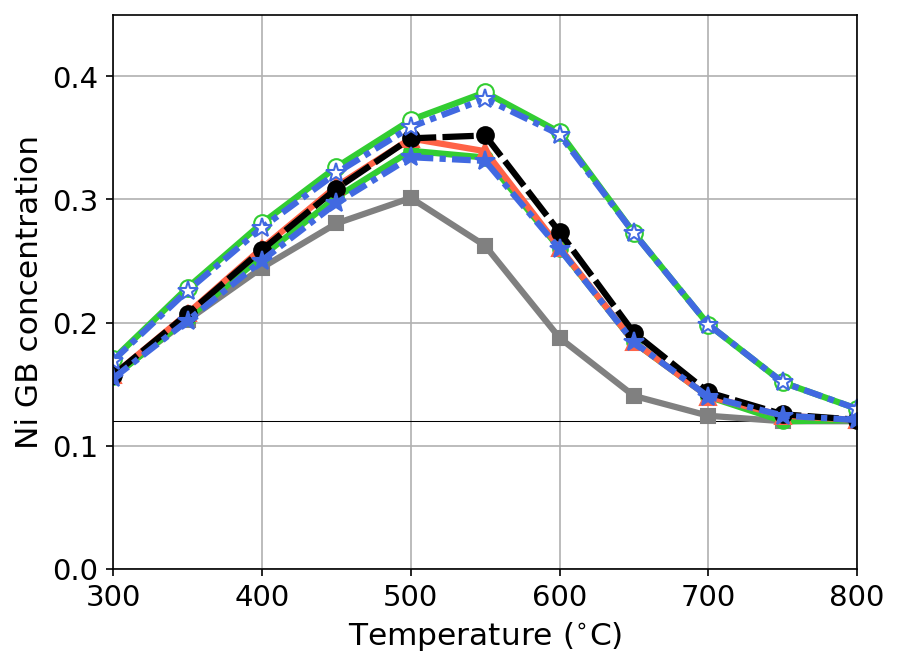}
    \caption{}
    \label{subfig:1D_verify_temp:Ni}
    \end{subfigure}
    ~
    \begin{subfigure}[t]{0.48\textwidth}
        \includegraphics[width=1\textwidth]{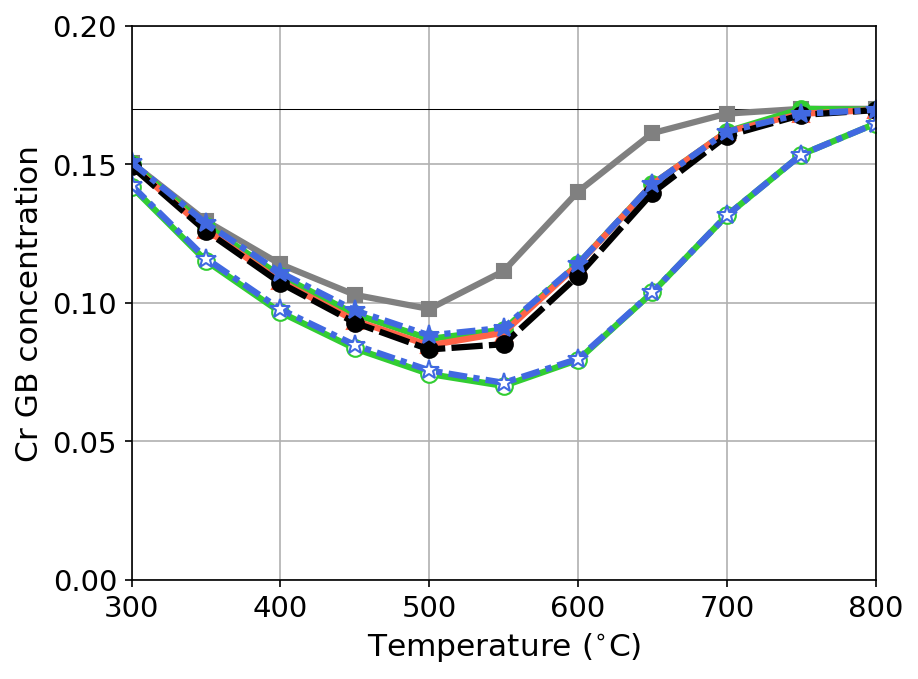}
    \caption{}
    \label{subfig:1D_verify_temp:Cr}
    \end{subfigure}
    ~~~ 
    \begin{subfigure}[t]{0.48\textwidth}
        \includegraphics[width=1\textwidth]{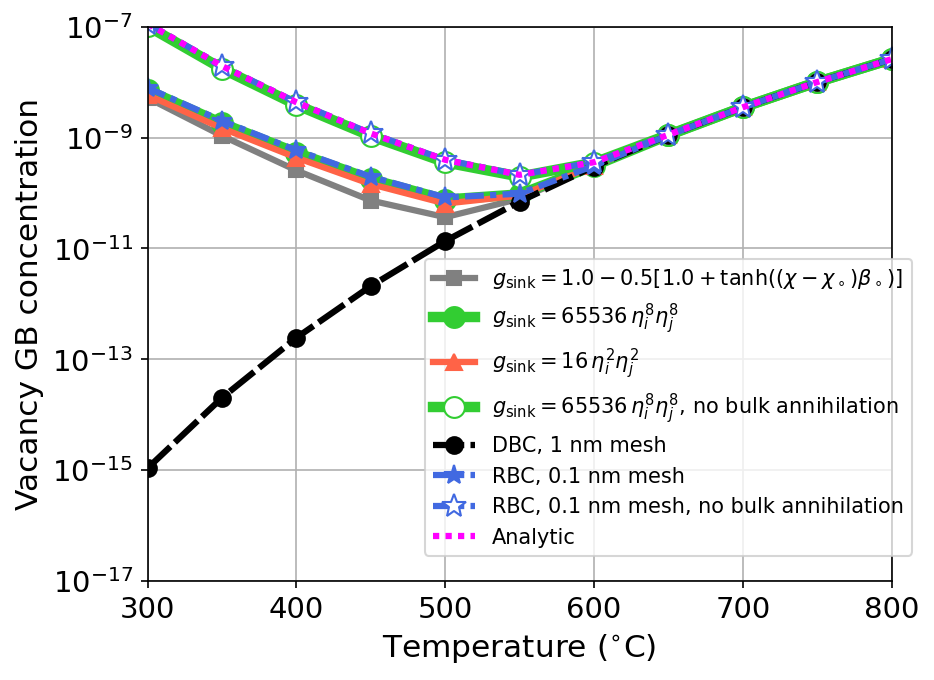}
    \caption{}
    \label{subfig:1D_verify_temp:vac}
    \end{subfigure}
    ~  
    \begin{subfigure}[t]{0.48\textwidth}
        \includegraphics[width=1\textwidth]{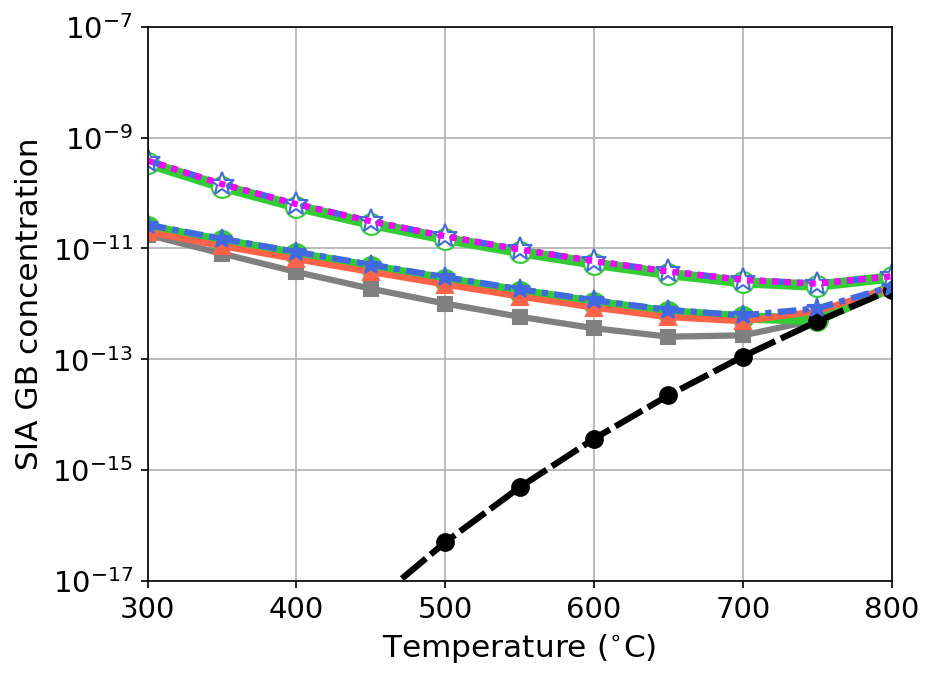}
    \caption{}
    \label{subfig:1D_verify_temp:sia}
    \end{subfigure}
    \caption{RIS at GB for (a) Ni, (b) Cr, (c) vacancy, and (d) SIA as a function of temperature as calculated from a 1D system of 1 $\mu$m grain size irradiated to 10 dpa or 60 days. Results are shown for different PF sink functions and compared against sharp-interface simulations with DBC and RBC. Analytic solutions for GB point defect concentrations are shown in (c,d).}
    \label{fig:1D_verify_temp}
\end{figure}

In Fig.~\ref{fig:1D_verify_temp}, the results for $\alpha^{-1}=5.5$ nm$^{-1}$, corresponding to a low-angle symmetric tilt GB between $7^\circ-15^\circ$ misorientation angle~\cite{duh2001numerical,gu2017point}, are presented as a function of temperature.
All the methods demonstrate ideal GB sink behavior at high temperatures: $T>600^\circ$C for vacancy and $T\approx800^\circ$C for SIA.
However, at lower temperatures, both the RBC and PF methods demonstrate point defect concentrations that deviate from ideal (DBC) values.
For both point defect and atomic concentrations, PF simulations employing the bell-shaped sink functions show excellent agreement with the RBC method. 
These results match well with Eq.~\ref{aeq:rbc_gb_conc} and with the DBC method employing a 1 nm mesh size.
The PF model with the smooth step-shaped function, however, shows deviation from the other methods. This can be attributed to the wider, flatter region over which the sink absorption term acts.
For analytic verification using Eq.~\ref{aeq:rbc_conc_bulk}, we also plotted (Fig.~\ref{fig:1D_verify_temp}) the results for simulations performed without bulk annihilation (i.e., neglecting recombination and dislocation absorption).
Good agreement is seen for the point defect concentrations.

\section*{ACKNOWLEDGMENTS}

PF model development was supported by the Laboratory Directed Research and Development (LDRD) program of Idaho National Laboratory (INL). 
AM microstructure modeling was sponsored by the U.S. Department of Energy Office of Nuclear Energy (DOE-NE)'s Advanced Materials and Manufacturing Technologies (AMMT) program.
This research made use of the resources of the High Performance Computing Center at Idaho National Laboratory, which is supported by DOE-NE and the Nuclear Science User Facilities under contract no. DE-AC07-05ID14517.  
This manuscript was authored in part by Battelle Energy Alliance, LLC under contract no.~DE-AC07-05ID14517.

 \bibliographystyle{elsarticle-num}
 \bibliography{references}





\appendix
\newpage
\setcounter{section}{0}
\setcounter{page}{1}
\setcounter{figure}{0}
\setcounter{equation}{0}
\renewcommand{\thesection}{S\arabic{section}}
\renewcommand{\theequation}{S\arabic{equation}}
\renewcommand{\thepage}{s\arabic{page}}
\renewcommand{\thetable}{S\arabic{table}}
\renewcommand{\thefigure}{S\arabic{figure}}
\numberwithin{equation}{section}

\setcounter{affn}{0}
\resetTitleCounters

\makeatletter
\let\@title\@empty
\makeatother

\title{\hspace{1.75cm} Supplementary Material \vspace{0.5cm} \newline Phase-field modeling of radiation-induced composition redistribution: An application to additively manufactured austenitic Fe-Cr-Ni}

\def\ps@pprintTitle{%
     \let\@oddhead\@empty
     \let\@evenhead\@empty
     \def\@oddfoot{\footnotesize\itshape
        Supplementary Data for \ifx\@journal\@empty Elsevier
       \else\@journal\fi\hfill\today}%
     \let\@evenfoot\@oddfoot}
\makeatother

\MaketitleBox
\hrule
\vskip24pt

\section{Atom--point defect flux coupling} \label{sm:fluxes}

Using Onsager transport and reciprocity relations, the partial fluxes of atomic elements $J^{{\upsilon}}_\phi$ are given in terms of the chemical potential gradients of atomic elements $\mu_k$ and point defects $\mu_{{\upsilon}}$ as:
\begin{subequations}
\begin{flalign}
    \boldsymbol{J}^{{\upsilon}}_{\phi} &= - \sum^K_{k = 1} L^{{\upsilon}}_{\phi k} \nabla \left[\mu_{k} + \text{sign}({{\upsilon}}) \mu_{{\upsilon}} \right], &&
\end{flalign}
\label{seq:total_flux}
\end{subequations}
where $\text{sign}(V)=-1$ and $\text{sign}(I)=+1$.
The point defect fluxes are given by $\boldsymbol{J}_{{\upsilon}} = \text{sign}({{\upsilon}}) \sum^K_{k=1} \boldsymbol{J}^{{\upsilon}}_k$.
By using the Gibbs-Duhem relation $\sum^K_{k=1} c_k \nabla \mu_k = 0$, and substituting $c_1 \approx 1-\sum^K_{k=2} c_k$, we can write $\nabla \mu_1 = - \sum^K_{k=2} c_k \nabla (\mu_k - \mu_1)$ and
$\nabla \mu_{k (\neq 1)} = (1-\sum^K_{k=2} c_k) \nabla (\mu_k - \mu_1) + \sum^K_{j=2} c_j \nabla (\mu_k - \mu_j)$.
The independent fluxes ($\phi = 2:K$) can then be written as:
\begin{subequations}
\begin{flalign}
    \nonumber &\boldsymbol{J}^{{\upsilon}}_{\phi} = - \sum^K_{k=2}   \left(L^{{\upsilon}}_{\phi k} - c_k \sum^K_{j=1} L^{{\upsilon}}_{\phi j}\right) \nabla (\mu_{k} - \mu_1) \\ 
    &\hspace{1.5cm}- \sum_{{{\upsilon}}=V,I} \left(\sum^K_{j=1} \text{sign}({{\upsilon}}) L^{{\upsilon}}_{\phi j}\right) \nabla \mu_{{\upsilon}} \\
    \nonumber &\boldsymbol{J}_{{\upsilon}} = - \sum^K_{k=2} \left[\text{sign}({{\upsilon}}) \left(\sum^K_{j=1} L^{{\upsilon}}_{kj} - c_k \sum^K_{i=1}\sum^K_{j=1} L^{{\upsilon}}_{ij}\right)\right] \nabla (\mu_k - \mu_1) \\
    &\hspace{1.5cm}- \left(\sum^K_{i=1}\sum^K_{j=1} L^{{\upsilon}}_{ij}\right) \nabla \mu_{{\upsilon}}. &&
\end{flalign}
\label{seq:total_red_flux}
\end{subequations}
The total atomic fluxes are $\boldsymbol{J}_\phi = \sum_{{{\upsilon}}=V,I} \boldsymbol{J}^{{\upsilon}}_\phi$.
By defining $L_{\phi {{\upsilon}}} = \sum^K_{j=1} \text{sign}({{\upsilon}}) L^{{\upsilon}}_{\phi j}$, $L_{{{\upsilon}}{{\upsilon}}} = \sum^K_{i=1}\sum^K_{j=1} L^{{\upsilon}}_{ij}$, $L^1_{{{\upsilon}} k} = \text{sign}({{\upsilon}}) \left( \sum^K_{j=1} L^{{\upsilon}}_{k j} - c_k L_{{{\upsilon}}{{\upsilon}}}\right)$, and $L^1_{\phi k} = \sum_{{{\upsilon}}} L^{{\upsilon}}_{\phi k} - \, c_k \sum_{{{\upsilon}}} \text{sign}({{\upsilon}}) L_{\phi {{\upsilon}}}$, we obtain the total atomic and point defect fluxes as:
\begin{subequations} \label{seq:final_flux}
\begin{flalign}
    \boldsymbol{J}_{\phi} &= - \sum^K_{k = 2} L^1_{\phi k} \nabla \mu_{k 1} - \sum_{{{\upsilon}}=V,I} L_{\phi {{\upsilon}}} \nabla \mu_{{\upsilon}}, \\
    \boldsymbol{J}_{{{\upsilon}}} &= - \sum^K_{k = 2} L^1_{{{\upsilon}} k} \nabla \mu_{k 1} - L_{{{\upsilon}} {{\upsilon}}} \nabla \mu_{{\upsilon}}. && 
\end{flalign}
\end{subequations}
Here, $\mu_{k1}=\mu_k - \mu_1 = \partial f_\text{C}(c_2,\dots,c_K)/\partial c_k$ is the diffusion potential of $k$ with respect to $1$.

\vspace{0.4cm}
\section{Density-based CALPHAD free energy} \label{sm:gb_fe}

The free energy density ($f_\text{C}$) is described as:
\begin{flalign} 
   f_\text{C} = G_m/V_m = \left(c_{Fe} {}^oG_{Fe} + c_{Cr} {}^oG_{Cr} + c_{Ni} {}^oG_{Ni} + G^{ideal}_{mix} + {}^EG_m\right)/V_m, &&
\end{flalign}
where $G_m$ is the molar Gibbs energy and $V_m$ is the molar volume. $ G^{ideal}_{mix}$ is the ideal configurational entropy, written as:
\begin{flalign}
    G^{ideal}_{mix} = R T (c_{Fe} \ln{c_{Fe}} + c_{Cr} \ln{c_{Cr}} + c_{Ni} \ln{c_{Ni}}). &&
\end{flalign}
Following the density-based CALPHAD free energy formulation described by Kamachali~\cite{kamachali2020model}, ${}^oG_{\phi}$ ($\phi$ = Fe,Cr,Ni) and $ {}^EG_m$ are expressed as functions of the relative atomic density ($\sigma$). 
Here, the pure component Gibbs free energies are given by:
\begin{flalign}
    {}^oG_{\phi} = {}^oH^{fcc}_{\phi} \sigma^2 - T\, {}^oS^{fcc}_{\phi} \sigma. &&
\end{flalign}
And the molar excess Gibbs energy of mixing is given by:
\begin{flalign} 
    \nonumber {}^EG_m =& \, c_{Fe} c_{Cr} L^{fcc}_{Fe,Cr} \sigma^2 + c_{Fe} c_{Ni} L^{fcc}_{Fe,Ni} \sigma^2 + c_{Cr} c_{Ni} L^{fcc}_{Cr,Ni} \sigma^2 \\ 
    &+ c_{Fe} c_{Cr} c_{Ni} L^{fcc}_{Fe,Cr,Ni} \sigma^2. &&
\end{flalign}
Per Dinsdale~\cite{dinsdale1991sgte}, the pure component enthalpies and entropies for the FCC phase in J/mol are:
\begin{subequations}
\begin{flalign}  
    \nonumber &{}^oH^{fcc}_{Fe} =  24.664\,T + 154717\,T^{-1} + 3.75752\times10^{-3}\,T^2 \\
    & \hspace{1.5cm} + 1.178538\times10^{-7} \,T^3 - 236.7, \\  
    \nonumber &{}^oH^{fcc}_{Cr} = 26.908\,T + 278500\,T^{-1} - 1.89435\times10^{-3}\,T^2 \\ 
    & \hspace{1.5cm} + 2.95442\times10^{-6}\,T^3 - 1572.94, \\
    &{}^oH^{fcc}_{Ni} = 4.8407\times10^{-3}\,T^2 + 22.096\,T - 5179.159 &&
\end{flalign}
\end{subequations}
and
\begin{subequations}
\begin{flalign}
    \nonumber &{}^oS^{fcc}_{Fe} = 7.51504\times10^{-3}\,T + 24.664\,\ln{T} + 77358.5\,T^{-2} \\ 
    & \hspace{1.5cm} + 1.767807\times10^{-7}\,T^2 - 107.7517, \\
    \nonumber &{}^oS^{fcc}_{Cr} = 26.908\,\ln{T} - 3.7887\times10^{-3}\,T + 139250\,T^{-2} \\ 
    & \hspace{1.5cm} + 4.43163\times10^{-6}\,T^2 - 130.735, \\
    &{}^oS^{fcc}_{Ni} = 9.6814\times10^{-3}\,T + 22.096\,\ln{T} - 95.758. &&
\end{flalign}
\end{subequations}
Per Miettinen~\cite{miettinen1999thermodynamic}, the optimized ternary interaction parameters of the FCC phase in J/mol are:
\begin{subequations}
\begin{flalign} 
    &L^{fcc}_{Fe,Cr} = (10833 - 7.477) + (-1410)(c_{Fe}-c_{Cr}), \\
    \nonumber &L^{fcc}_{Fe,Ni} = (-12054 + 3.274\,T) + (11082 - 4.45\,T)(c_{Fe}-c_{Ni}) \\ 
    & \hspace{1.5cm} +(-726)(c_{Fe}-c_{Ni})^2, \\ 
    &L^{fcc}_{Cr,Ni} = (8030 - 12.88\,T) + (33080 - 16.036\,T)(c_{Cr}-c_{Ni}) \\
    &L^{fcc}_{Fe,Cr,Ni} = (-6500)c_{Fe}+(10000-10\,T)c_{Cr} + (48000)c_{Ni}. &&
\end{flalign}
\end{subequations}

\vspace{0.4cm}
\section{PF framework for GB} \label{sm:pf_gb}

Using the chain rule of differentiation, the gradient of atomic concentration $c_{\phi}(\boldsymbol{\mu}_{\phi 1},\boldsymbol{\eta})$ at any point can be written in terms of the gradients of $\mu_{\phi 1}$ and $\eta_n$ as:
\begin{flalign} \label{seq:chain_rule_conc}
    \nabla c_{\phi} &= \sum^K_{j=2} \frac{\partial c_\phi}{\partial \mu_{j1}} \nabla \mu_{j1} + \sum^N_{n=1} \frac{\partial c_\phi}{\partial \eta_n} \nabla \eta_n.  &&
\end{flalign}
Using Eq.~\ref{eq:tot_conc_atomic}, we have $\frac{\partial c_{\phi}}{\partial \eta_n} = \left(c^g_{\phi} - c^b_{\phi} \right)\frac{\partial g_{\text{mw}}}{\partial \eta_n}$. Substituting in Eq.~\ref{seq:chain_rule_conc}, we get:
\begin{flalign} \label{seq:grad_conc_2}
\nabla c_{\phi} &= \sum^K_{j=2} \chi_{\phi j} \nabla \mu_{j1} + \sum^N_{n=1} \left(c^g_{\phi} - c^b_{\phi} \right)\frac{\partial g_{\text{mw}}}{\partial \eta_n}  \nabla \eta_n,  &&
\end{flalign}
where $\chi_{\phi j}$ is the susceptibility.
For convenience, the above system of equations for $\phi = 2:K$ can be rewritten in the matrix form and rearranged as:
\begin{flalign}  \label{eq:Cijkl_crystal}
    \begin{pmatrix}
    \nabla c_2 - \sum^N_{n=1} \left(c^g_2 - c^b_2 \right)\frac{\partial g_{\text{mw}}}{\partial \eta_n}  \nabla \eta_n \\
    \vdots \\
    \nabla c_K - \sum^N_{n=1} \left(c^g_K - c^b_K \right)\frac{\partial g_{\text{mw}}}{\partial \eta_n}  \nabla \eta_n
    \end{pmatrix} = 
    \begin{pmatrix}
    \chi_{22} & \dots & \chi_{2K} \\
    \vdots & \ddots & \vdots \\
    \chi_{K1} & \dots & \chi_{KK}
    \end{pmatrix}
   \begin{pmatrix}
    \nabla \mu_{21} \\
    \vdots \\
    \nabla \mu_{K1} 
    \end{pmatrix}, &&
\end{flalign}
where the first term on the right-hand side of the equation is the atomic susceptibility matrix $\boldsymbol{\chi}$. 
The gradients of diffusion potentials can then be expressed as:
\begin{flalign}  \label{seq:diff_pot_gradient_matrix}
    \begin{pmatrix}
    \nabla \mu_{21} \\
    \vdots \\
    \nabla \mu_{K1} 
    \end{pmatrix} = 
    \begin{pmatrix}
    \chi_{22} & \dots & \chi_{2K} \\
    \vdots & \ddots & \vdots \\
    \chi_{K1} & \dots & \chi_{KK}
    \end{pmatrix}^{-1} 
    \begin{pmatrix}
    \nabla c_2 - \sum^N_{n=1} \left(c^g_2 - c^b_2 \right)\frac{\partial g_{\text{mw}}}{\partial \eta_n}  \nabla \eta_n \\
    \vdots \\
    \nabla c_K - \sum^N_{n=1} \left(c^g_K - c^b_K \right)\frac{\partial g_{\text{mw}}}{\partial \eta_n}  \nabla \eta_n
    \end{pmatrix}, &&
\end{flalign}
where the inverse of the susceptibility matrix is the thermodynamic factor matrix, given by:
\begin{flalign}  \label{seq:theta_matrix}
    \boldsymbol{\theta} = 
    \begin{pmatrix}
    \theta_{22} & \dots & \theta_{2K} \\
    \vdots & \ddots & \vdots \\
    \theta_{K1} & \dots & \theta_{KK}
    \end{pmatrix} =
    \boldsymbol{\chi}^{-1} =
    \begin{pmatrix}
    \chi_{22} & \dots & \chi_{2K} \\
    \vdots & \ddots & \vdots \\
    \chi_{K1} & \dots & \chi_{KK}
    \end{pmatrix}^{-1}. &&
\end{flalign}
The system of equations for $\nabla \mu_{\phi1}$ is:
\begin{flalign} \label{seq:diff_pot_gradient_system}
    \nabla \mu_{\phi1} &= \sum^K_{j=2} \theta_{\phi j} \nabla c_j - \sum^K_{j=2} \sum^N_{n=1} \theta_{\phi j} \left(c^g_{\phi} - c^b_{\phi} \right)\frac{\partial g_{\text{mw}}}{\partial \eta_n} \nabla \eta_n.  &&
\end{flalign}
Using Eq.~\ref{eq:tot_conc_atomic}, the local susceptibility at any point is realized as the interpolation between the phase susceptibilities as:
\begin{flalign} \label{seq:susc_interp}
    \chi_{ij} &= \frac{\partial c_i}{\partial \mu_{j1}} = \frac{\partial [c^b_i (1-\bar{g}_\text{mw}) + c^g_i \bar{g}_\text{mw}]}{\partial \mu_{j1}} = \chi^b_{ij}(1-\bar{g}_\text{mw}) + \chi^g_{ij}\bar{g}_\text{mw}.  &&
\end{flalign}
Here, the phase susceptibilities $\chi_{ij}^{b/g}$ are obtained from the thermodynamic factors of the CALPHAD free energy $f_\text{C}$ as $\chi^{-1}_{ij} = \theta_{ij} = \frac{\partial^2 f_\text{C}}{\partial c_i \partial c_j}$.
For a ternary component alloy ($1$ being Fe, $2$ being Cr and $3$ being Ni), by using Eq.~\ref{seq:theta_matrix} and $\bar{g}^b_\text{mw}=1-\bar{g}_\text{mw}$, we can write the thermodynamic factors as:
\begin{subequations}
\begin{flalign} \nonumber
    \theta_{22} &= \frac{\chi^b_{33}\bar{g}^b_\text{mw} + \chi^g_{33} \bar{g}_\text{mw}}{\left(\chi^b_{22}\bar{g}^b_\text{mw} + \chi^g_{22} \bar{g}_\text{mw}\right)\left(\chi^b_{33}\bar{g}^b_\text{mw} + \chi^g_{33} \bar{g}_\text{mw}\right)-\left(\chi^b_{23}\bar{g}^b_\text{mw} + \chi^g_{23} \bar{g}_\text{mw}\right)^2}, \\ \nonumber
    \theta_{23} &= \frac{-\left(\chi^b_{23}\bar{g}^b_\text{mw} + \chi^g_{23} \bar{g}_\text{mw}\right)}{\left(\chi^b_{22}\bar{g}^b_\text{mw} + \chi^g_{22} \bar{g}_\text{mw}\right)\left(\chi^b_{33}\bar{g}^b_\text{mw} + \chi^g_{33} \bar{g}_\text{mw}\right)-\left(\chi^b_{23}\bar{g}^b_\text{mw} + \chi^g_{23} \bar{g}_\text{mw}\right)^2}, \\
    \theta_{33} &= \frac{\chi^b_{22}\bar{g}^b_\text{mw} + \chi^g_{22} \bar{g}_\text{mw}}{\left(\chi^b_{22}\bar{g}^b_\text{mw} + \chi^g_{22} \bar{g}_\text{mw}\right)\left(\chi^b_{33}\bar{g}^b_\text{mw} + \chi^g_{33} \bar{g}_\text{mw}\right)-\left(\chi^b_{23}\bar{g}^b_\text{mw} + \chi^g_{23} \bar{g}_\text{mw}\right)^2}. &&
\label{seq:thetas}
\end{flalign}
\end{subequations}
To prevent atomic concentrations from taking nonphysical values ($c_k<0$ or $c_k>1$) during RIS, we introduce concentration dependence to the phase-specific thermodynamic factors in Eqs.~\ref{eq:alt_time_evol} and~\ref{eq:am_time_evol} as: ${\theta}_{23}=\bar{\theta}^\circ_{23}/c_{1}$, ${\theta}_{22}=\bar{\theta}^\circ_{22}(1-c_{3})/c_{1}$ and ${\theta}_{33}=\bar{\theta}^\circ_{33}(1-c_{2})/c_{1}$.
Here, $\bar{\theta}^\circ$ are the normalized factors calculated at nominal concentrations as $\bar{\theta}^\circ_{23}={\theta}^\circ_{23}c^\circ_{1}$, $\bar{\theta}^\circ_{22}={\theta}^\circ_{22}c^\circ_{1}/(1-c^\circ_{3})$ and $\bar{\theta}^\circ_{33}={\theta}^\circ_{33}c^\circ_{1}/(1-c^\circ_{2})$. Here, all quantities $\theta$ and $c_k$ are specific to the phase $b$ or $g$.

\section{Solute-SIA binding} \label{sm:sec:solute-SIA}
\begin{figure}[htp!]
    \centering
    \begin{subfigure}[t]{0.48\textwidth}
        \includegraphics[width=1\textwidth]{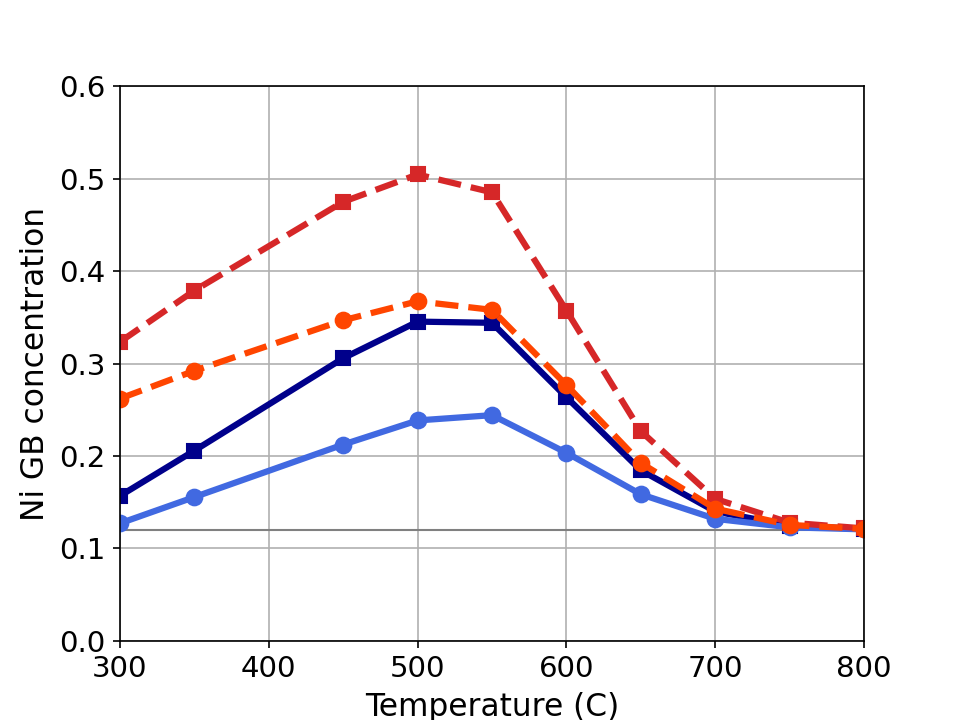}
    \caption{}
    \label{subfig:1D_ris_temp_SIA_binding_Ni}
    \end{subfigure}
    ~  
    \begin{subfigure}[t]{0.48\textwidth}
        \includegraphics[width=1\textwidth]{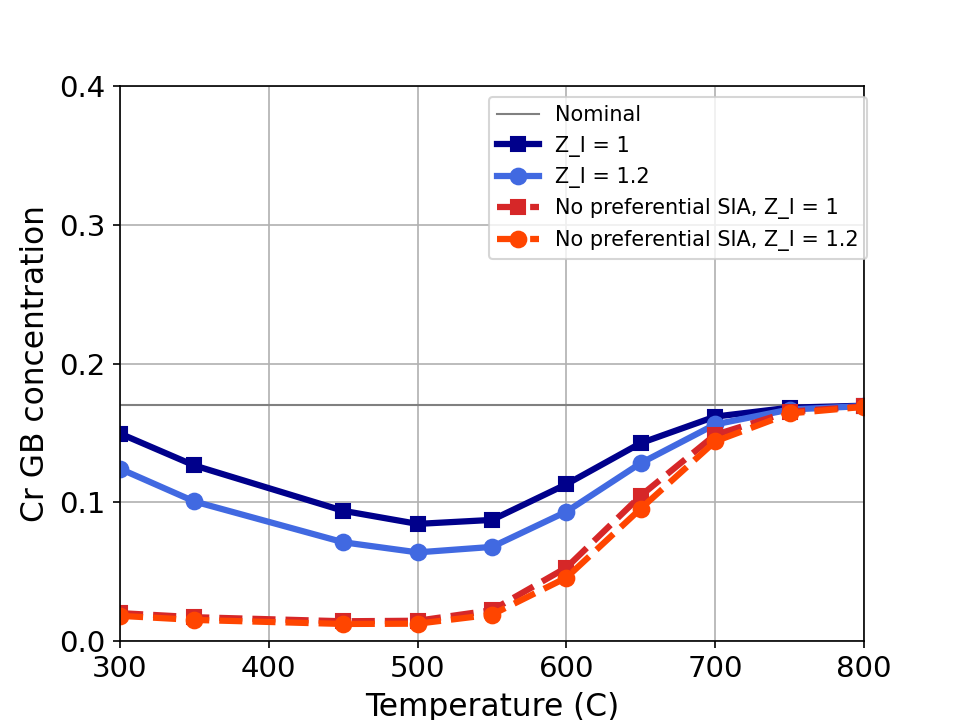}
    \caption{}
    \label{subfig:1D_ris_temp_SIA_binding_Cr}
    \end{subfigure}
    \caption{RIS at GB for (a) Ni and (b) Cr versus temperature calculated from a 1D system of 1 $\mu$m grain size irradiated to 10 dpa or 60 days. The DBC method with a 1 nm mesh was used for simulation. Corresponding to a bulk dislocation density of $\rho_b = 10^{14}$ m$^{-2}$, results are shown for cases with and without preferential SIA binding and for $Z_I=1$ and $Z_I=1.2$ (biased absorption). All other parameters default to Table~\ref{tab:myfirstlongtable}.}
    \label{fig:1D_ris_temp_SIA_binding}
\end{figure}

RIS simulations in the main paper were performed using preferential Solute-SIA binding, with a positive binding for Cr and negative binding for Ni. 
In Fig.~\ref{fig:1D_ris_temp_SIA_binding}, we compare these results with the simulations performed without preferential Solute-SIA binding (i.e. $E^b_{CrI}-E^b_{FeI}=0$ and $E^b_{NiI}-E^b_{FeI}=0$ yielding binding factors of $\beta_k=1$ for Fe, Cr and Ni). 
With the assumption that there is no difference in the binding energies as well as migration energies for transport of the elements via SIA, RIS to GB is primarily determined by the preferential transport of the elements via vacancies. Due to the absence preferential Cr flux via SIA transport, Cr depletion is seen to be significantly greater, especially at lower temperatures. Correspondingly, Ni enrichment is also significantly greater. 
These observations are in agreement with similar changes in RIS observed by Yang et al.~\cite{yang2016roles} for simulations performed (at 320 \textdegree C and $8\times10^{-7}$ dpa/s) with and without SIA-atom coupling.
These results show that the RIS model based on Yang et al.~\cite{yang2016roles} involves significant contributions from both vacancy and SIA fluxes.
For the default bulk dislocation density $\rho_b=10^{-14}$ m$^{-2}$, if a strong bias of $Z_I=1.2$ for SIA absorption is assumed, RIS results (without Solute-SIA binding) show a significant decrease in Ni enrichment compared to the case of $Z_I=1$ without absorption bias.
This can be explained by the increased absorption of SIAs and thus the reduced flux of Ni via SIA to the GB.
Since Cr concentration is already low due to preferential vacancy-Cr exchange, reduced flux of Cr is not seen to significantly alter Cr RIS.
The effect of bias ($Z_I=1.2$) on RIS for the case with Solute-SIA binding (as discussed in the main paper) is seen to decrease Ni enrichment and slightly enhance Cr depletion. 
Since the total Cr outflux from GB is the sum of vacancy and SIA outfluxes, the reduction in Cr influx for $Z_I=1.2$ results in an effectively greater Cr outflux via vacancy-Cr exchange.

\vspace{0.4cm}
\section{Equilibrium GB properties} \label{sm:eq_gb}

To relate the PF model parameters with the material properties, we consider a planar interface. The following relations then hold under chemical equilibrium and stationary interface conditions:
\begin{subequations}
\begin{flalign} \label{eq_gp:1D_eq_conc}
    &\mu_{k1}(x) \equiv {\mu}^e_{k1} (\text{const.}(x)), \\ \label{eq_gp:1D_eq_phi}
    &\frac{\delta \Omega}{\delta \eta_i} = \frac{\partial \omega}{\partial \eta^e_i} - \kappa \frac{d^2 \eta^e_i}{dx^2} = 0.  &&
\end{flalign}
\end{subequations}
Here, the superscript ``$e$" denotes the equilibrium condition and $x$ is the spatial coordinate normal to the planar GB.
The equilibrium concentrations at any point are given by:
\begin{flalign} \label{eq_gp:eq_conc}
     c^e_k =& (1-\bar{g}^e_\text{mw})\,c^{b,e}_k + \bar{g}^e_\text{mw} c^{g,e}_k,  &&
\end{flalign}
where the phase concentrations are constant with respect to $x$, owing to Eq.~\ref{eq_gp:1D_eq_conc}.
As such, the phase grand potential densities are also constant with respect to $x$, and are thus decoupled from the order parameters.
Eq.~\ref{eq_gp:1D_eq_phi} follows as:
\begin{flalign} 
    (m_0 + \Delta \omega^e_g) \frac{d\bar{g}^e_\text{mw}}{d\eta^e_i} - \kappa\frac{d^2\eta^e_i}{dx^2} = 0, &&
\end{flalign}
where $\Delta\omega^e_g = \omega^g(\mu^e)-\omega^b(\mu^e)$ captures the parallel tangent distance for equilibrium GB segregation.
Following Moelans et al.~\cite{moelans2008aniso-grain}, we can write the integrated equation involving the two order parameters $\eta_i$ and $\eta_j$ defining the GB:
\begin{flalign} \label{eq:equi_partition}
    (m_0 + \Delta \omega^e_g) \bar{g}^e_\text{mw} - \frac{\kappa}{2}\left[\left(\frac{d\eta^e_i}{dx}\right)^2+\left(\frac{d\eta^e_j}{dx}\right)^2\right] = 0. &&
\end{flalign}
The following far field boundary conditions for $\eta_i$ and $\eta_j$ are utilized here: $\eta_i = 1$ and $\eta_j = 0$ for $x\rightarrow-\infty$; $\eta_i=0$ and $\eta_j=1$ for $x\rightarrow+\infty$; and $d\eta_i/dx=d\eta_j/dx=0$ for $x\rightarrow\pm\infty$.
Rearrangement of the above equation yields:
\begin{flalign} \label{eq:grad_eta_equil_1}
    \frac{d\eta^e_i}{dx} = -\frac{\sqrt{2(m_0+\Delta \omega^e_g)\bar{g}_\text{mw}}}{\sqrt{\kappa\left[1+\left(\frac{d\eta^e_j}{d\eta^e_i} \right)^2\right]}} \hspace{0.2cm} \text{and} \hspace{0.2cm} \frac{d\eta^e_j}{dx} = -\frac{\sqrt{2(m_0+\Delta \omega^e_g)\bar{g}_\text{mw}}}{\sqrt{\kappa\left[1+\left(\frac{d\eta^e_i}{d\eta^e_j} \right)^2\right]}}. &&
\end{flalign}
Since $\frac{d\eta^e_j}{d\eta^e_i} = -1$, the above reduces to:
\begin{flalign} \label{eq:grad_eta_equil_2}
    \frac{d\eta^e_i}{dx} = -\frac{\sqrt{(m_0+\Delta \omega^e_g)\bar{g}_\text{mw}}}{\sqrt{\kappa}} \hspace{0.2cm} \text{and} \hspace{0.2cm} \frac{d\eta^e_j}{dx} = \frac{\sqrt{(m_0+\Delta \omega^e_g)\bar{g}_\text{mw}}}{\sqrt{\kappa}}. &&
\end{flalign}
Defining the GB width $\delta$ as the gradient of $\eta_i$ at $x=0$ and substituting $\eta^e_i(x=0)=\eta^e_j(x=0)=0.5$ or $\bar{g}^e_\text{mw}(x=0)=1$ yields:
\begin{flalign} 
    \delta = \frac{1}{\left|\left({d\eta^e_i}/{dx}\right)_{x=0}\right|} = \frac{\sqrt{\kappa}}{\sqrt{m_0+\Delta \omega^e_g}}. &&
\end{flalign}
Substituting for $\bar{g}_\text{mw}$ and $\delta$, Eq.~\ref{eq:grad_eta_equil_2} can be written as:
\begin{flalign} \label{eq:grad_eta_equil_3} 
    \frac{d\eta^e_i}{dx} = -\frac{4}{\delta}\eta_i(1-\eta_i) \hspace{0.2cm} \text{and} \hspace{0.2cm} \frac{d\eta^e_j}{dx} = \frac{4}{\delta} \eta_j(1-\eta_j). &&
\end{flalign}
Following Ref.~\cite{moelans2008aniso-grain}, the solutions for the order parameters at equilibrium are:
\begin{flalign} \label{eq:eta_equil_tanh} 
    \eta_i(x)=\frac{1}{2}\left[1-\tanh{\left(\frac{2x}{\delta}\right)}\right] \hspace{0.2cm} \text{and} \hspace{0.2cm} \eta_j(x)=\frac{1}{2}\left[1+\tanh{\left(\frac{2x}{\delta}\right)}\right]. &&
\end{flalign}
The GB energy is defined as the excess grand potential per unit area and is obtained for the stationary planar interface at chemical equilibrium as:
\begin{flalign} \label{eq:gb_energy_functional} \nonumber
    \gamma =& \int_{-\infty}^{\infty}
    \left[\omega^e_b + (m_0 + \Delta \omega^e_g)\bar{g}_{\text{mw}} + \frac{\kappa}{2} \left(\frac{d\eta_i}{dx}\right)^2 + \frac{\kappa}{2} \left(\frac{d\eta_i}{dx}\right)^2 \right]  dx \\
    &- \int_{-\infty}^{\infty} \omega^e_b  dx. &&
\end{flalign}
Employing the condition for equipartition of energy derived in Eq.~\ref{eq:equi_partition} gives us:
\begin{flalign}
    \gamma =& \int_{-\infty}^{\infty} 2(m_0 + \Delta \omega^e_g)\bar{g}_{\text{mw}}\,  dx. &&
\end{flalign}
Changing the variable of integration from $x$ to $\eta^e_i$
and substituting Eq.~\ref{eq:grad_eta_equil_2} gives us:
\begin{flalign} \nonumber
    \gamma =& \int_0^1 2(m_0 + \Delta \omega^e_g)\bar{g}_{\text{mw}}\, \frac{dx}{d\eta^e_i} d\eta^e_i \\
    =& \int_0^1 2\sqrt{(m_0 + \Delta \omega^e_g)\kappa \bar{g}_{\text{mw}}}\, d\eta^e_i. &&
\end{flalign}
Substituting $\bar{g}_\text{mw}=16\eta^2_i (1-\eta_j)^2$ and taking the spatially independent terms out of the integral, we get:
\begin{flalign}
    \gamma =& 8\sqrt{(m_0 + \Delta \omega^e_g)\kappa} \int_0^1 \eta^e_i (1-\eta^e_j) \, d\eta^e_i. &&
\end{flalign}
With the integral converging to $1/6$, the GB energy is finally given as:
\begin{flalign}
    \gamma =& \frac{4}{3} \sqrt{(m_0 + \Delta \omega^e_g)\kappa}. &&
\end{flalign}

\section{Equilibrium PF properties for GB} \label{appendix_C}

\begin{figure}[htp!]
    \centering
    \begin{subfigure}[t]{0.48\textwidth}
        \includegraphics[width=1\textwidth]{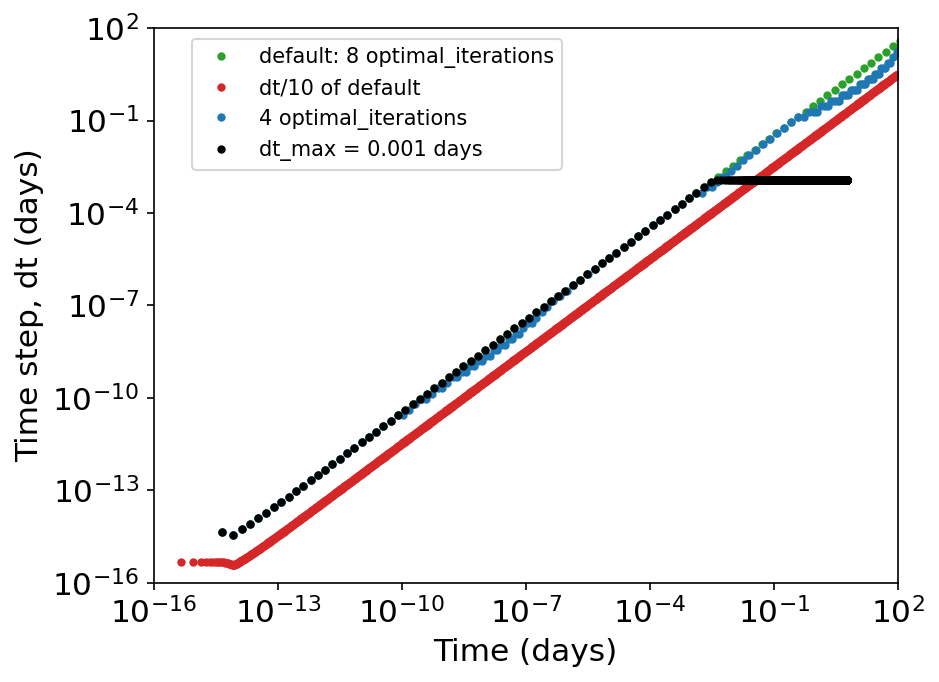}
    \caption{}
    \end{subfigure}
    ~  
    \begin{subfigure}[t]{0.48\textwidth}
        \includegraphics[width=1\textwidth]{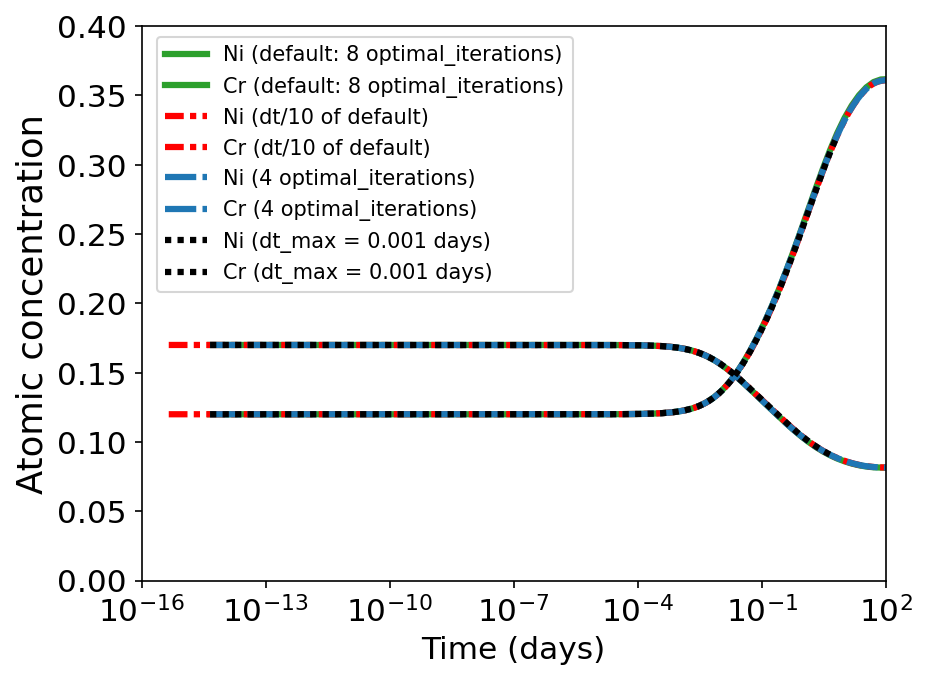}
    \caption{}
    \end{subfigure}
    ~  
    \begin{subfigure}[t]{0.48\textwidth}
        \includegraphics[width=1\textwidth]{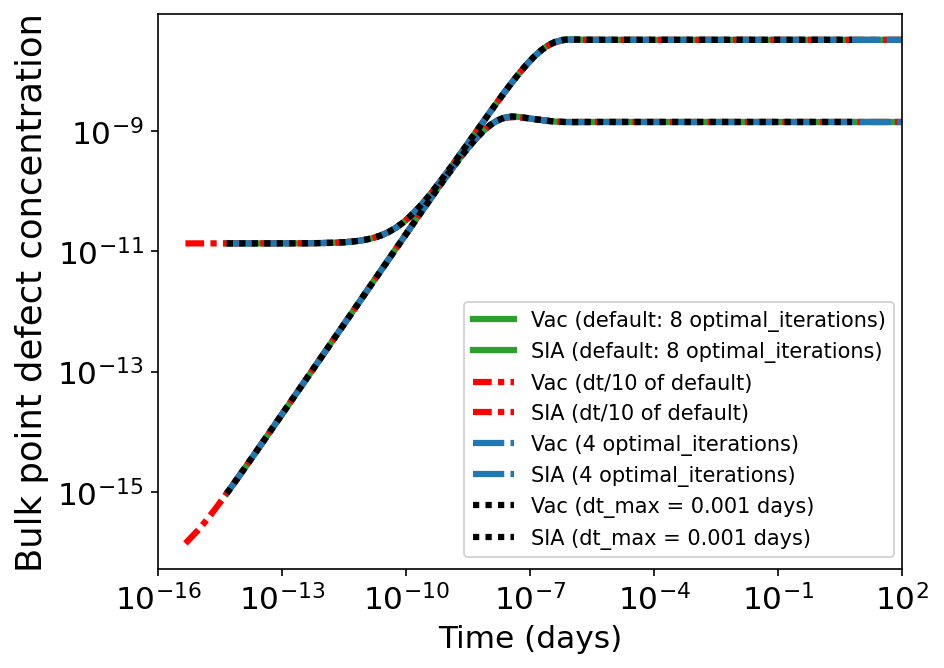}
    \caption{}
    \end{subfigure}
    ~  
    \begin{subfigure}[t]{0.48\textwidth}
        \includegraphics[width=1\textwidth]{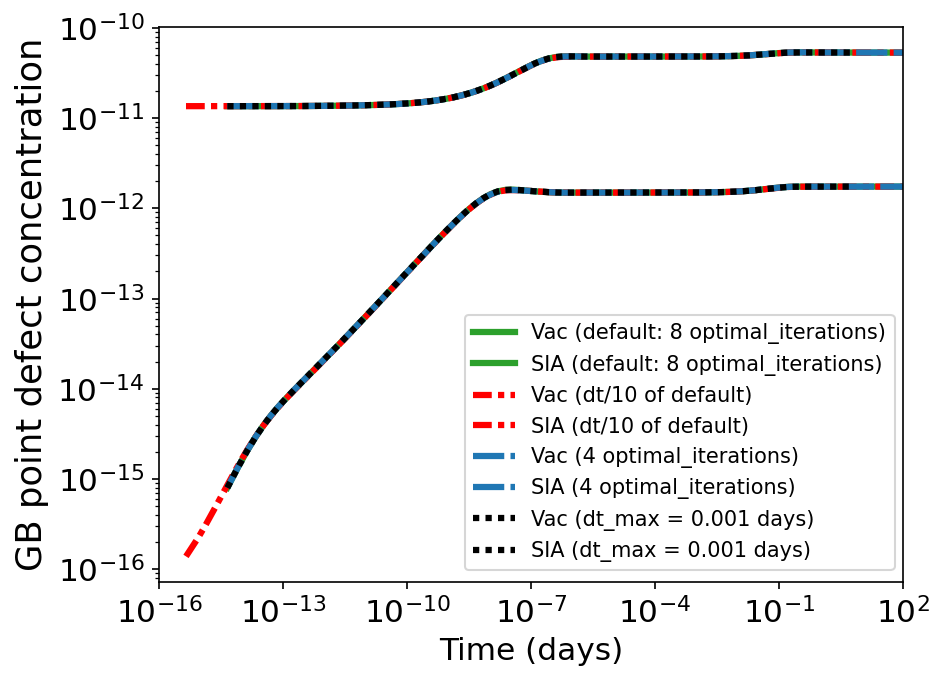}
    \caption{}
    \end{subfigure}
    \caption{Results from different settings in MOOSE's IterationAdaptiveDT time stepper.(a) Simulation time step versus time. Evolution of (b) Ni and Cr RIS at GB. Evolution of vacancy (vac.) and SIA concentrations at (c) bulk and (d) GB.}
    \label{fig:time_stepper}
\end{figure}

For a planar PF interface at equilibrium, the excess GB energy $\gamma$ and GB width $\delta$ are obtained as (see Sec.~\ref{sm:eq_gb}):
\begin{flalign} \label{eq:gb_energy}
    \gamma = \frac{4}{3} \sqrt{(m_0 + \Delta \omega^e_g)\kappa} &&
\end{flalign}
and
\begin{flalign} \label{eq:gb_width}
    \delta = \frac{\sqrt{\kappa}}{\sqrt{m_0+\Delta \omega^e_g}}, &&
\end{flalign}
where $\Delta \omega^e_g = \omega^e_g-\omega^e_b$. 
Here, ``$e$" denotes that the quantities are evaluated for the condition of a stationary GB, $\partial \eta_i/\partial t = 0$, and a constant diffusion potential $\mu_{k1}$ across the system. 
For a large system, the equilibrium concentration in the bulk can be assumed to be close to nominal. 
Thus, considering ``$e$'' and ``$\circ$'' (determined in Sec.~\ref{sec:impl:equil}) to be equivalent gives us $\Delta \omega^e_g = f^{g,o} - f^{b,o} - \mu^o_{CrFe} (c^{g,\circ}_{Cr}-c^{b,\circ}_{Cr}) - \mu^o_{NiFe} (c^{g,\circ}_{Ni}-c^{b,\circ}_{Ni})$. 
At 500\,$^\circ$C and $\sigma = 0.8$, this yields $\Delta \omega^e_g = 2.5\times 10^8$ J/m$^3$.
Thus, for a GB energy of $\gamma = 1$ J/m$^2$, $m_0=0$ could be set to uniquely determine $\kappa=2.2\times10^{-9}$ J/m and $\delta=2.9$ nm by using the above relations.
However, the temperature dependence of $\gamma$ of $\Delta \omega^e_g$ would lead to significant GB width variation with temperature.
Moreover, $\Delta \omega^e_g$ tends to vanish as $\sigma$ approaches unity (i.e., as the free energy of the GB approaches that of the bulk), yielding unphysical widths.
Therefore, to model GBs without TS, a finite $m_0$ must be determined by setting a value for $\delta$ per the convention in PF models. 
To simplify the parameterization and interpretation of results, we omit $\omega_g - \omega_b$ from Eq.~\ref{eq:eta_evol} and set $\Delta \omega^e_g=0$ in Eqs.~\ref{eq:gb_energy} and~\ref{eq:gb_width}. 
We assume $\gamma = 1$ J/m$^2$ and $\delta=1$nm, thus yielding the parameters $m_0=7.5\times10^8$ J/m$^3$ and $\kappa=7.5\times10^{-10}$ J/m.
{
\section{Results of testing the adaptive time stepper} \label{sm_time_stepper_results}
}
Here, we present the results of convergence study of 1D RIS simulations performed for different parameters of the time stepping scheme, \texttt{IterationAdaptiveDT}, in the MOOSE framework.
The time stepper increases or decreases the time step to maintain a certain number of nonlinear iterations, as specified by the parameter \texttt{optimal\_iterations} and \texttt{iteration\_window}. For the simulations in this work, we set \texttt{optimal\_iterations = 8} and \texttt{iteration\_window = 2} as the default. As high values of \texttt{optimal\_iterations} can result in greater time steps that could potentially increase discretization error of the simulation results, we performed a simulation for a lower value of \texttt{optimal\_iterations = 4} in Fig.~\ref{fig:time_stepper}.
A simulation was also performed with \texttt{optimal\_iterations = 8} by cutting the time step sizes (\texttt{dt}) realized by the default simulation by a factor of 10.
Another simulation was performed by limiting the maximum time step size (\texttt{dt\_max}) to $10^{-3}$ days.
The results for atomic and point defect concentrations do not change significantly for the different parameters values of the time stepper tested here; the error between the concentrations at any time is less than 1\%.

\newpage
{
\section{Supplementary results on atomic and point defect concentrations} \label{sm_results}
}

\begin{figure}[htp!]
    \centering
    \begin{subfigure}[t]{0.48\textwidth}
        \includegraphics[width=1\textwidth]{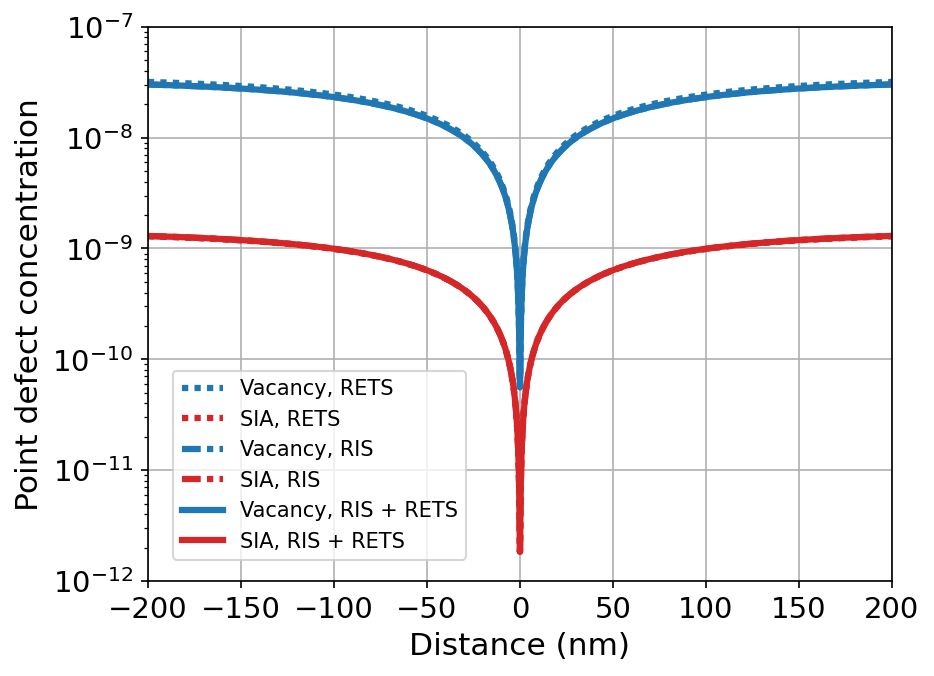}
    \end{subfigure}
    \caption{Steady-state point defect concentrations near the GB in a 1D system of 1 $\mu$m grain size when under irradiation at $2\times10^{-6}$ dpa/s and 500\textdegree C. Profiles corresponding to the RIS (Fig.~\ref{subfig:1D_ris_zoom}), RETS (Fig.~\ref{subfig:1D_rets_zoom}) and RIS+RETS (Fig.~\ref{subfig:1D_ris_and_rets_strong_zoom}) mechanisms are nearly identical.}
    \label{fig:1D_defect_profile_500C}
\end{figure}

\begin{figure}[htp!]
    \centering
    \begin{subfigure}[t]{0.48\textwidth}
        \includegraphics[width=1\textwidth]{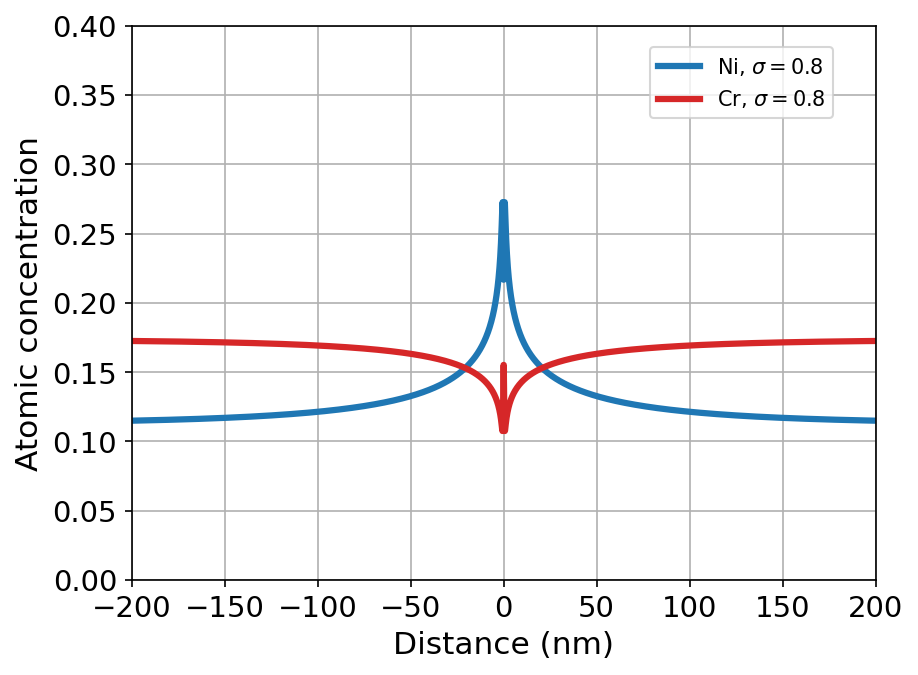}
    \end{subfigure}
    ~
    \begin{subfigure}[t]{0.48\textwidth}
        \includegraphics[width=1\textwidth]{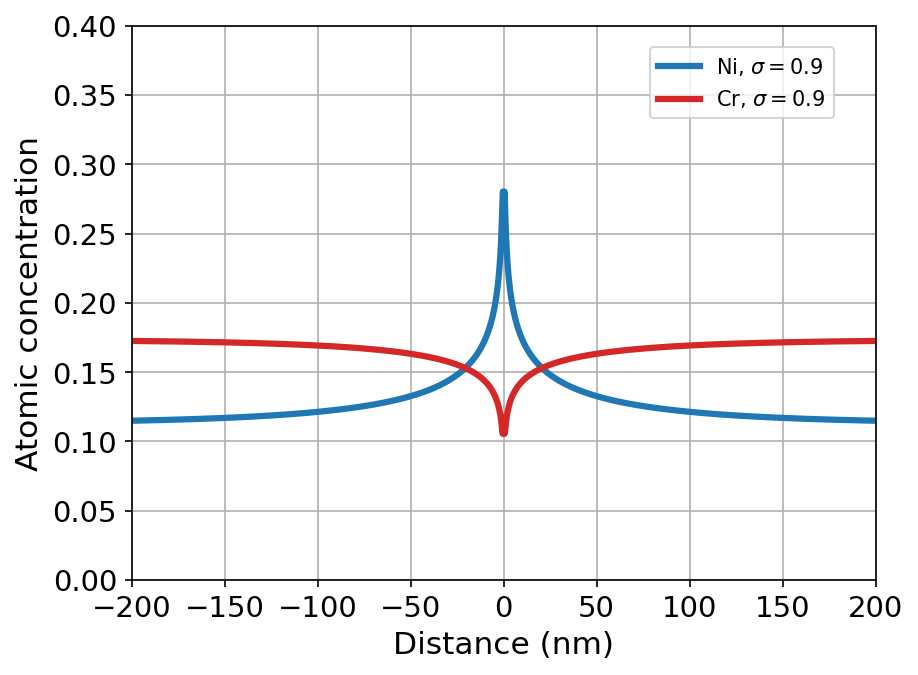}
    \end{subfigure}
    \caption{Low magnification plot of segregation profiles from Fig.~\ref{fig:1D_ris_and_rets} corresponding to RIS and TS mechanisms in a 1D system of 1 $\mu$m grain size when under irradiation at $2\times10^{-6}$ dpa/s and 500\textdegree C. (a) $\sigma = 0.8$ and (b) $\sigma = 0.9$. The non-monotonic segregation at the GB is not discernible for Ni in (a) and for Cr and Ni in (b).}
    \label{fig:1D_ris_and_ts_profile_500C_wide}
\end{figure}

\begin{figure}[htp!]
    \centering
    \begin{subfigure}[t]{0.48\textwidth}
        \includegraphics[width=1\textwidth]{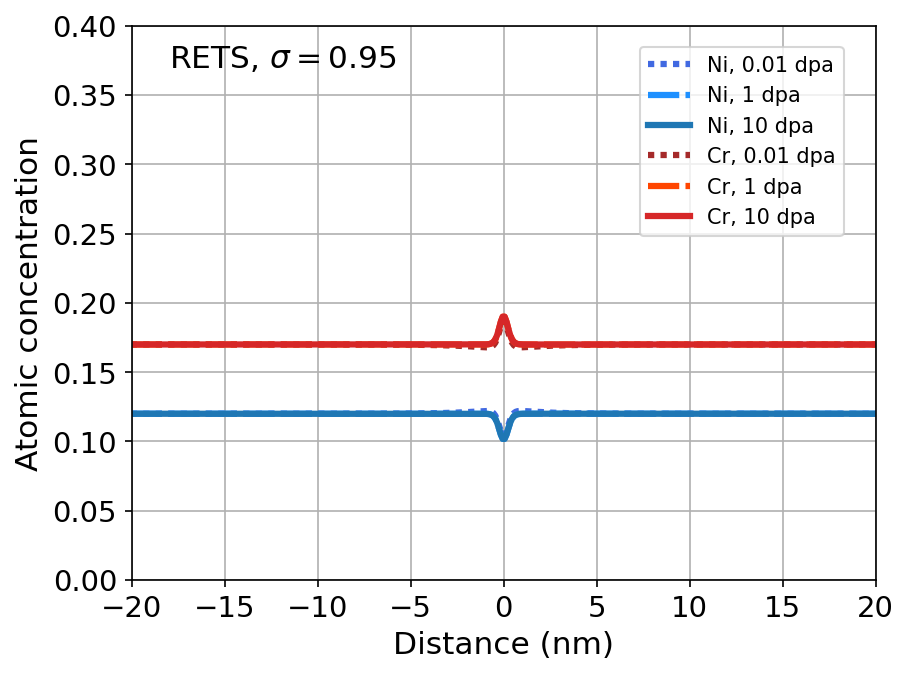}
    \end{subfigure}
    ~
    \begin{subfigure}[t]{0.48\textwidth}
        \includegraphics[width=1\textwidth]{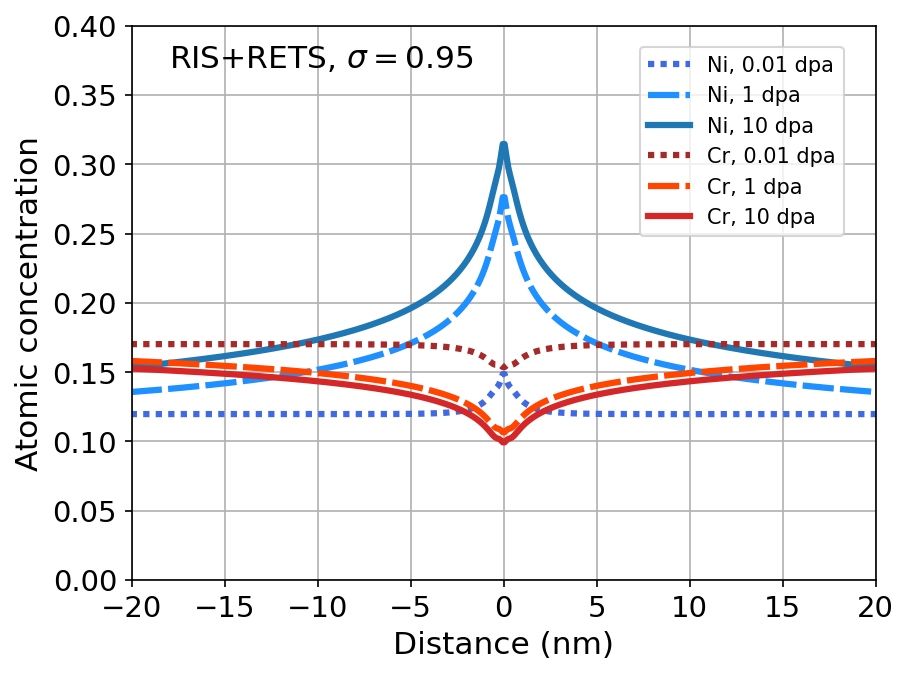}
    \end{subfigure}
    \caption{Segregation due to very weak TS modeled by $\sigma = 0.95$ in a 1D system of 1 $\mu$m grain size when under irradiation at $2\times10^{-6}$ dpa/s and 500\textdegree C. (a) The RETS mechanism. (b) The combined RIS and TS mechanisms.}
    \label{fig:1D_ris_and_weak_ts_profile_500C}
\end{figure}

\begin{figure}[htp!]
    \centering
    \begin{subfigure}[t]{0.48\textwidth}
        \includegraphics[width=1\textwidth]{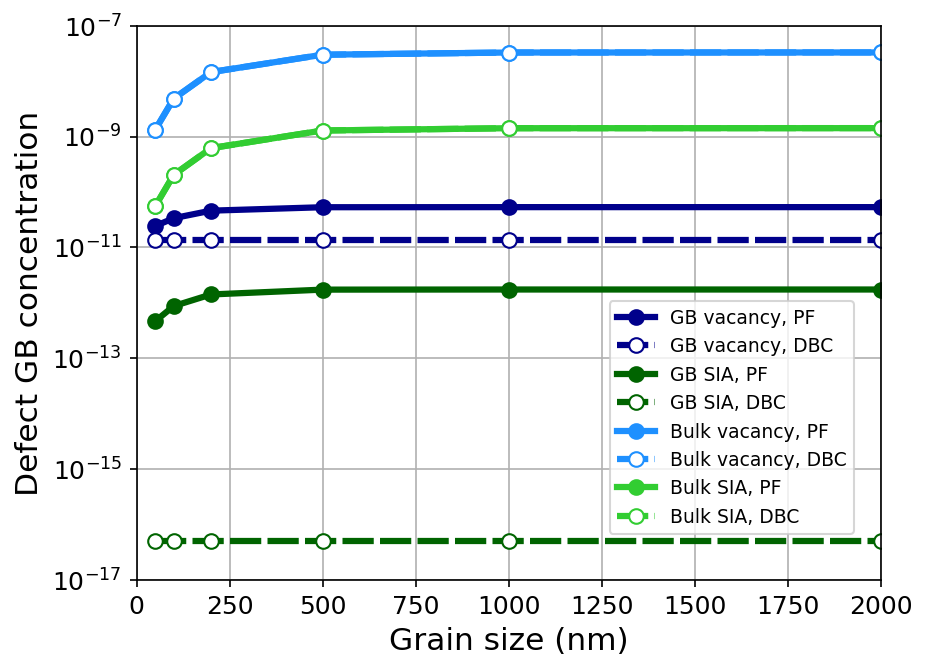}
    \end{subfigure}
        \caption{Point defect concentrations as a function of grain size from 1D PF simulations of RIS in Fig.~\ref{fig:1D_2D_RIS_vs_grain} with bulk dislocation density $\rho_b=10^{14}$ m$^{-2}$ and dislocation absorption efficiency of $Z_I=1$ for SIA. Bulk concentrations are seen to be identical between the PF and DBC (1 nm mesh) implementation, and decrease with decrease in grain size. GB concentrations are higher in the PF implementation.}
    \label{fig:1D_ris_defect_vs_grain_size_500C}
\end{figure}

\begin{figure}[htp!]
    \centering
    \begin{subfigure}[t]{0.48\textwidth}
        \includegraphics[width=1\textwidth]{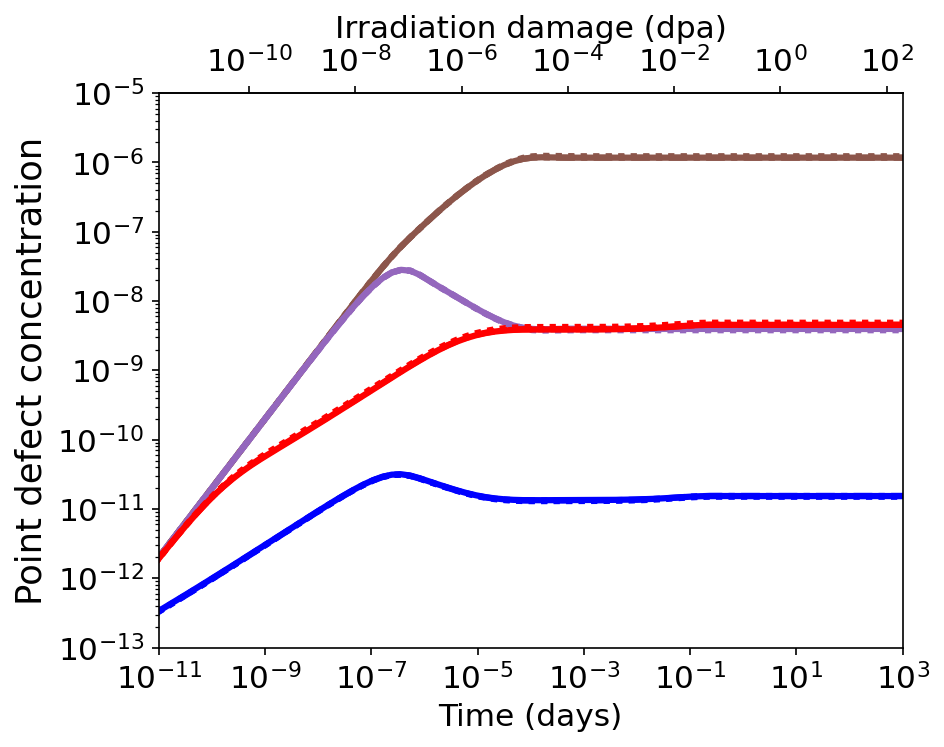}
    \caption{300\textdegree C}
    \end{subfigure}
    ~  
    \begin{subfigure}[t]{0.48\textwidth}
        \includegraphics[width=1\textwidth]{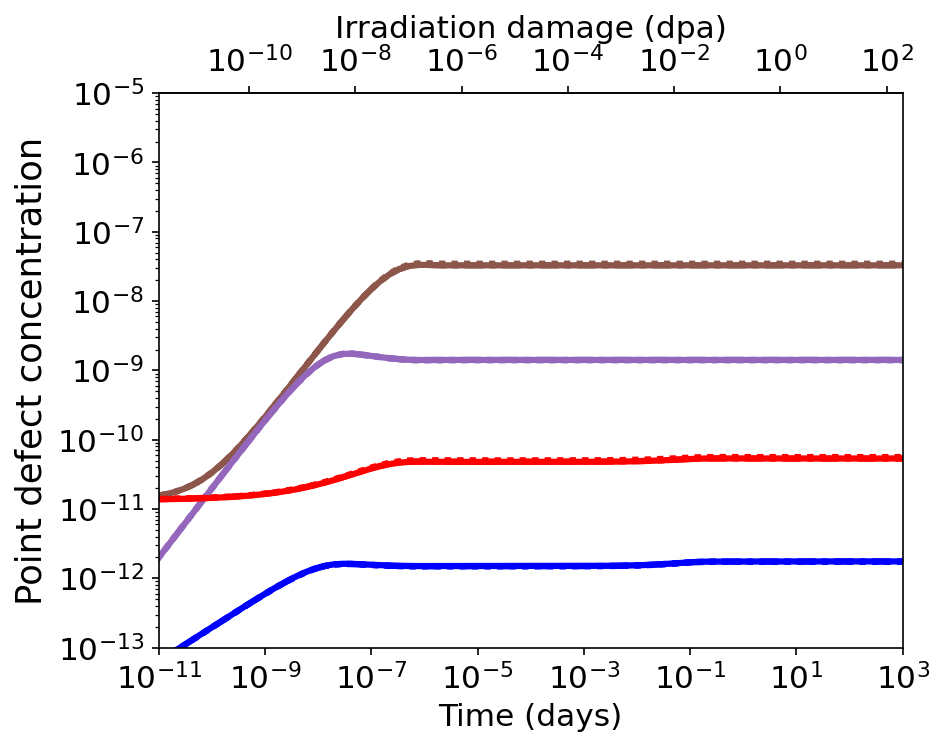}
    \caption{500\textdegree C}
    \end{subfigure}
    ~  
    \begin{subfigure}[t]{0.63\textwidth}
        \includegraphics[width=1\textwidth]{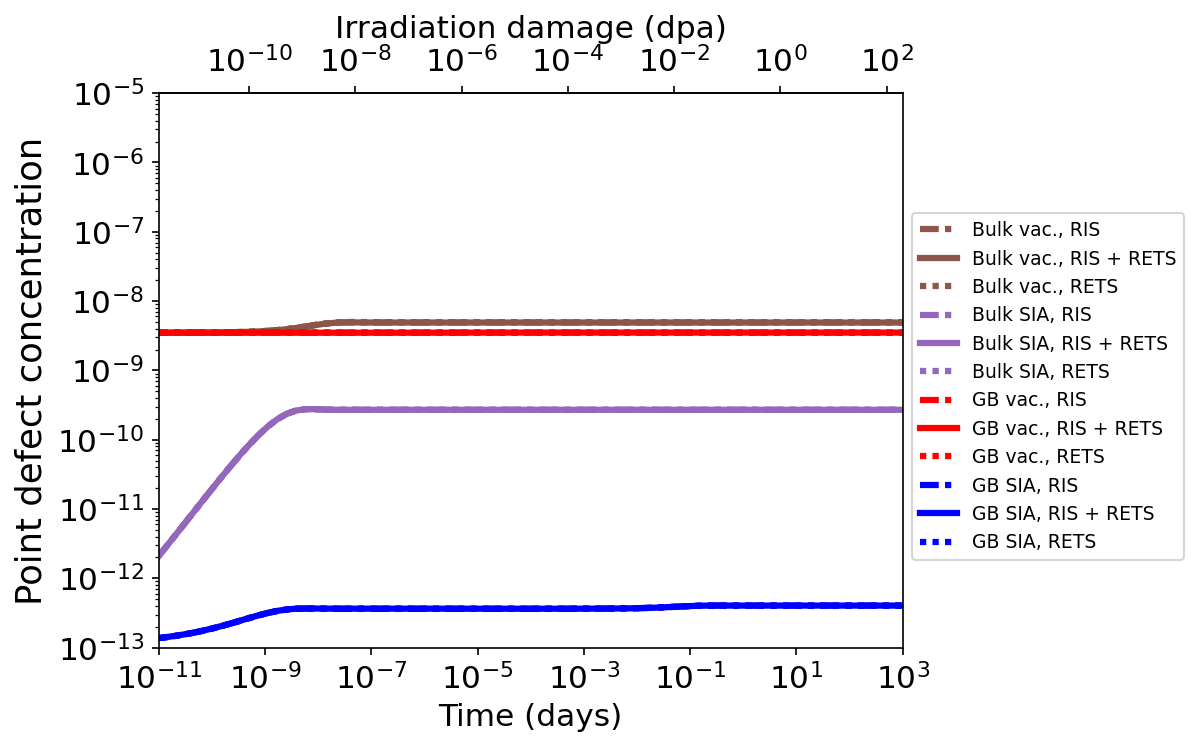}
    \caption{700\textdegree C}
    \end{subfigure}
    \caption{Evolution of bulk and GB concentrations of vacancy (vac.) and SIA in a 1D system of 1 $\mu$m grain size when under irradiation at $2\times10^{-6}$ dpa/s at different temperatures. For each temperatures, plots for the RIS, RETS and RIS+RETS mechanisms are found to be nearly identical. Results for 500\textdegree C corresponds to Fig.~\ref{fig:1D_conc_vs_time}.}
    \label{fig:defect_time_evol}
\end{figure}

\begin{figure}[htp!]
    \centering
    \begin{subfigure}[t]{0.48\textwidth}
        \includegraphics[width=1\textwidth]{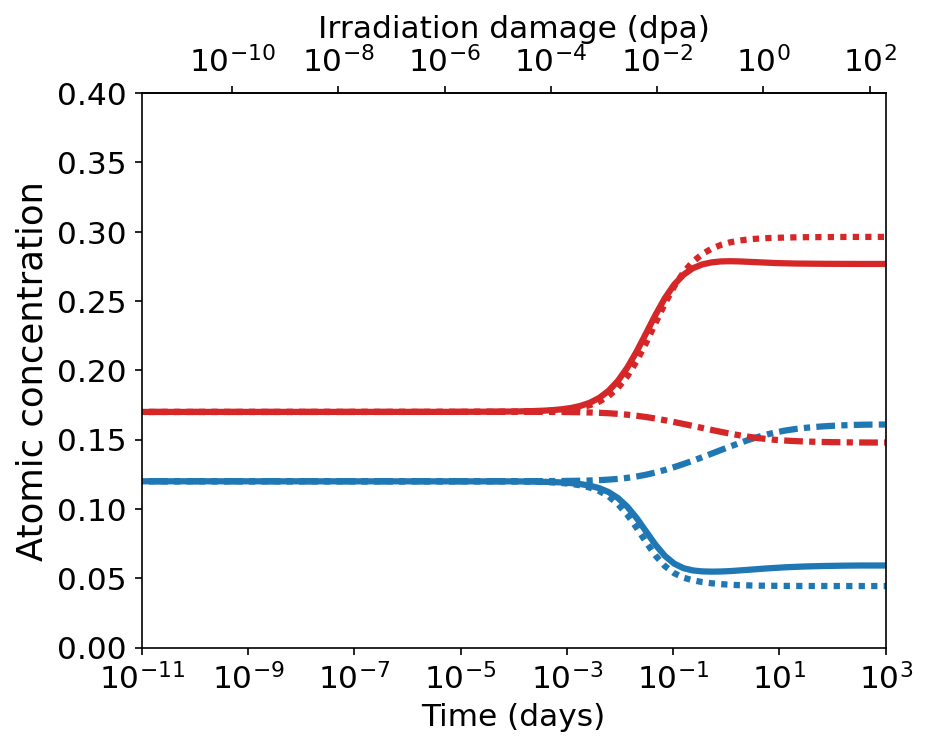}
    \caption{300\textdegree C}
    \end{subfigure}
    ~  
    \begin{subfigure}[t]{0.48\textwidth}
        \includegraphics[width=1\textwidth]{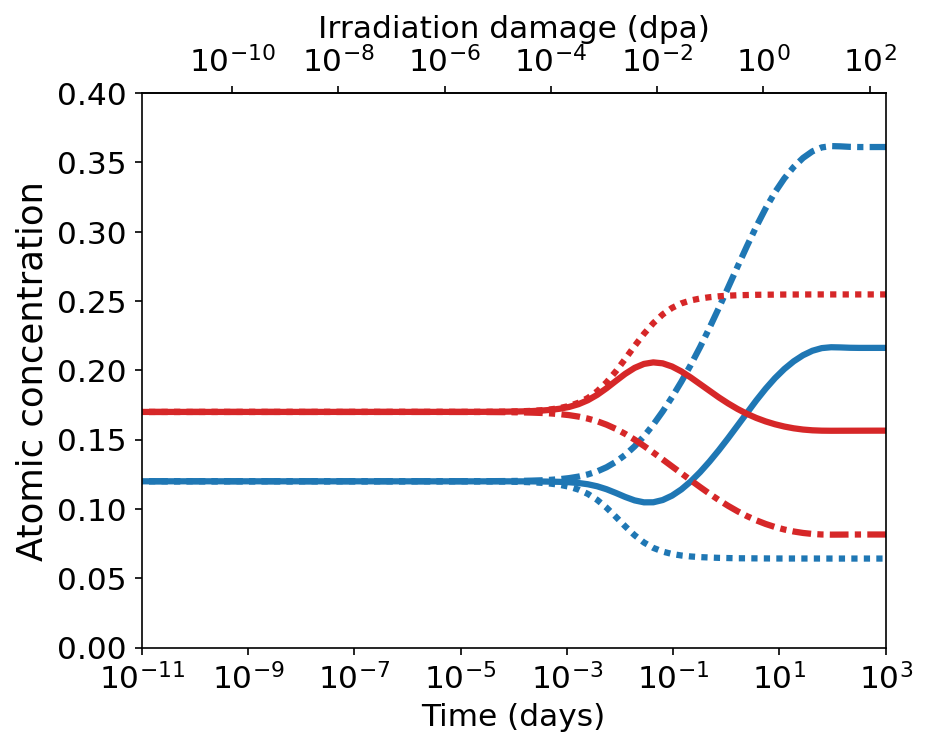}
    \caption{500\textdegree C}
    \end{subfigure}
    ~  
    \begin{subfigure}[t]{0.63\textwidth}
        \includegraphics[width=1\textwidth]{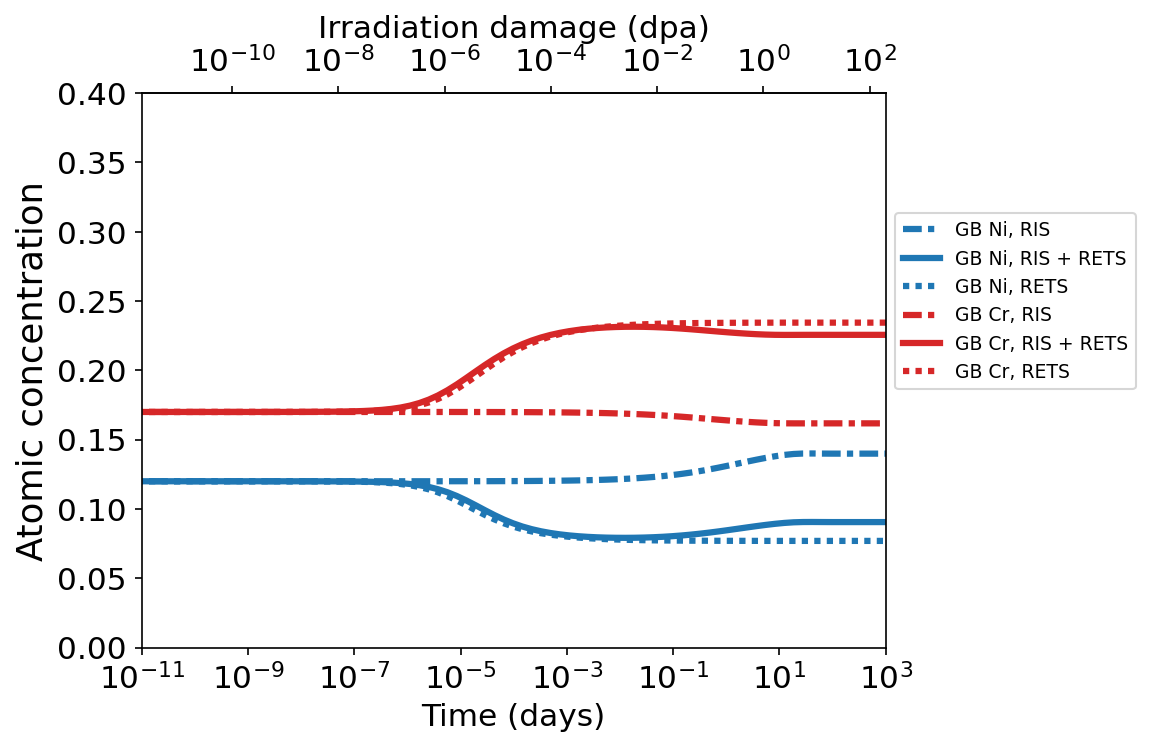}
    \caption{700\textdegree C}
    \end{subfigure}
    \caption{Evolution of bulk and GB concentrations of Ni and Cr corresponding to Fig.~\ref{fig:defect_time_evol}. Both RIS and RETS are seen to evolve after the point defect concentrations in Fig.~\ref{fig:defect_time_evol} reach a quasi-steady state determined by point defect generation, recombination, bulk dislocation absorption and GB sink. For 300\textdegree C and 700\textdegree C, RETS is seen to dominate, whereas for 500\textdegree C, RIS is seen to dominate. Atomic concentrations are seen to reach equilibrium RETS faster than steady-state RIS. Results for 500\textdegree C are identical to Fig.~\ref{fig:1D_conc_vs_time}.}
    \label{fig:atom_time_evol_compare}
\end{figure}
\begin{figure}[htp!]
    \centering
    \begin{subfigure}[t]{0.6\textwidth}
        \includegraphics[width=1\textwidth]{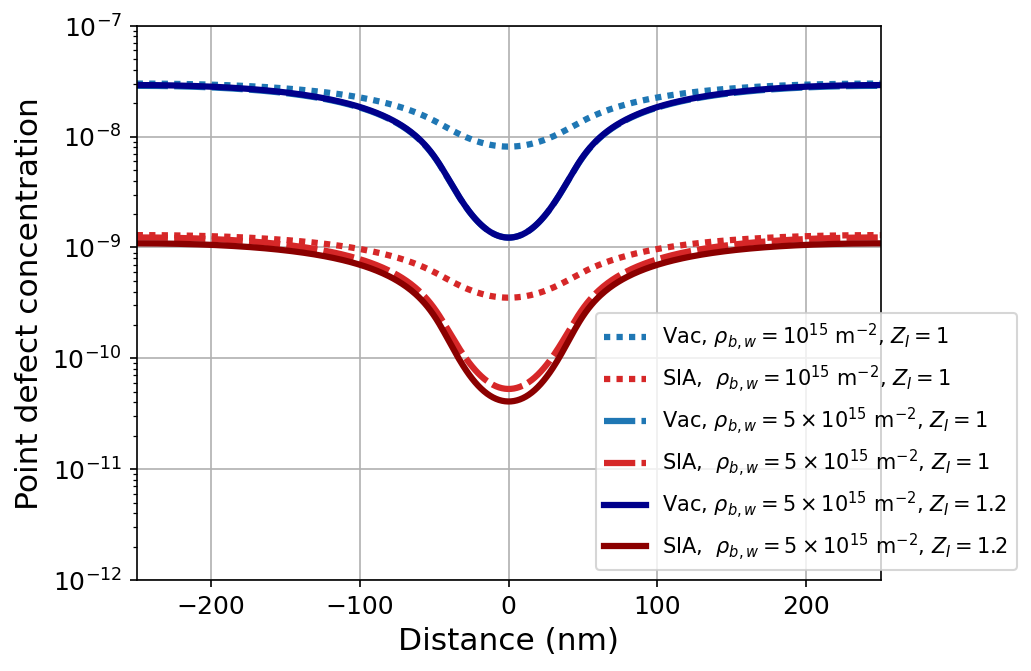}
    \end{subfigure}
    \caption{Steady-state point vacancy (vac.) and SIA concentrations in a 1D AM microstructure of 500 nm dislocation cell size irradiated at 500\textdegree C. The AM system corresponds to conditions similar to Fig.~\ref{subfig:1D_am_noic_profile}. Higher CW dislocation density $\rho_{b,w}$ results in significantly lower point defect concentrations at the CW. Biased absorption ($Z_I=1.2$) of SIA by dislocations results in a slightly lower SIA concentration.}
    \label{fig:1D_am_defect_profile_500C_500nm}
\end{figure}

\begin{figure}[htp!]
    \centering
    \begin{subfigure}[t]{0.48\textwidth}
        \includegraphics[width=1\textwidth]{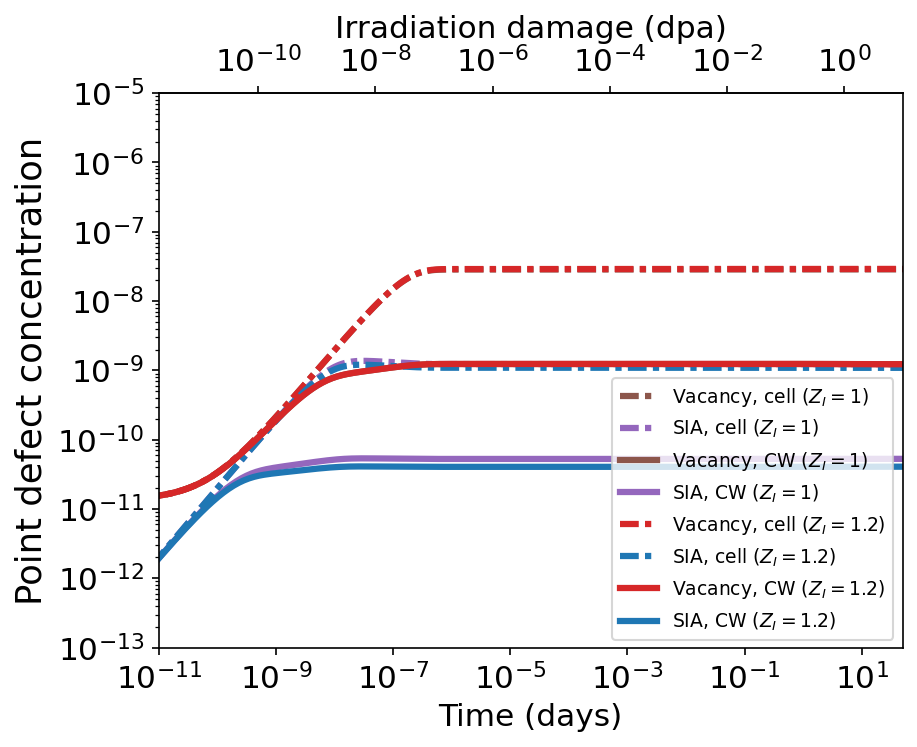}
    \caption{}
    \end{subfigure}
    ~
    \begin{subfigure}[t]{0.473\textwidth}
        \includegraphics[width=1\textwidth]{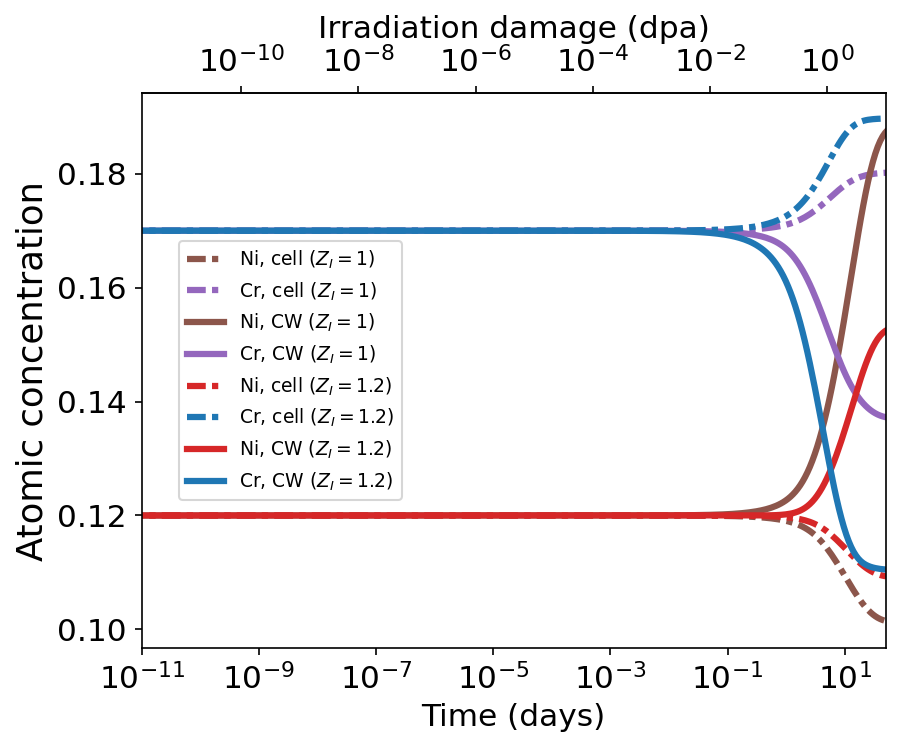}
    \caption{}
    \end{subfigure}
    \caption{Evolution of point defect concentrations at the CW and cell centers in the 1D AM microstructure of Fig.~\ref{fig:1D_am_defect_profile_500C_500nm}. Results are shown for different $Z_I$ and a CW dislocation density of $\rho_{b,w}=5\times10^{15}$ m$^{-2}$. }
    \label{fig:1D_am_defect_time_evaol_500C}
\end{figure}
\begin{figure*}[htp!]
    \centering
    \begin{subfigure}[t]{0.48\textwidth}
        \includegraphics[width=1\textwidth]{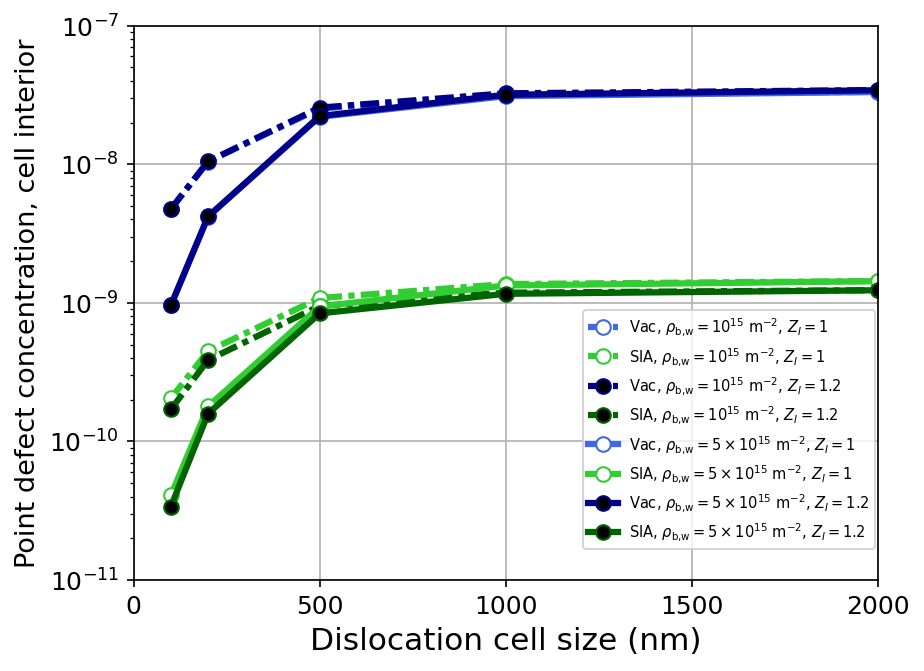}
    \caption{}
    \end{subfigure}
    \begin{subfigure}[t]{0.48\textwidth}
        \includegraphics[width=1\textwidth]{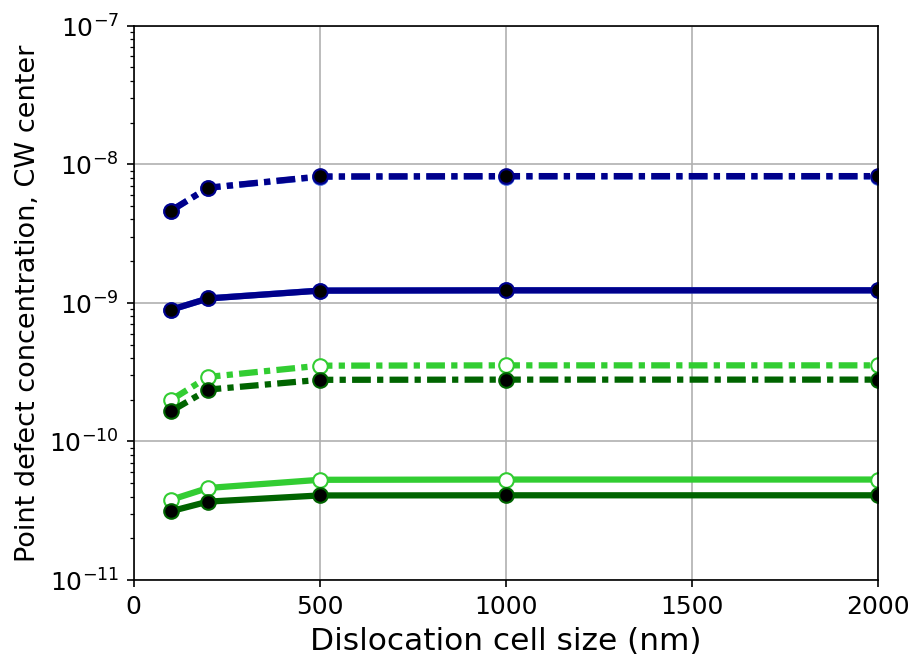}
    \caption{}
    \end{subfigure}
    \caption{Point defect concentrations at (a) the dislocation cell center and (b) the dislocation CW center as a function of dislocation cell size from RIS simulations of 1D AM microstructure corresponding to Fig.~\ref{fig:1D_CW_seg_vs_size}. Increasing the dislocation density $\rho_{b,w}$ or the absorption efficiency for SIA $Z_I$ for $\rho_{b,w}=5\times10^{15}$ m$^{-2}$ is seen to decrease the point defect concentration at the CW. Significant decrease in the concentrations at the cell interior is only observed for low dislocation cell sizes.}
    \label{fig:1D_ris_defect_vs_am_cell_size_500C}
\end{figure*}

\end{document}